\documentclass[twocolumn,aps,superscriptaddress,floatfix]{revtex4}
\usepackage{amsmath}
\usepackage{amssymb}
\usepackage{graphicx}
\usepackage{color}

\newcommand{\be}{\begin{equation}}
\newcommand{\ee}{\end{equation}}
\newcommand{\bea}{\begin{eqnarray}}
\newcommand{\eea}{\end{eqnarray}}
\newcommand{\bse}{\begin{subequations}}
\newcommand{\ese}{\end{subequations}}

\newcommand{\tcs}{${\rm ThCr_2Si_2}$}
\newcommand{\cca}{CaCo$_{2-y}$As$_2$}
\newcommand{\sca}{${\rm SrCo_2As_2}$}

\newcommand{\csca}{Ca$_{1-x}$Sr$_x$Co$_{2-y}$As$_2$}
\newcommand{\bsfa}{Ba$_{1-x}$Na$_x{\rm Fe_2As_2}$}
\newcommand{\bpfa}{Ba$_{1-x}$K$_x{\rm Fe_2As_2}$}
\newcommand{\ssfa}{Sr$_{1-x}$Na$_x{\rm Fe_2As_2}$}
\newcommand{\csfa}{Ca$_{1-x}$Na$_x{\rm Fe_2As_2}$}

\begin{document}

\title{Anomalous Composition-Induced Crossover in the Magnetic Properties of the Itinerant-Electron Antiferromagnet Ca$_{1-x}$Sr$_x$Co$_{2-y}$As$_2$}

\author{N. S. Sangeetha}
\affiliation{Ames Laboratory, Iowa State University, Ames, Iowa 50011, USA}
\author{V. Smetana}
\author{A.-V. Mudring}
\affiliation{Ames Laboratory, Iowa State University, Ames, Iowa 50011, USA}
\affiliation{Department of Materials Science and Engineering, Iowa State University, Ames, Iowa 50011, USA}
\author{D. C. Johnston}
%\altaffiliation{johnston@ameslab.gov}
\affiliation{Ames Laboratory, Iowa State University, Ames, Iowa 50011, USA}
\affiliation{Department of Physics and Astronomy, Iowa State University, Ames, Iowa 50011, USA}

\date{\today}

\begin{abstract}

The inference of Ying et al.\ [EPL {\bf 104}, 67005 (2013)] of a composition-induced change from $c$-axis ordered-moment alignment in a collinear A-type AFM structure (AFMI) at small~$x$ to $ab$-plane alignment in an unknown AFM structure (AFMII) at larger~$x$ in \csca\ with the body-centered tetragonal \tcs\ structure is confirmed.  Our major finding is an anomalous magnetic behavior in the crossover region $0.2\lesssim x \lesssim 0.3$ between these two phases.  In this region the magnetic susceptibility versus temperature~$\chi_{ab}(T)$ measured with  magnetic fields~{\bf H} applied in the $ab$~plane exhibit typical AFM behaviors with cusps at the N\'eel temperatures of \mbox{$\sim 65$~K}, whereas $\chi_c(T)$ and the low-temperature isothermal magnetization $M_c(H)$ with {\bf H} aligned along the $c$~axis exhibit extremely soft ferromagnetic-like behaviors.

\end{abstract}

\maketitle

Much research since 2008 has focused on studies of iron-based layered pnictides and chalcogenides due to their unique lattice, electronic, magnetic and superconducting properties \cite{Johnston2010, Stewart2011, Dagotto2013, Fernandes2014, Dai2015, Si2016}.  An important family of these materials is comprised of doped and undoped body-centered tetragonal parent compounds $A{\rm Fe_2As_2}$ ($A$ = Ca, Sr, Ba, Eu) with the  \tcs-type structure  (122-type compounds).  The undoped and underdoped $A{\rm Fe_2As_2}$ compounds exhibit a tetragonal to orthorhombic distortion of the crystal structure at $T_{\rm S} \lesssim 200$~K\@.  They also exhibit itinerant collinear antiferromagnetic (AFM) spin-density-wave ordering at a temperature~$T_{\rm N}$ the same or slightly lower than $T_{\rm S}$.  The ordered moments in the stripe AFM structure of the orthorhombic phase are oriented in the $ab$~plane.  In 2014 a temperature~$T$-induced AFM spin-reorientation transition to a new AFM $C_4$ phase was discovered in the hole-underdoped \bsfa\ system upon cooling below $T_{\rm N}$ that can coexist with superconductivity \cite{Avci2014}.  A subsequent investigation by polarized and unpolarized neutron diffraction determined that the ordered moments in the new phase are aligned along the $c$~axis instead of along the $ab$~plane as in the stripe AFM structure \cite{Wasser2015}. The work on the \bsfa\ system was followed by the observation of the same AFM $C_4$ phase in the \bpfa\ \cite{Bohmer2015, Mallett2015a, Mallett 2015b}, \ssfa\ \cite{Allred2016, Taddei2016}, and \csfa\ \cite{Taddei2017} systems.  These results are important to understanding the mechanism of superconductivity and other aspects of the hole-doped iron arsenides \cite{Kang2015, Gastiasoro2015, Christensen2015, Fernandes2016, Hoyer2016, Scherer2016, Christensen2016}.

A similar but composition-induced moment realignment was suggested to occur in the isostructural  CoAs-based system \csca\ \cite{Ying2013}.  \cca\ has a so-called collapsed-tetragonal (cT) structure with As--As bonding along the $c$~axis between adjacent CoAs layers \cite{Anand2012} and has $\approx 7$\% vacancies on the Co sites \cite{Quirinale2013, Anand2014}.  It exhibits itinerant A-type AFM ordering below $T_{\rm N} = 52$--77~K, depending on the sample, with the ordered moments of $\approx 0.3$--0.4$~\mu_{\rm B}$/Co ($\mu_{\rm B}$ is the Bohr magneton) within an $ab$-plane Co layer aligned ferromagnetically (FM) along the $c$~axis and with AFM alignments between moments in adjacent Co planes  \cite{Quirinale2013, Anand2014, Ying2012, Cheng2012,  Jayasekara2017}.  The dominant interactions are found to be FM from the positive Weiss temperature obtained by fitting the magnetic susceptility $\chi$ versus~$T$ measurements above~$T_{\rm N}$ by the Curie-Weiss law \cite{Ying2012, Cheng2012, Anand2014}, indicating that interplane AFM interactions responsible for the A-type AFM ordering are much weaker than the intraplane FM ones.  Electronic structure calculations are consistent with the itinerant A-type AFM ground state and the small ordered moment \cite{Korotin2015}.  Spin-flop transitions in single crystals of \cca\ with the magnetic field~$H$ applied parallel to the $c$~axis occur at $H_{\rm SF}\sim 3.5$--3.7~T \cite{Ying2012, Anand2014, Cheng2012b, Zhang2015}.  On the other hand, metallic \sca\ exhibits no magnetic transitions versus~$T$ \cite{Pandey2013}.  However, single-crystal $\chi(T)$ data for both $c$-axis and $ab$-plane magnetic field alignments show broad maxima at $T\approx 115$~K, suggesting the presence of dynamic AFM correlations.  Indeed, inelastic neutron scattering measurements on single crystals revealed strong AFM correlations with the same stripe wave vector as seen in the parent and doped Fe-based $A{\rm Fe_2As_2}$ compounds \cite{Jayasekara2013}.

In view of the A-type AFM of \cca\ where strong FM correlations dominate and the contrasting strong AFM correlations in paramagnetic (PM) \sca\ detected by neutron scattering, studies of the magnetic properties of \csca\ crystals have the potential to reveal additional interesting physics. This system is metallic with the \tcs-type structure over the entire composition range $0\leq x\leq 1$ \cite{Ying2013}. On the basis of magnetization versus magnetic field $M(H)$ and electrical resistivity measurements on \csca\ crystals, the authors of Ref.~\cite{Ying2013} inferred A-type $c$-axis AFM for $0\leq x \lesssim 0.15$ (AFMI), a $c$-axis FM phase for $0.15\lesssim x\lesssim 0.27$, $ab$-plane AFM ordering for $0.27\lesssim x \lesssim 0.42$ (AFMII), and a PM phase for $0.42\lesssim x\leq 1$, with one crystal defining each of the FM and AFMII phase regions.  A first-order transition from the cT to the uncollapsed-tetragonal (ucT) structure at $x\approx 0.4$ was suggested from the variation in the $c$-axis lattice parameter with~$x$, which the authors suggested was important to the evolution of the magnetic structure with~$x$ \cite{Ying2013}.

Here we present a detailed study of the magnetic properties and phase diagram of the \csca\ system that was carried out using ten single crystals with Sr compositions in the range $0\leq x\leq 0.52$.  For crystals with $x$ in the intermediate crossover regime $0.2\lesssim x \lesssim 0.3$ between the $c$-axis and $ab$-plane AFM moment orientations of the AFMI and AFMII phases, respectively, we discovered that $\chi_{ab}(T)$ exhibits a cusp at the magnetic ordering temperature typical of an $ab$-plane AFM transition, whereas the response of $\chi_c(T)$ and $M_c(H)$ to a $c$-axis field is extremely soft, suggesting instead a $c$-axis FM structure in this region as proposed in Ref.~\cite{Ying2013}.  This divergence between the magnetic responses in the two field directions is highly anomalous.  We also established that the moment realignment between the AFMI and AFMII phases results from a continuous composition-induced evolution of the magnetocrystalline anisotropy field from $c$-axis to $ab$-plane orientations; and that the structural parameters versus~$x$, $T_{\rm N}(x)$ [defined as the cusp temperature in $\chi_{ab}(T)$],  the low-$T$ saturation moment $\mu_{\rm sat}(x)$, and the effective moment $\mu_{\rm eff}(x)$ in the PM state at $T >T_{\rm N}$ all vary continously with~$x$ for $0\leq x \leq 0.45$.  Details about the crystal growth together with elemental analyses and single-crystal structure determinations are given in Ref.~\cite{SupplMat}.  $M(H,T)$ and $C_{\rm p}(T)$ data were obtained using a Quantum Design Magnetic Properties Measurement System and Physical Properties Measurement System, respectively.

\begin{figure}
\includegraphics[width=4in]{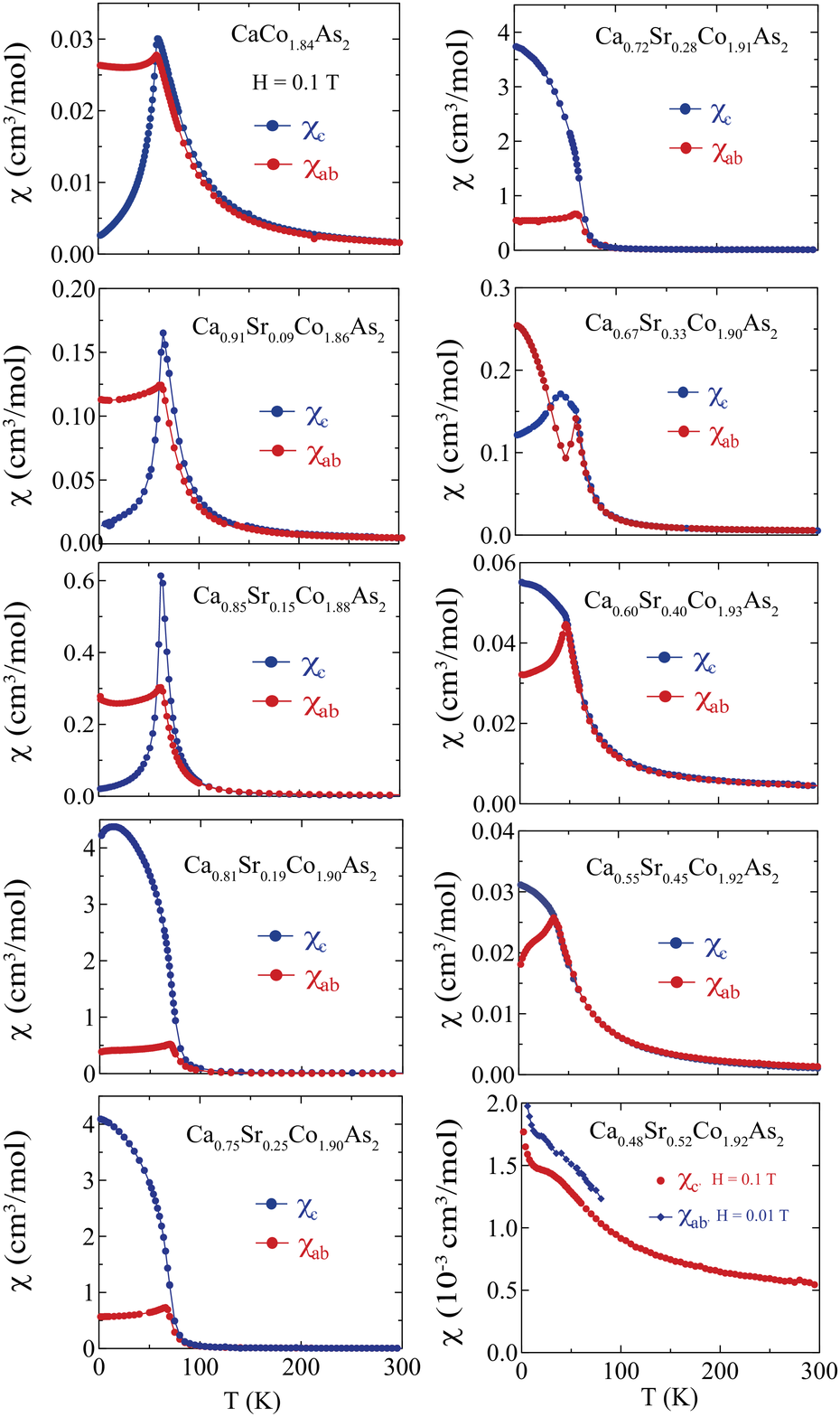} 
\caption{(Colour online)  Zero-field-cooled magnetic susceptibility $\chi$ of Ca$_{1-x}$Sr$_x$Co$_{2-y}$As$_2$ crystals versus $T$ in $H$ = 0.1~T applied in the $ab$~plane ($\chi_{ab}$) and along the $c$~axis ($\chi_c$).  Note the large changes in the ordinate scales versus~$x$.}
\label{Fig:SrxCa1-x_MT_1KOe}
\end{figure}

\begin{figure}
\includegraphics[width=3.4in]{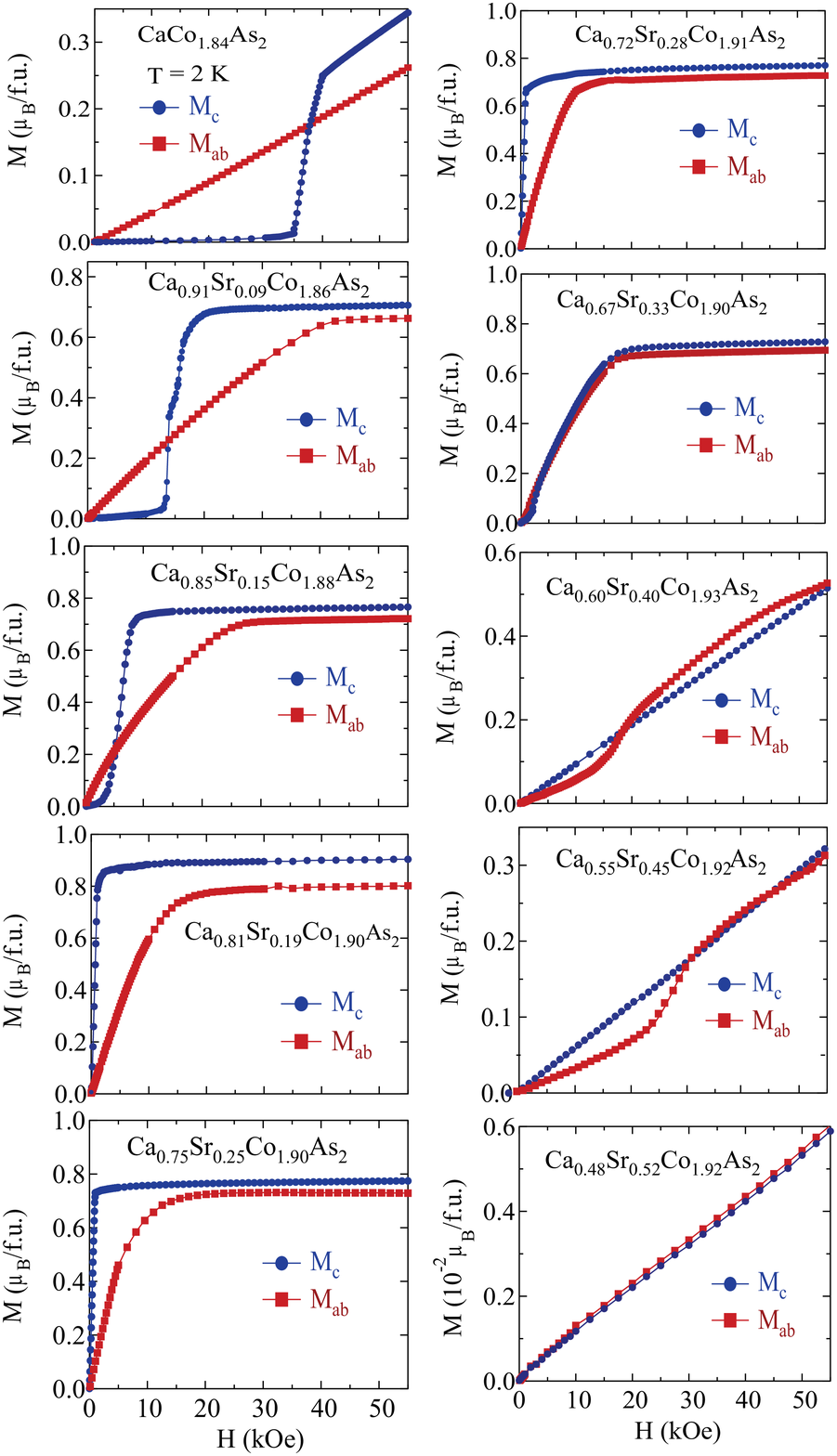} 
\caption{(Colour online)  $M(H)$ data for \csca\ crystals at $T = 2$~K for $H$ applied in the $ab$~plane ($M_{ab}$) and along the $c$~axis ($M_c$).}
\label{Fig:SrxCa1-x_MH_2K}
\end{figure}

Shown in Fig.~\ref{Fig:SrxCa1-x_MT_1KOe} are zero-field-cooled $\chi(T)\equiv M(T)/H$ data in $H=0.1$~T applied both parallel and perpendicular to the $c$~axis for $0\leq x\leq 0.52$.  For $x=0$ to 0.15 one sees that $\chi_c\ll\chi_{ab}$ for $T\ll T_{\rm N}\approx 63$~K, suggesting A-type collinear $c$-axis ordering as already established for $x=0$. Here $\chi_{ab}$ exhibits a cusp at $T\equiv T_{\rm N}$ and is approximately independent of~$T$ below $T_{\rm N}$, as expected for an AFM transition.  Even though $T_{\rm N}$ is nearly independent of~$x$ in this region, the magnetic system becomes very soft against $c$-axis fields as reflected by the sharp peak in $\chi_c$ for $x\leq 0.15$ that appears to be diverging with increasing~$x$.  This is also reflected by the order of magnitude increase in the maximum $\chi_c$ from $x=0$ to $x = 0.19$ in Fig.~\ref{Fig:SrxCa1-x_MT_1KOe}.  For $x = 0.33$ a $T$-induced spin-reorientation transition appears to occur at about 50~K\@. This plot is similar to that for $x=0.34$ reported in Ref.~\cite{Ying2013}. For $x = 0.40$ and~0.45, the $\chi$ values at low~$T$ decrease to small values, and the $\chi(T)$ data suggest $ab$-plane AFM ordering.  Finally, for $x=0.52$, a PM behavior is observed. As a point of reference, the magnetic dipole interaction between local moments on a simple-tetragonal lattice with $c/a \approx 3.65$ as in \cca\ predicts that the ordered moments should lie in the $ab$~plane \cite{Johnston2016}.

To clarify the origins of the behaviors in Fig.~\ref{Fig:SrxCa1-x_MT_1KOe} and the magnetic ground state versus~$x$, we carried out isothermal $M(H)$ measurements on the same set of ten crystals at $T=2$~K and the results are shown in Fig.~\ref{Fig:SrxCa1-x_MH_2K}.  As expected for A-type $c$-axis collinear AFM ordering for small~$x$ \cite{Quirinale2013}, the crystals with $x=0$, 0.09, and 0.15 show clear evidence for first-order field-induced spin-flop (SF) transitions for $H\parallel c$, where the SF fields decrease from $H_{\rm SF} \approx 3.7$~T for $x=0$ to $H_{\rm SF} \approx 0.6$~T for $x=0.15$.  The measured saturation moment  $\mu_{\rm sat}$ values at high fields are in the narrow range of $\approx 0.35$ to $\approx 0.42~\mu_{\rm B}$/Co for $0.09\leq x \leq 0.33$.  Additional magnetic data in Ref.~\cite{SupplMat} show that the effective moment $\mu_{\rm eff}$ per Co atom in the PM state is also nearly independent of $x$ for $0\leq x \leq 0.45$.

\begin{figure}
\includegraphics[width=3.in]{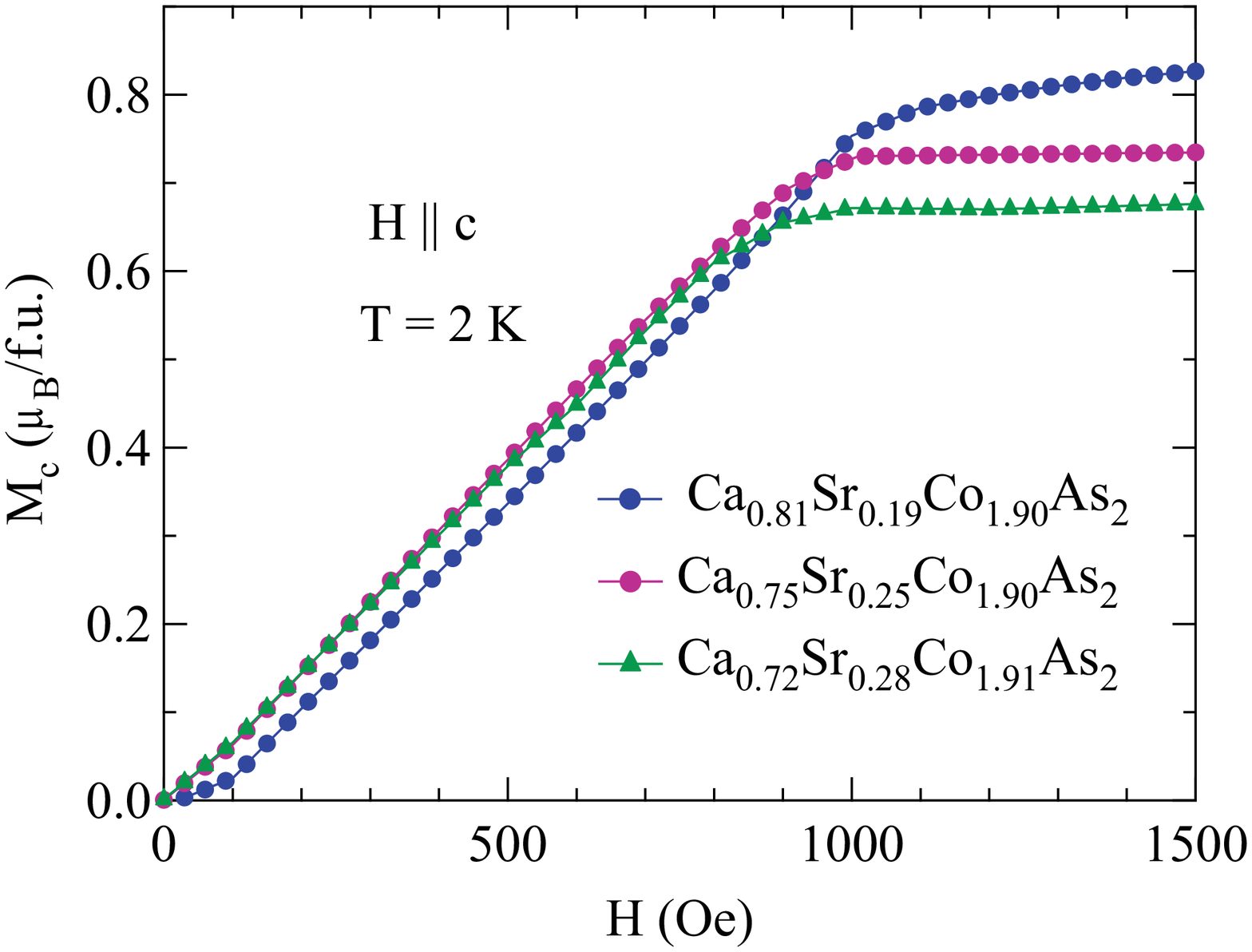}
\caption{(Colour online)  $M_c(H)$ data for $H\leq 1500$~Oe at $T=2$~K for three crystals of \csca.}
\label{Fig:MH_2K_Sr19_Sr25_expand}
\end{figure}

Figure~\ref{Fig:SrxCa1-x_MH_2K} indicates that at 2~K, $H_{\rm SF} \approx 0$ for $x=0.19$, 0.25 and~0.28, where $T_{\rm N}$ from Fig.~\ref{Fig:SrxCa1-x_MT_1KOe} and Ref.~\cite{SupplMat} decreases  continuously from 71 to 63~K over this composition range.  At the same time, the slope of $M_c(H)$ at low fields for these compositions remains high.  To examine this low-field dependence in more detail, expanded plots of $M_c(H)$ for $0\leq H\leq 1500$~Oe are shown in Fig.~\ref{Fig:MH_2K_Sr19_Sr25_expand}.  For $x=0.19$ a weak SF transition is detected at $H_{\rm SF} \approx 100$~Oe.  One also sees in Fig.~\ref{Fig:MH_2K_Sr19_Sr25_expand} that the rapid increase in $M_c$ with $H$ at low fields for $H\parallel c$ in Fig.~\ref{Fig:SrxCa1-x_MH_2K} is linear in $H$ with nearly the same slope for the three compositions.  This effect is expected if the slope is limited by demagnetization effects.  The dimensionless volume susceptibility of the $M_c(H)$ data in Fig.~\ref{Fig:MH_2K_Sr19_Sr25_expand} up to $H=800$~Oe for all three compositions is $\chi_c \equiv dM_c/dH\approx 0.077$.  This value is about the same as the value $\chi_c=1/4\pi\approx 0.080$ expected if $M_c/H$ in the absence of demagnetization effects is so large  that the observed value is limited by the demagnetization factor~\cite{Johnston2016}. Hence the magnetic response along the $c$~axis is extremely soft for $x=0.19$, 0.25 and~0.28, which may be expected if these compositions are in a crossover region in the magnetocrystalline anisotropy field from being parallel to perpendicular to the $c$~axis.  The rapid growth with increasing~$x$ of the sharp peak in $\chi_c(T_{\rm N})$ for $x = 0.09$ and~0.15 in Fig.~\ref{Fig:SrxCa1-x_MT_1KOe} is an additional indication of the growth of strong FM fluctuations.

\begin{figure}
\includegraphics[width=3.in]{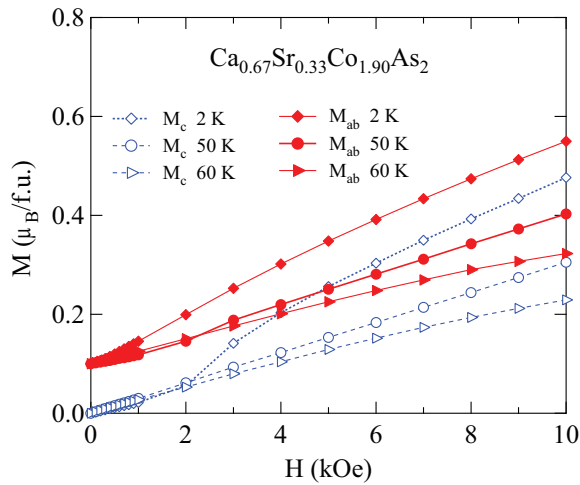} 
\caption{(Colour online)  Magnetization~$M$ versus applied magnetic field~$H$ for fields aligned along the $c$~axis (open symbols) and $ab$~plane (filled symbols), each at 2~K, 50~K and~60~K\@. The $ab$-plane data are offset vertically by $0.1~\mu_{\rm B}$/f.u.\ for clarity.}
\label{Fig:MH33}
\end{figure}

The crystal with Sr composition $x=0.33$ is unique among the ten \csca\ crystals, because it exhibits a $T$-induced spin-reorientiation transition as revealed by the $\chi$ data for this composition in Fig~\ref{Fig:SrxCa1-x_MT_1KOe}.  On cooling below its $T_{\rm N} = 60$~K, the $\chi$ anisotropy indicates that the ordered moments are oriented within the $ab$~plane.  But then on further cooling to the range $T = 45$ to~50~K, the ordered moment direction switches from the $ab$~plane to the $c$~axis and remains so down to 2~K\@.  This spin-reorientation transition is confirmed by the anisotropy of the magnetization isotherms in Fig.~\ref{Fig:MH33}, which show a weak metamagnetic transition at about 2.5~kOe at 50~K for {\bf H} aligned in the $ab$~plane, and instead a spin-flop transition at $H_{\rm SF} \approx 2.4$~kOe at 2~K for {\bf H} aligned along the $c$~axis.

The $M_c(H)$ behavior for $x=0.2$ in Ref.~\cite{Ying2013} is about the same as we see for $x=0.19$, 0.25 and~0.28 in Fig.~\ref{Fig:SrxCa1-x_MH_2K}, which those authors interpreted as a $c$-axis FM region of the phase diagram.  However, the $\chi_{ab}(T)$ behaviors in Fig.~\ref{Fig:SrxCa1-x_MT_1KOe} for $x=0.19$, 0.25 and~0.28 instead suggest that the magnetic structure is AFM in this region.  The magnetic behaviors of ${\rm ThCo_{1.2}Cu_{0.8}Sn_2}$ with a different crystal structure \cite{Rosa2015} are similar to our data for \csca\ in the crossover regime with $x\approx 0.2$ to~0.3, but no discussion or interpretation of the data were given.

At larger $x$ values of 0.40 and~0.45, metamagnetic transitions occur at 2~K with $H$ applied in the $ab$~plane instead of along the $c$~axis which indicates $ab$-plane AFM ordering (AFMII) as suggested for $x = 0.34$ in Ref.~\cite{Ying2013}.  We interpret this as resulting from a crossover in the anisotropy field from being parallel to the $c$~axis for $0\leq x \lesssim 0.2$ to perpendicular to the $c$~axis for $0.35\lesssim x \lesssim 0.45$.  If this ordering is collinear, one expects AFM domains to occur in the $ab$~plane with orthogonal easy axes, corresponding to an extrinsic noncollinear AFM structure.  The $\chi_{ab}(T)$ data for $x = 0.40$ and 0.45 in Fig.~\ref{Fig:SrxCa1-x_MT_1KOe}, which remain relatively large with respect to $\chi_c(T)$ at $T\ll T_{\rm N}$, suggest that the AFM structure in this $x$~range is either extrinsically or intrinsically noncollinear \cite{Johnston2012, Johnston2015, Sangeetha2016}.  In either case, the high-field $M_{ab}(H)$ behavior is not straightforward to interpret.  Therefore we denote the metamagnetic transition field for $x= 0.40$ and~0.45 as $H_{\rm mm}$ instead of $H_{\rm SF}$.  One sees from Fig.~\ref{Fig:SrxCa1-x_MH_2K} that $H_{\rm mm}$ increases from $\approx 1.7$~T for $x=0.40$ to  $\approx 2.8$~T for $x=0.45$ before becoming irrelevant in the PM state for $x \geq 0.52$.

We have also carried out  $C_{\rm p}(T)$ measurements on the \csca\ crystals in $H=0$ \cite{SupplMat}.  As in the $C_{\rm p}(T)$ measurements for $x=0$ in Ref.~\cite{Anand2014}, we see no distinct anomalies in $C_{\rm p}$ at $T_{\rm N}$ for any of the \csca\ crystals. The low-$T$ data indicate substantial Sommerfeld coefficients~$\gamma$, suggesting enhanced densities of states at the Fermi energy.

\begin{figure}
\includegraphics[width=3.3in]{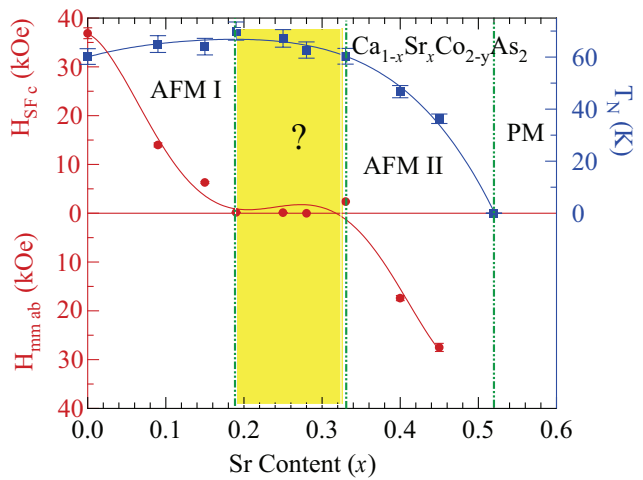} 
\caption{(Colour online)  Magnetic phase diagram of the \csca\ system versus~$x$ at $T=2$~K, containing the $c$-axis AFMI (A-type), $ab$-plane AFMII and PM phases.  The crossover region in yellow and the AFMII phase have unknown structures.  Plotted are the spin-flop field $H_{\rm SF}$ with $H\parallel c$ for $x<0.2$ and the metamagnetic transition field $H_{\rm mm}$ with $H\parallel ab$ for $x>0.3$ (left-hand ordinates). Also shown is $T_{\rm N}(x)$ (right-hand ordinate), defined as the cusp temperatures in the $\chi_{ab}(T)$ data in Fig.~\ref{Fig:SrxCa1-x_MT_1KOe}.}
\label{Fig:phase_diagram}
\end{figure}

The ground-state magnetic phase diagram of the \csca\ system derived from the measurements at $T=2$~K of $H_{\rm SF}$ and $H_{\rm mm}$ versus~$x$ is shown in Fig.~\ref{Fig:phase_diagram}, where $T_{\rm N}(x)$ is also plotted.  The AFMI phase for $0\leq x \lesssim 0.2$ is an A-type AFM with $c$-axis moment alignment.  The AFM structure in the anisotropy crossover region $0.2\lesssim x \lesssim 0.3$ is unknown.  Then an AFMII phase with unknown AFM structure with the ordered moments aligned in the $ab$ plane occurs for $x=0.40$ and~0.45, followed by a PM region for $x \geq 0.52$.  From the $\chi(T)$ data for $x=0.40$ and~0.45 in Fig.~\ref{Fig:SrxCa1-x_MT_1KOe} we infer that the AFMII structure is intrinsically noncollinear or else extrinsically noncollinear due to multiple AFM domains aligned in the $ab$~plane.  Since $T_{\rm N}$, $\mu_{\rm sat}$ and $\mu_{\rm eff}$ over the region $0.2\lesssim x \lesssim 0.3$ do not change appreciably \cite{SupplMat}, one might infer that a continuous tilting of the ordered moment from $c$-axis to $ab$-plane orientation occurs over this region.  However, this seems unlikely since the magnetic data for this region do not provide clear evidence for it and the magnetism is itinerant.

In summary, the magnetic and thermal properties of \csca\ single crystals with $0\leq x\leq 0.52$ were studied and the magnetic phase diagram at \mbox{$T=2$~K} was constructed.  We confirm a collinear $c$-axis AFM phase at small~$x\lesssim 0.2$ and an $ab$-plane AFM phase at large $x = 0.40$ and~0.45 \cite{Ying2013}.  Our major result is the observation of AFM-like transitions from $\chi_{ab}(T)$ coexisting with strong FM-like $c$-axis correlations from $\chi_c(T)$ and $M_c(H)$ data in the crossover region $0.2\lesssim 0.3$ between these two phases, an anomalous dichotomy. We also find continuous evolutions with~$x$ from $x=0$ to~0.52 of the crystal structure,  of the anisotropy field at $T=2$~K which results in the composition-induced ordered-moment reorientation between the $c$-axis AFMI and $ab$-plane AFMII phases, and of the ordered and effective moments \cite{SupplMat}.  An important feature of this system is that it is isoelectronic and isostructural over the entire composition range $0\leq x \leq 1$, which may simplify theoretical analyses.  The itinerant-electron AFM \csca\ system provides fertile ground for additional experimental and theoretical investigations that may also shed light on the origin of the temperature-induced ordered-moment realignments of the AFM phases in the hole-underdoped 122-type FeAs-based high-$T_{\rm c}$ superconductors.  

\acknowledgments

We thank Vivek Anand for his contributions to the early stages of this work.  Helpful discussions with Andreas Kreyssig, Robert McQueeney, and Makariy Tanatar are gratefully acknowledged.  This research was supported by the U.S. Department of Energy, Office of Basic Energy Sciences, Division of Materials Sciences and Engineering.  Ames Laboratory is operated for the U.S. Department of Energy by Iowa State University under Contract No.~DE-AC02-07CH11358.

\clearpage

\begin{widetext}

\appendix*

\section{Supplemental Information}

\section*{Single-Crystal Growth}

Single crystals of  \csca\ with nominal compositions ($x$ = 0, 0.1, 0.2, 0.3, 0.4, 0.5, 0.6 and 0.8) were grown using Sn flux. The starting materials were high-purity elemental Ca (99.98\%), Co (99.998\%), As (99.99999\%) and Sn (99.999\%) from Alfa Aesar, and Sr (99.95\%) from Sigma-Aldrich. The sample and Sn flux taken in a 1:5 molar ratio were placed in an alumina crucible that was sealed under $\approx$ 1/4 atm high purity argon in a silica tube. The sealed samples were preheated at 600~$^{\circ}$C for 5~h, and then heated to 1150~$^{\circ}$C at the rate of 50~$^{\circ}$C/h and held there for 20~h for homogenization. Then the furnace was slowly cooled at the rate of 2.5~$^{\circ}$C/h to 700~$^{\circ}$C\@. The single crystals were separated by decanting the molten Sn flux with a centrifuge at that temperature.  Large (4$-$6~mm size) shiny platelike single crystals were obtained from each growth.  Energy-dispersive x-ray spectoscopy (EDS) was used to determine the compositions of the crystals. We selected several crystals from each batch and performed EDS measurments on both sides of each crystal. We observed a significant inhomogeneity in each crystal except for CaCo$_{1.84}$As$_2$. Hence, we cleaved each crystal and performed EDS mesurements on both sides of each cleaved crystal until we obtained homogeneous crystals. The result was that the crystals studied in the main text are small platelike crystals with masses of order only a few mg.  Measurements such as neutron diffraction requiring much larger samples were therefore not possible.  The Sr content~$x$ and Co~content~$2-y$ in the homogeneous \csca\ crystals were determined by EDS and single-crystal x-ray structural analysis below assuming the As site was fully occupied. The results of the composition analyses are given in Table~\ref{Tab:ChemAnal} below.  The same crystals were used to perform the physical property measurements reported in the main text and in the sections below.

%\clearpage

\section*{\label{Sec:EDSXRD} Energy-Dispersive X-ray Spectroscopy and Single-Crystal Structure Determinations}

Chemical analysis of the \csca\ crystals was performed using a JEOL scanning electron microscope (SEM), equipped with an energy dispersive x-ray spectroscopy (EDS) analyzer. The compositions of each side of a platelike crystal were measured at about ten positions on each face, and the results averaged.  If the compositions on the two sides were significantly different, the crystal was cleaved and the measurements repeated until the compositions on both sides were the same.

Single-crystal X-ray diffraction (XRD) measurements were performed at room temperature on a Bruker D8 Venture diffractometer operating at 50~kV and 1~mA equipped with a Photon 100 CMOS detector, a flat graphite monochromator and a Mo~K$\alpha$ I$\mu$S microfocus source ($\lambda = 0.71073$~\AA). The raw frame data were collected using the Bruker APEX3 program [40], while the frames were integrated with the Bruker SAINT software package [41] using a narrow-frame algorithm integration of the data and were corrected for absorption effects using the multiscan method (SADABS) [42]. The occupancies of the Ca\/Sr and Co atomic sites were refined assuming random occupancy of the Ca\/Sr sites and are presented below assuming complete occupancy of the As site.  The atomic thermal factors were refined anisotropically.  Initial models of the crystal structures were first obtained with the program SHELXT-2014 [43] and refined using the program SHELXL-2014 [44] within the APEX3 software package.

The chemical compositions obtained from the EDS and XRD analyses for our crystals with $0\leq x \leq 0.52$ are listed in Table~\ref{Tab:ChemAnal}, and the crystal data are given in Table~\ref{CrystalData}.  In the latter table the compositions listed in the first column are duplicated from Table~\ref{Tab:ChemAnal}, together with additional such data for $x=0.74$ and~0.87 for which we did not carry out physical property measurements.  The crystal data for ${\rm SrCo_2As_2}$ were obtained from Ref.~[32].  

\begin{table*}
\caption{\label{Tab:ChemAnal} Composition analysis results obtained using energy-dispersive x-ray spectroscopy (EDS) and single-crystal x-ray diffraction (XRD) measurements of the crystal structure of the ten crystals studied in the main text.  The last column gives the compositions of the ten crystals quoted in the main paper, together with the error bars on the Sr and Co concentrations obtained from the combined EDS and XRD data.}

\begin{ruledtabular}
\begin{tabular}{ccccccc}

\multicolumn{3}{c}{Atomic Ratio from XRD} & \multicolumn{3}{c}{Atomic Ratio from EDS} &  Composition  \\

Ca & Sr & Co & Ca & Sr & Co\\

\hline

1& 0 & 1.82(2) & 1& 0 & 1.86(2) & CaCo$_{1.84(3)}$As$_2$\\  
0.92(2)& 0.08(2) & 1.84(2) & 0.90(1)& 0.10(1) & 1.88(2)  & Ca$_{0.91(2)}$Sr$_{0.09(2)}$Co$_{1.86(3)}$As$_2$ \\
0.83(2)&0.17(2)&1.89(2) & 0.86(1)&0.14(1)&1.87(3) & Ca$_{0.85(3)}$Sr$_{0.15(2)}$Co$_{1.88(4)}$As$_2$ \\
0.81(1)&0.19(1)&1.90(1) & 0.81(1)&0.19(1)&1.90(2) & Ca$_{0.81(1)}$Sr$_{0.19(1)}$Co$_{1.90(2)}$As$_2$ \\
0.76(2)&0.24(2)&1.92(2) & 0.75(1)&0.25(1)&1.89(5) & Ca$_{0.75(2)}$Sr$_{0.25(2)}$Co$_{1.90(5)}$As$_2$ \\
0.71(1)&0.29(1)&1.92(1) & 0.73(1)&0.27(1)&1.91(2) & Ca$_{0.72(1)}$Sr$_{0.28(1)}$Co$_{1.91(2)}$As$_2$ \\
0.68(1)&0.32(1)&1.90(1) & 0.66(1)&0.34(1)&1.91(2) & Ca$_{0.67(1)}$Sr$_{0.33(1)}$Co$_{1.90(2)}$As$_2$ \\
0.62(1)&0.38(1)&1.93(1) & 0.58(1)&0.42(1)&1.92(3) & Ca$_{0.60(2)}$Sr$_{0.40(2)}$Co$_{1.93(3)}$As$_2$ \\
0.55(2)&0.45(2)&1.92(2) & 0.56(3)&0.44(3)&1.92(1) & Ca$_{0.55(4)}$Sr$_{0.45(4)}$Co$_{1.92(2)}$As$_2$ \\
0.48(1)&0.52(1)&1.91(2) & 0.48(1)&0.52(2)&1.92(2) & Ca$_{0.48(1)}$Sr$_{0.52(2)}$Co$_{1.92(3)}$As$_2$ \\

\end{tabular}
\end{ruledtabular}
\end{table*}

\begin{table*}
\caption{\label{CrystalData} Crystallographic data for \csca\ single crystals at room temperature, including the fractional $c$-axis position $z_{\rm AS}$ of the As site, the tetragonal lattice parameters $a$ and~$c$, the unit cell volume $V_{\rm cell}$ containing two formula units of \csca, and the $c/a$ ratio. The data for ${\rm SrCo_2As_2}$ are room-temperature data taken from Ref.~[32].  The compositions in the first column except for  ${\rm SrCo_2As_2}$ were all obtained from combined EDS and single-crystal XRD analyses.}
\begin{ruledtabular}
\begin{tabular}{ clllll }

 Compound  & $z_{\rm As}$ & $a$ (\AA)  & $c$ (\AA) & $V_{\rm cell}$ (\AA$^3$) & $c/a$ \\
\hline 
CaCo$_{1.84(3)}$As$_2$  &   0.3669(1)  &  3.9900(7)  &  10.2972(19) &163.93(6) & 2.5807(5) \\
Ca$_{0.91(2)}$Sr$_{0.09(2)}$Co$_{1.86(3)}$As$_2$  & 0.3660(3) & 3.9937(8) & 10.372(2) &  165.43(8) & 2.597(1) \\
Ca$_{0.85(3)}$Sr$_{0.15(2)}$Co$_{1.88(4)}$As$_2$  & 0.3650(2) &  3.9889(2) & 10.4770(7) &166.70(2) & 2.6265(3) \\
Ca$_{0.81(1)}$Sr$_{0.19(1)}$Co$_{1.90(2)}$As$_2$  & 0.3649(1)  & 3.9874(3)  & 10.5363(8)& 167.52(3) & 2.6424(4)\\
Ca$_{0.75(2)}$Sr$_{0.25(2)}$Co$_{1.90(5)}$As$_2$  & 0.3642(3)  & 3.9852(5)  & 10.6406(11)& 168.99(4) & 2.6700(3) \\
Ca$_{0.72(1)}$Sr$_{0.28(1)}$Co$_{1.91(2)}$As$_2$  & 0.3638(2)  & 3.980(5)  &  10.696(14) & 169.4(4) &2.687(3)\\
Ca$_{0.67(1)}$Sr$_{0.33(1)}$Co$_{1.90(2)}$As$_2$  & 0.3636(2) & 3.9730(4) &  10.8113(14) & 170.65(4) & 2.7211(1)  \\
Ca$_{0.60(2)}$Sr$_{0.40(2)}$Co$_{1.93(3)}$As$_2$  & 0.3628(1) & 3.9721(8) &  10.920(4)&172.3(1) & 2.749(1)   \\
Ca$_{0.55(4)}$Sr$_{0.45(4)}$Co$_{1.92(2)}$As$_2$  & 0.3621(2) &3.962(2)  &11.058(8) & 173.5(3) & 2.791(3) \\
Ca$_{0.48(1)}$Sr$_{0.52(2)}$Co$_{1.92(3)}$As$_2$  & 0.3614(1) & 3.9593(5) & 11.2024(16)& 175.61(5) & 2.8294(4) \\
Ca$_{0.26(2)}$Sr$_{0.74(2)}$Co$_{1.92(2)}$As$_2$  & 0.3595(2) & 3.949(2) &  11.574(6) &  180.5(2) & 2.931(4)\\
Ca$_{0.13(3)}$Sr$_{0.87(3)}$Co$_{1.96(2)}$As$_2$  & 0.3588(2) & 3.9441(6)  &  11.6725(18) & 181.58(6) & 2.9594(5) \\
SrCo$_{2}$As$_2$  & 0.3587(3) &3.9466(2)  &  11.773(1) & 183.37(3) & 2.9831(4) \\
\end{tabular}
\end{ruledtabular}

\end{table*}

Plots of the composition~$x$ dependences of $a$ and~$c$ of our crystals of \csca\ are given in Fig.~\ref{Fig:a_and_c_vs_x}(a), of $c/a$ and the unit cell volume $V_{\rm cell}=a^2c$ in Fig.~\ref{Fig:a_and_c_vs_x}(b), and of $z_{\rm As}$ in Fig.~\ref{Fig:a_and_c_vs_x}(c).  All parameters show a smooth crossover from the collapsed-tetragonal structure at $x=0$ to the uncollapsed-tetragonal structure at $x=1$.

\begin{figure}[h]
\includegraphics[width=4in]{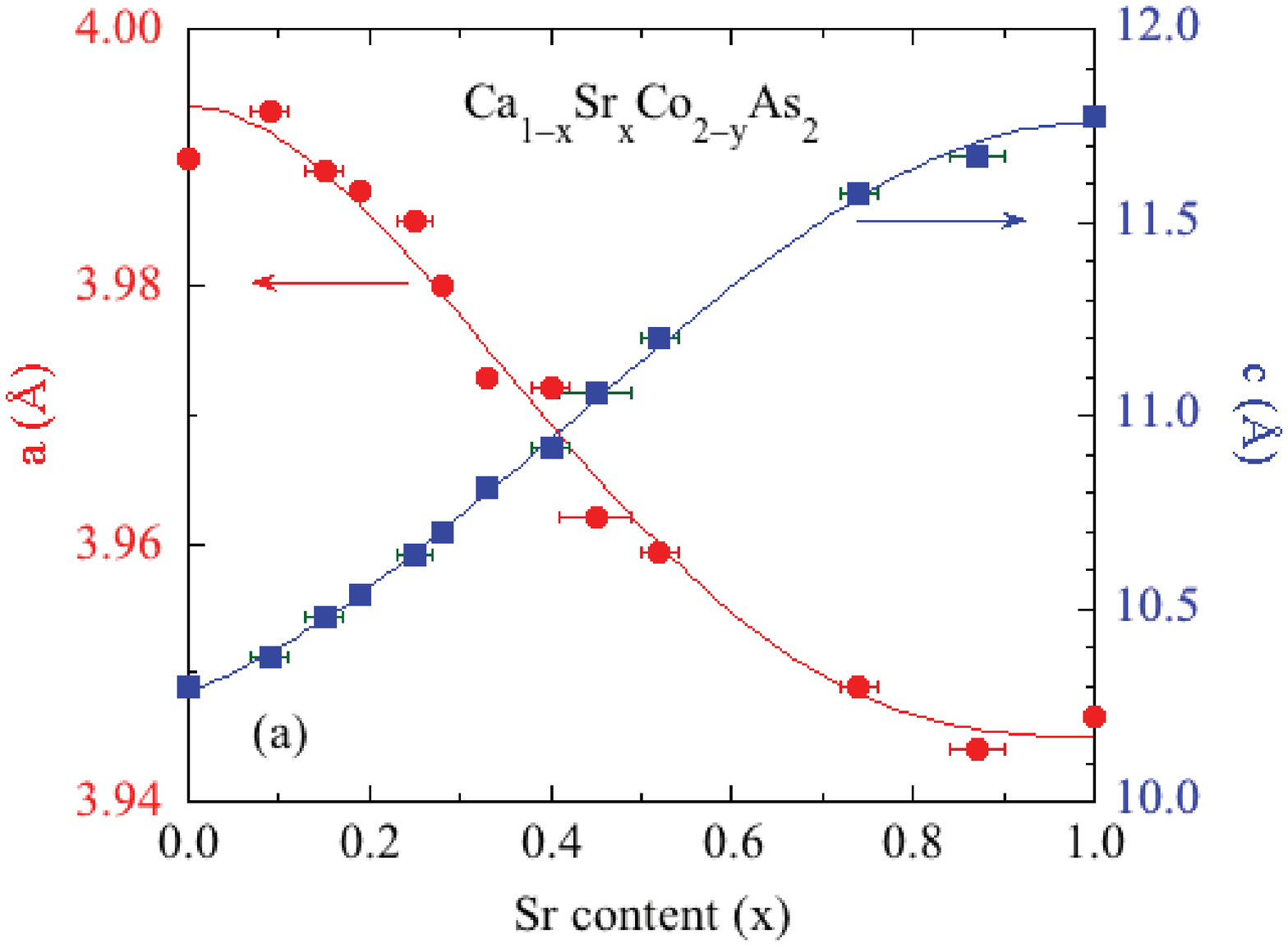}
\includegraphics[width=4in]{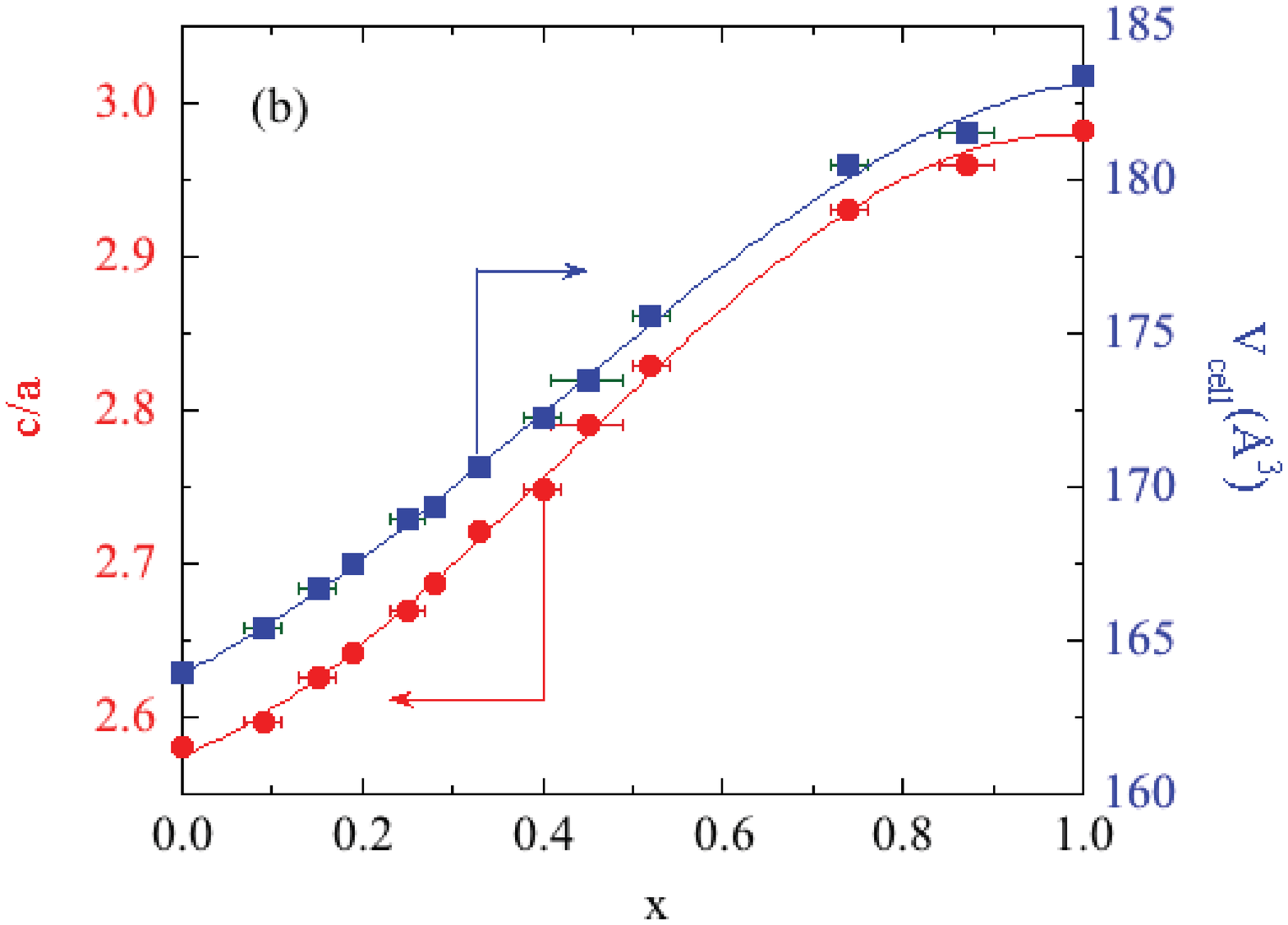} 
\includegraphics[width=3.9in,viewport= 0 0 550 400,clip]{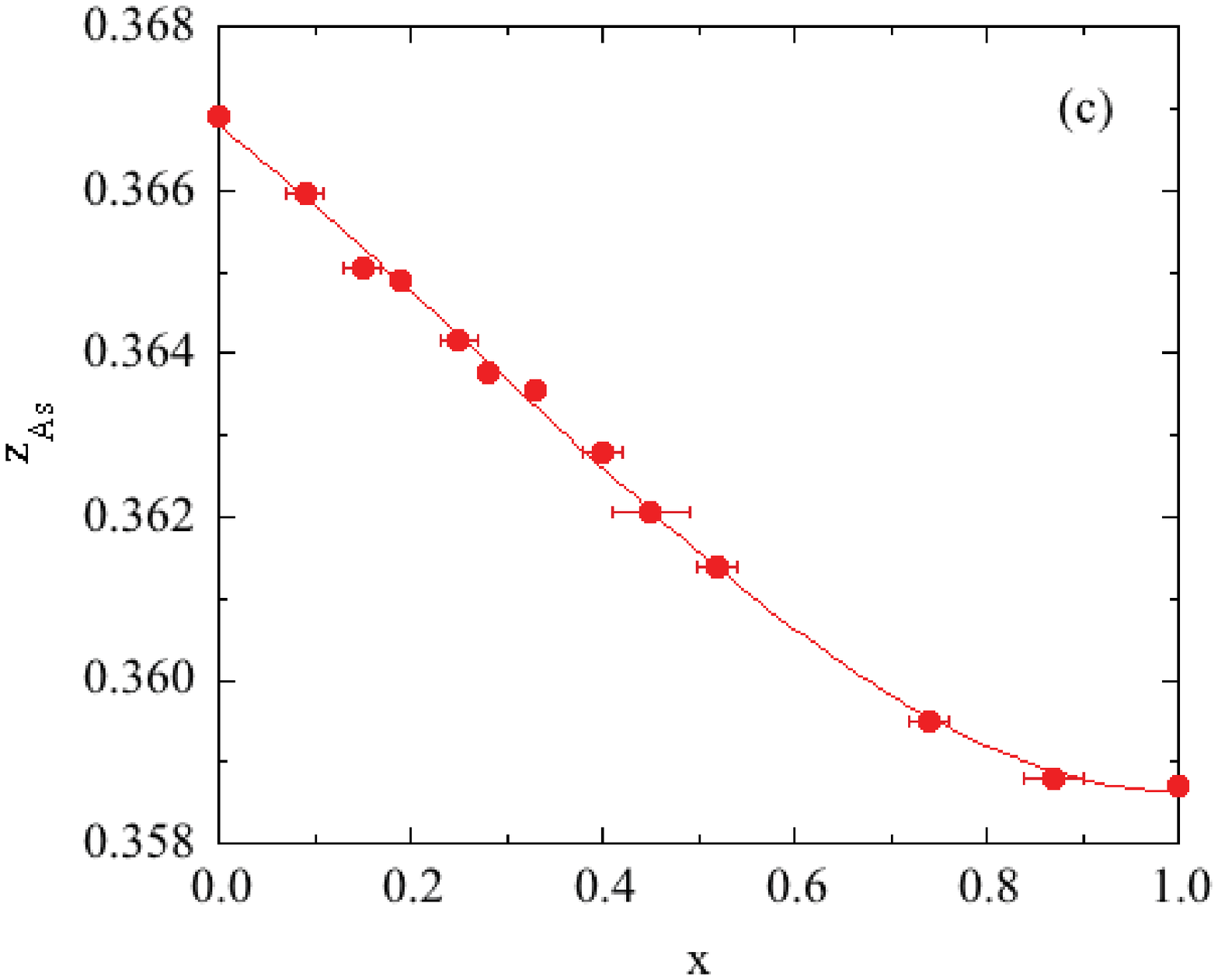} 
\caption{(Colour online)  Crystallographic parameters for \csca\ versus composition~$x$, including (a)~the $a$ and~$c$ lattice parameters, (b)~the $c/a$ ratio and unit-cell volume $V_{\rm cell}$, and (c)~the As $c$-axis parameter~$z_{\rm As}$.  The solid curves are guides to the eye.}
\label{Fig:a_and_c_vs_x}
\end{figure}

%\clearpage

\begin{table*}
\caption{\label{Tab:Uii} Anisotropic thermal displacement parameters $U_{\rm {11}} = U_{\rm {22}}$ and $U_{33}$ and equivalent parameters $U_{\rm eq} = (U_{11}+U_{22}+U_{33})/3$ for  Ca$_{1-x}$Sr$_x$Co$_{2-y}$As$_2$ crystals, as obtained from single-crystal x-ray diffraction analysis.  The off-diagonal components $U_{12},\ U_{13}$ and~$U_{23}$ are zero. }
\begin{ruledtabular}
\begin{tabular}{ c c| c c |c c| c c |c| c| c }

\multicolumn{2}{c|}{Composition} & \multicolumn{2}{c|} {Ca/Sr} & 
\multicolumn{2}{c|}{Co} & 
\multicolumn{2}{c|}{As} &
\multicolumn{3}{c}{$U_{\rm {eq}}$(\AA$^2$)}
\\

  $x$ & $y$    &  $U_{\rm {11}}$~(\AA$^2$) & $U_{\rm {33}}$~(\AA$^2$)   & $U_{\rm {11}}$~(\AA$^2$) & $U_{\rm {33}}$~(\AA$^2$) & $U_{\rm {11}}$~(\AA$^2$)& $U_{\rm {33}}$~(\AA$^2$)    & Ca/Sr & Co &As\\
\hline 
          0 & 1.84(3)  &  0.0090(4)   &  0.0117(7) 	&	0.0077(3)	& 	0.0101(4) 	&0.0091(2)	&0.0131(3)	& 0.0099(3)	&0.0085(2)	&0.01042(18)\\
 0.09(2) & 1.86(3)  &  0.0071(12) &  0.0087(17)	&	0.0089(7)	&	0.0126(10)	&0.0084(5)	&0.0140(8)	&0.0077(10)	&0.0101(6)	&0.0102(4)\\
 0.15(2) & 1.88(4)  &  0.0085(11) &  0.0097(13)	&	0.0083(8)	&	0.0138(9)	&0.0093(6)	&0.0164(8)	&0.0089(10)	&0.0101(6)	&0.0116(5)\\ 
0.19(1) & 1.90(2)   &  0.0065(5)   &  0.0105(7)	&	0.0064(3)	&	0.0144(4)	&0.0071(2)	&0.0158(4)	&0.0078(4)	&0.0091(3)	&0.0100(2)\\
0.25(2) & 1.90(5)   &  0.0089(8)   &  0.0117(14)	&	0.0086(5)	&	0.0169(9)	&0.0084(4)	&0.0172(7)	&0.0099(8)	&0.0114(5)	&0.0114(4)\\
0.28(1) & 1.91(2)   &  0.0084(5)   &  0.0154(8)	&	0.0081(3)	&	0.0206(5)	&0.0081(2)	&0.0213(4)	&0.0107(4)	&0.0123(3)	&0.0125(2)\\
0.33(1) & 1.90(2)   & 0.0078(5)    &  0.0167(7)	&	0.0066(3)	&	0.0078(3)	&0.0078(3) 	& 0.0232(5)	&0.0108(4)	&0.0118(3)	&0.0129(3)\\
0.40(2) & 1.93(3)   &  0.0096(4)   &  0.0099(6)	&	0.0076(3)	&	0.0179(5)	&0.0086(2)	&0.0170(4)	&0.0097(4)	& 0.0110(3)	&0.0114(2)\\
0.45(4) & 1.92(2)   &  0.0095(6)   &  0.0140(9)	&	0.0077(4)	&	0.0211(7)	&0.0084(3)	&0.0213(6)	&0.0110(5)	&0.0121(4)	&0.0127(3)\\
0.52(2) & 1.92(3)   &  0.0091(3)   &  0.0175(5)	&	0.0081(2)	&	0.0232(5) 	&0.0093(2)	&0.0232(4)	&0.0119(3)	&0.0131(2)	&0.0139(2)\\
0.74(2) & 1.92(2)   &  0.0073(8)   &  0.0138(13)	&	0.0049(7)	&	0.0161(11)	&0.0063(5)	&0.0174(9)	&0.0095(8)	&0.0086(6)	&0.0100(5)\\
0.87(3) & 1.96(2)   &  0.0068(6)   &  0.0134(8)	&	0.0058(5)	&	0.0154(7)	&0.0067(4)	&0.0146(6)	&0.0090(5)	&0.0090(5)	&0.0093(4)\\
 \end{tabular}
\end{ruledtabular}

\end{table*}

\begin{figure}[h]
\includegraphics[width=3.3in]{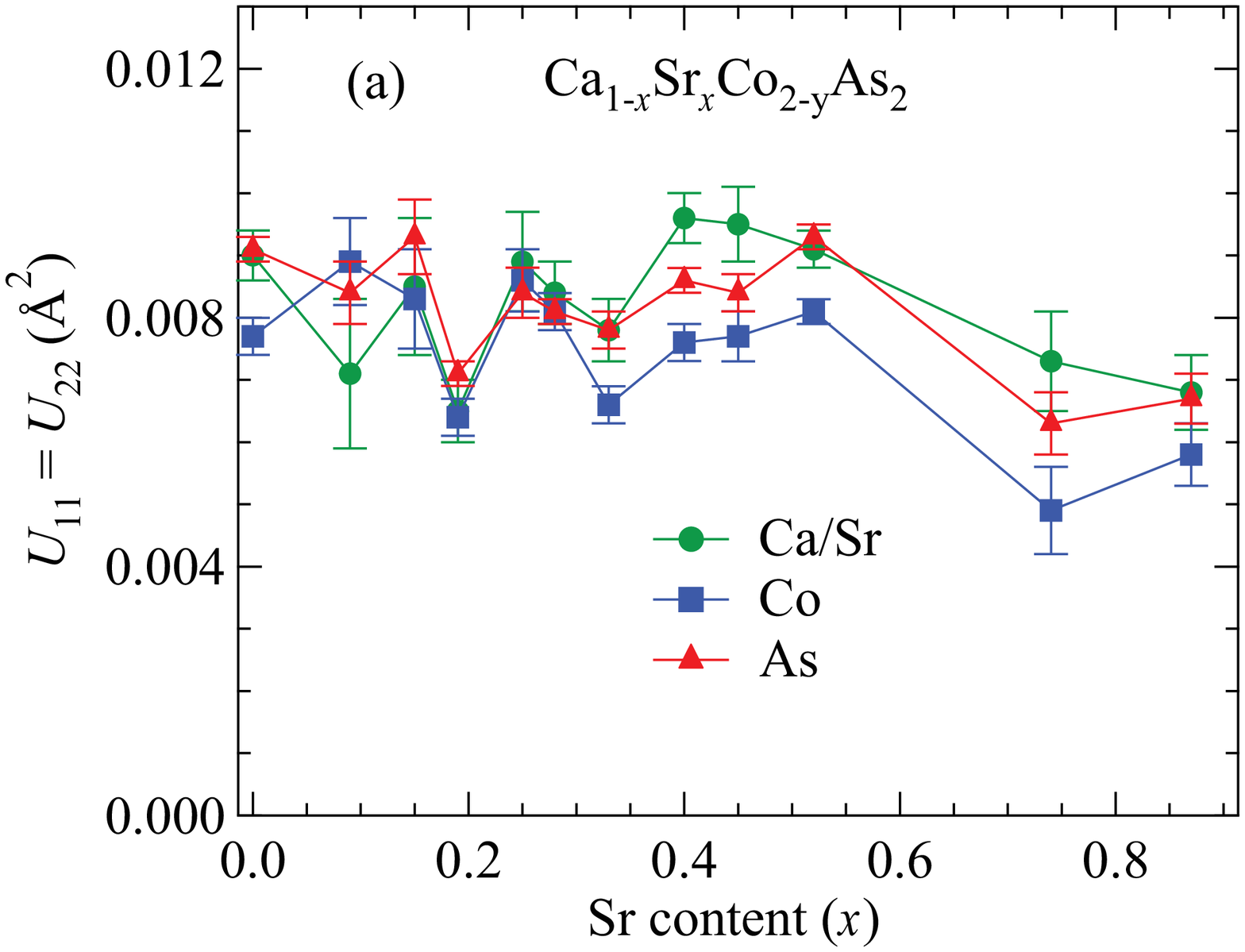}
\includegraphics[width=3.3in]{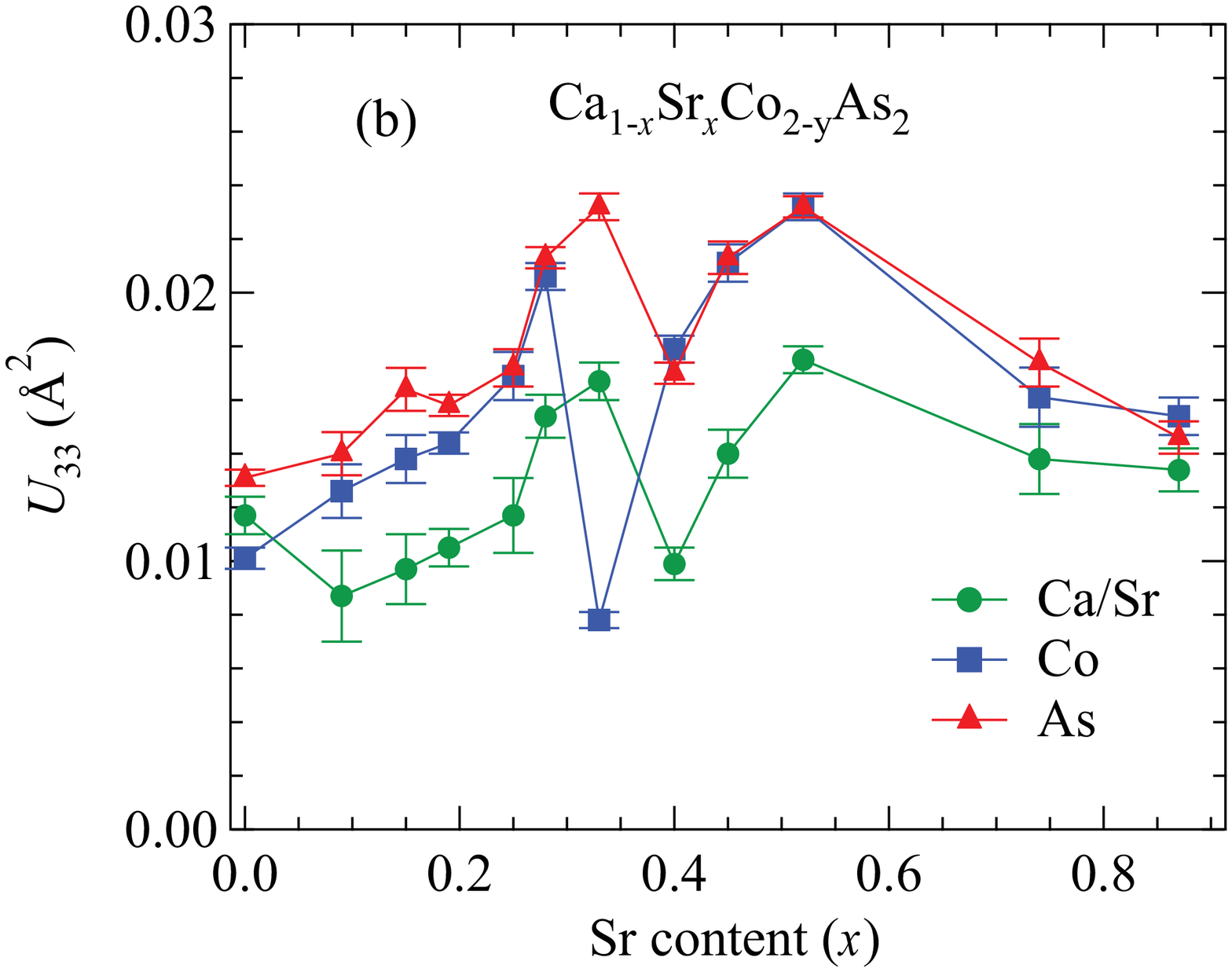}
\includegraphics[width=3.3in]{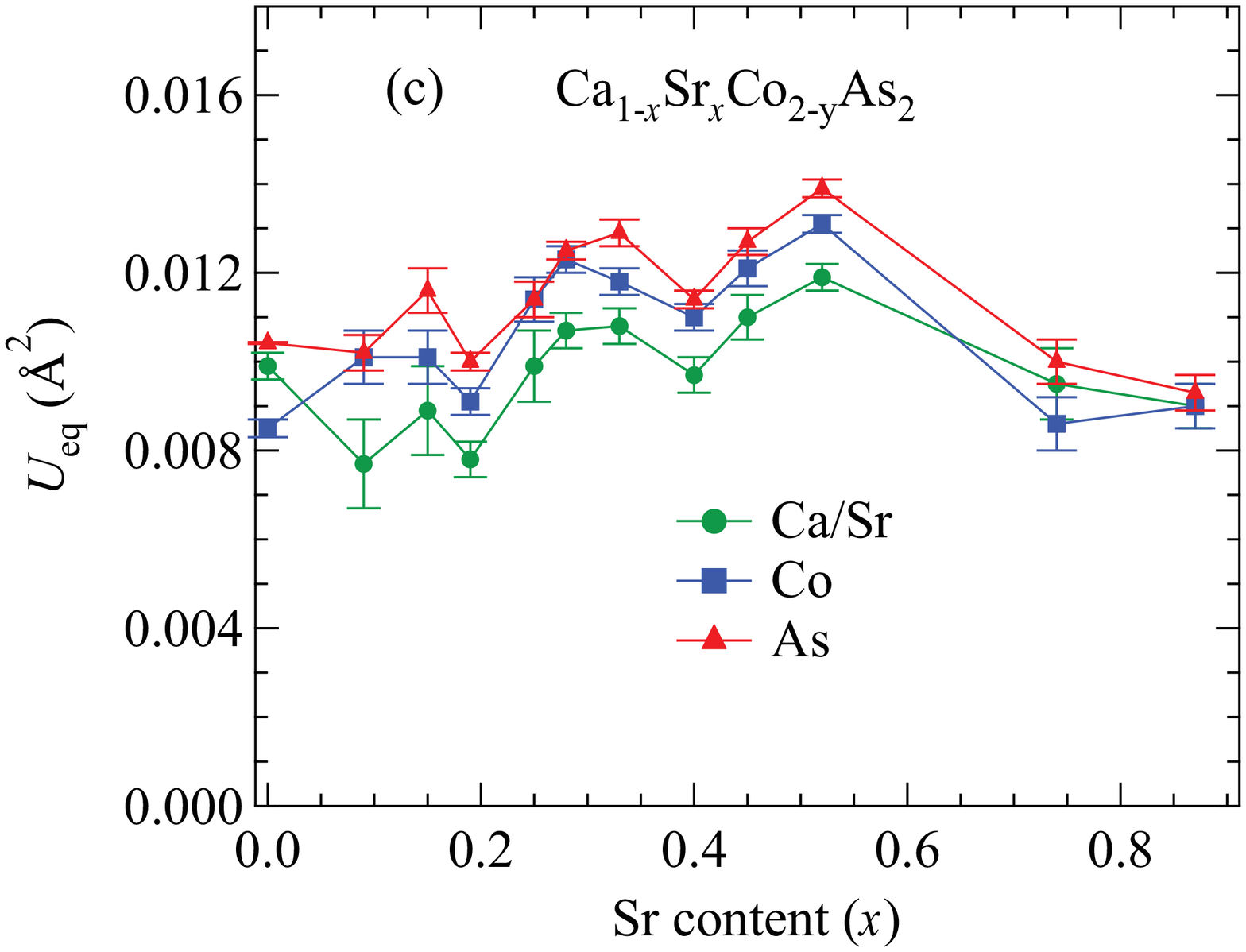}
\caption{(Color online) Anisotropic room-temperature thermal parameters (a)~$U_{11}=U_{22}$, (b)~$U_{33}$, and (c)~$U_{\rm eq}$ versus Sr composition~$x$ in \csca\ crystals from the data in Table~\ref{Tab:Uii}.  The lines connecting the points in each panel are guides to the eye.}
\label{Fig:U_SXRD}
\end{figure}

\clearpage

\section*{\label{Sec:Cp} Heat Capacity Data}

\begin{figure*}[h]
\includegraphics[width=5.75in]{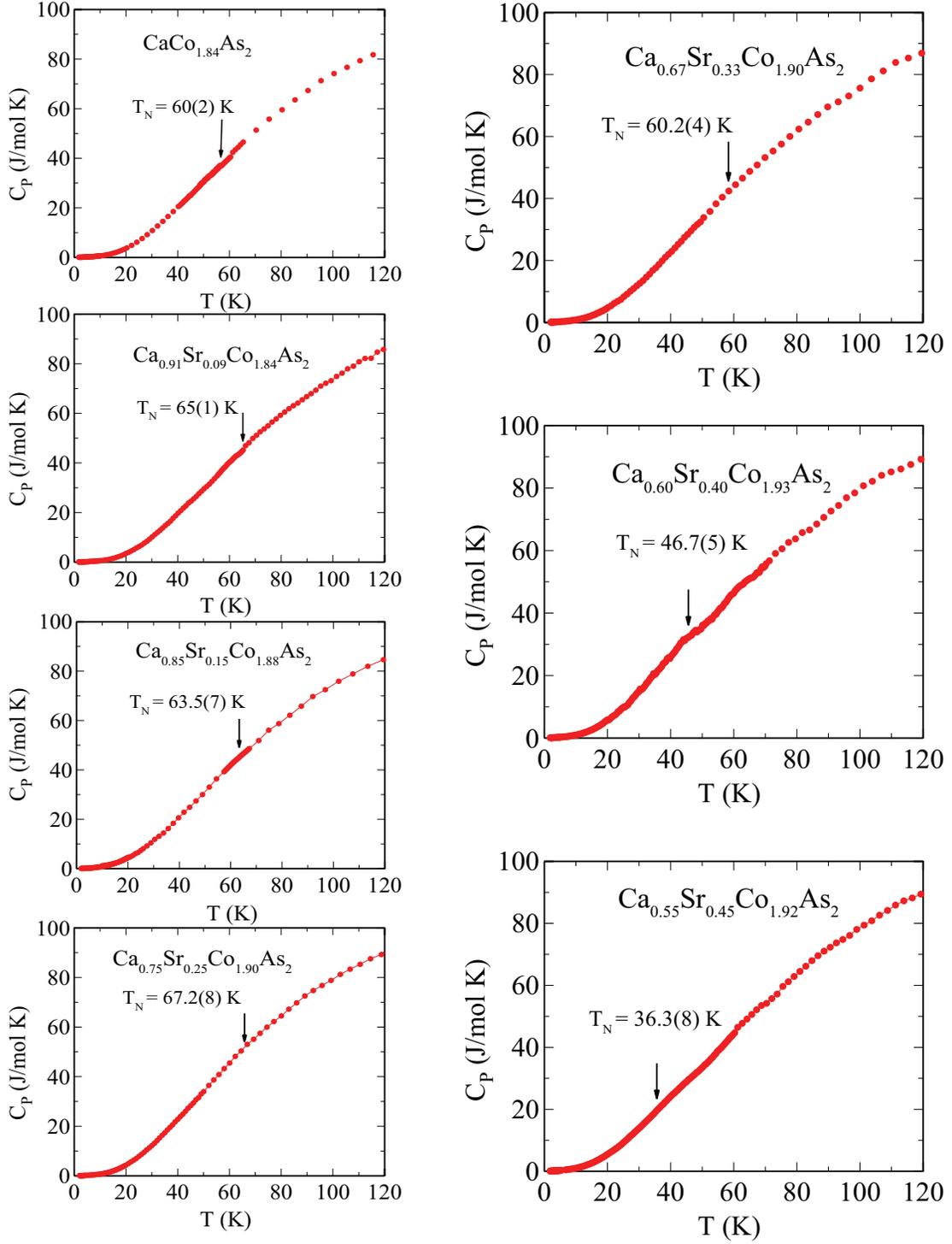}
\caption{(Color online) Heat Capacity $C_{\rm p}$ of seven \csca\ single crystals versus temperature~$T$ measured in zero magnetic field.  No clear anomalies are seen at the indicated N\'eel temperatures~$T_{\rm N}$, except possibly for $x=0.40$.  }
\label{Fig:Cp_vs_T}
\end{figure*}

\begin{figure*}[h]
\includegraphics[width=6.5in]{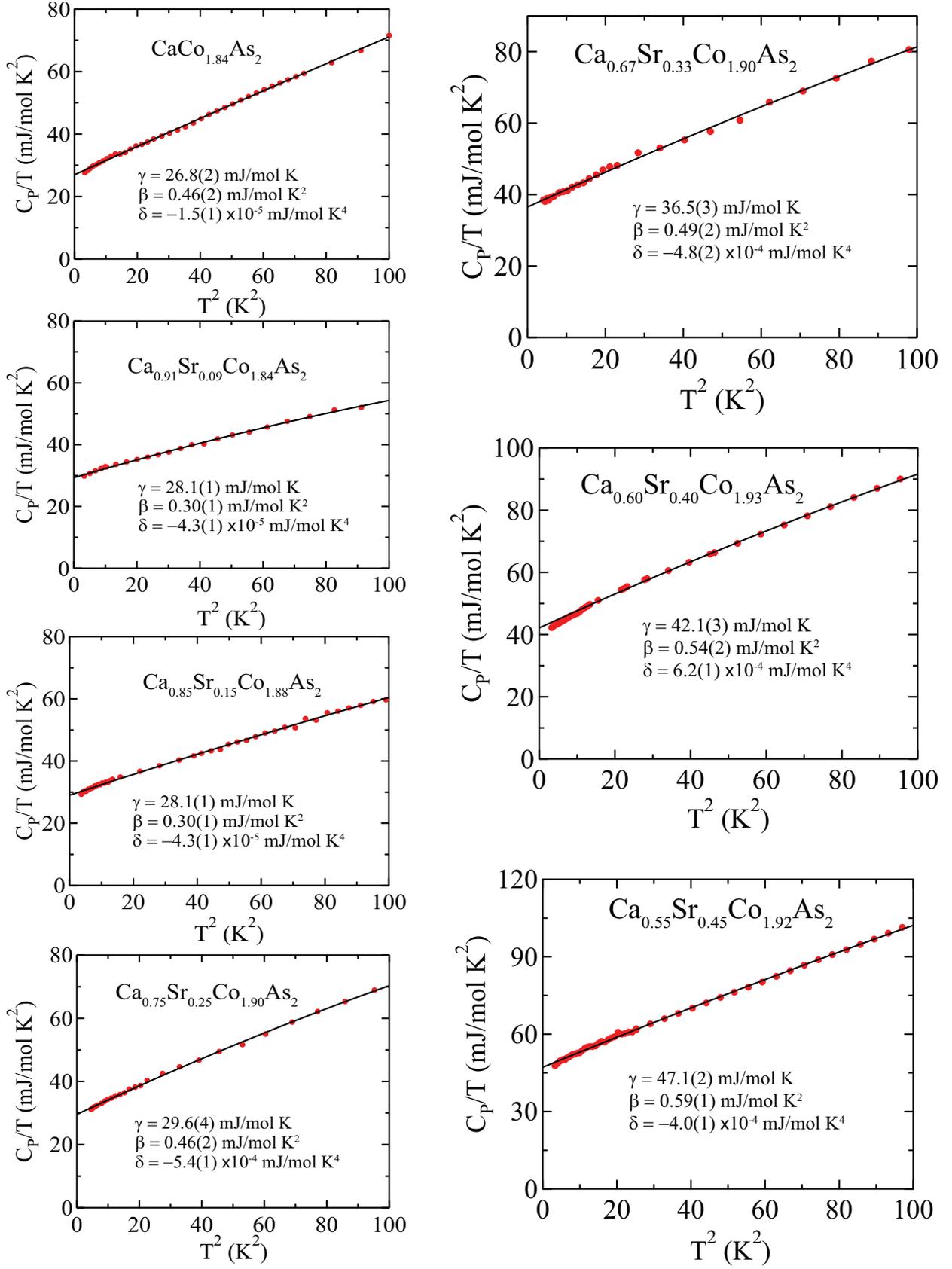}
\caption{(Color online) Low-temperature heat capacities $C_{\rm p}$ divided by temperature~$T$ of seven \csca\ single crystals versus~$T$ measured in zero magnetic field.  The data were fitted by $C_{\rm p}/T = \gamma +\beta T^2 + \delta T^4$, and the fitted parameters are listed in each figure, respectively.  The densities of states at the Fermi energy for both spin directions calculated from the $\gamma$ values are in the range 11 to 20 states/eV~f.u. for $x=0$ and~0.45, respectively, where f.u.\ means formula unit.}
\label{Fig:Cp_div_T}
\end{figure*}

\clearpage

\section*{\label{Sec:MHChi} Additional Magnetization and Magnetic Susceptibility Data}

In this section, the Tesla~(T) is an abbreviation for 10~kOe.

\begin{table*}[b]
\caption{\label{Tab:MagProps} Summary of magnetic data for the \csca\ crystals discussed in the main text and in this section. Included are the N\'eel temperature $T_{\rm N}$; the following transition fields at $T=2$~K: the spin-flop field $H_{{\rm SF}c}$ for $c$-axis fields, the metamagnetic transition field $H_{{\rm mm}ab}$ for $ab$-plane fields, the critical fields $H_{{\rm c},c}$ and $H_{{\rm c},ab}$ for fields aligned along the $c$~axis and $ab$~plane, respectively, at which the $M(H)$ isotherms reach saturation with increasing field; and the effective moments per Co $\mu_{{\rm eff}c}$ and $\mu_{{\rm eff}ab}$ and Weiss temperatures $\theta_c$ and $\theta_{ab}$ in the PM state obtained from fitting the $c$-axis and $ab$-plane susceptibilities in the PM state by the modified Curie-Weiss law over the temperature range 100--300~K\@.  The estimated systematic error bar for $\mu_{{\rm eff}\alpha}$ or $\theta_\alpha$ is the difference between the respective fitting parameter from the 100--300~K fit and a 150--300~K fit.  The $c$-axis SF field listed for $x=0.33$ at 2~K transitions to an $ab$-plane metamagnetic transition field of about the same magnitude at 50~K: see Fig.~\ref{Fig:Sr33MH} below and Fig.~1 in the main text.}
\begin{ruledtabular}
\begin{tabular}{ c  c   c  c cccccc }

Compound   & $T_{\rm {N}}$ 	& $H_{{\rm SF}c}$ 	& $H_{{\rm mm}ab}$  &  $H_{{\rm c},c}$	& $H_{{\rm c},ab}$  &  $\mu_{{\rm eff}c}$	& $\mu_{{\rm eff}ab}$  	&  $\theta_c$	& $\theta_{ab}$ \\
		&	(K)			& 	(T)			&   		(T)		&  	 (T) 		&		(T)		&  $(\mu_{\rm B}$/Co)	& $(\mu_{\rm B}$/Co)	&	(K)		& 	(K) \\
\hline 
CaCo$_{1.84(3)}$As$_2$  						&  60(2)  & 3.69  	&		&  8.9 	& 11.2 			&	1.40(1)			&	1.36(1)			&	70(5)		&	70(8)	\\
Ca$_{0.91(2)}$Sr$_{0.09(2)}$Co$_{1.86(3)}$As$_2$  & 65.5(3) & 1.4 	&		& 1.80 	& 4.31   			&	1.52(18)			&	1.57(10)			&	82(10)		&	79(9)	\\
Ca$_{0.85(3)}$Sr$_{0.15(2)}$Co$_{1.88(4)}$As$_2$  & 64.6(2)	& 0.63 	&		& 0.875 	& 2.57   			&	1.76(11)			&	1.65(19)			&	86(6)		&	83(14)	\\
Ca$_{0.81(1)}$Sr$_{0.19(1)}$Co$_{1.90(2)}$As$_2$  & 70.5(5)	& 0.015   &		& 0.143	& 1.40  			&					&					&				&		\\
Ca$_{0.75(2)}$Sr$_{0.25(2)}$Co$_{1.90(5)}$As$_2$  & 67(1)  	& 0.01  	&		& 0.094	& 0.95   			&	1.64(12)			&	1.64(9)			&	86(7)		&	85(9)	\\
Ca$_{0.72(1)}$Sr$_{0.28(1)}$Co$_{1.91(2)}$As$_2$  & 62.7(7) & 0  	&  		&  0.084 	& 1.06  			&	1.72(19)			&	1.61(19)			&	79(13)		&	80(10)	\\
Ca$_{0.67(1)}$Sr$_{0.33(1)}$Co$_{1.90(2)}$As$_2$  & 60.3(5)	&	0.24	&  	&  1.62 	& 1.69   			&	1.45(12)			&	1.39(7)			&	71(10)		&	72(10)	\\
Ca$_{0.60(2)}$Sr$_{0.40(2)}$Co$_{1.93(3)}$As$_2$  &46.7(9) 	&		& 1.74 	&	  	& 5.25   			&	1.47(8)			&	1.41(4)			&	44(10)		&	44(5)	\\
Ca$_{0.55(4)}$Sr$_{0.45(4)}$Co$_{1.92(2)}$As$_2$  &36.4(3) 	&		& 2.8		& 		& 7.35   			&	1.46(1)			&	1.43(3)			&	26(6)		&	27(7)	\\
Ca$_{0.48(1)}$Sr$_{0.52(2)}$Co$_{1.92(3)}$As$_2$  &	0 	&		&  		& 		&    			&	0.47(1)			&	0.44(4)			&	$-1$(6)		&	$-5$(8) \\
SrCo$_2$As$_2$ [32]				&		&		&		&		&				&	1.8(2)\footnotemark[1]&	&	$-$140(40)\footnotemark[1]&		\\
\end{tabular}
\end{ruledtabular}
\footnotetext[1]{Spherically-averaged values obtained in Ref.~[32] from fits of the modified Curie-Weiss law to the spherically-averaged single-crystal susceptibilities $\chi_c$ and $\chi_{ab}$ in the temperature range 200--300~K.}
\end{table*}

%\clearpage

Due to the two-orders-of-magnitude variation in the ordinate scales of the $\chi_\alpha(T)$ $(\alpha = c,~ab)$ plots for the ten crystals of \csca\ in Fig.~1 of the main text, in Fig.~\ref{Fig:ChiLog_vs_T} are shown plots of $\log_{10}\chi_\alpha$ versus~$T$ for the same ten crystals on single plots for each of $\chi_c(T)$ and~$\chi_{ab}(T)$.

The $\chi_\alpha(T)$ data for the crystals in the PM state above $T_{\rm N}$ were fitted from 100 to 300~K by the modified Curie-Weiss law $\chi_\alpha = \chi_{0\alpha} + C_\alpha/(T-\theta_\alpha)$, where $C_\alpha$ is the Curie constant and $\theta_\alpha$ is the Weiss temperature.  The effective moment $\mu_{\rm{eff}\alpha}$ was calculated from $\mu_{{\rm eff}\alpha}(\mu_{\rm B}/{\rm Co}) = \sqrt{8C_\alpha/(2-y)}$. The results for $\mu_{{\rm eff}\alpha}$ and $\theta_\alpha$ are listed in Table~\ref{Tab:MagProps} and plotted in Fig.~\ref{Fig:mu_eff_vs_x}.  

\begin{figure}[h]
\includegraphics[width=3.5in]{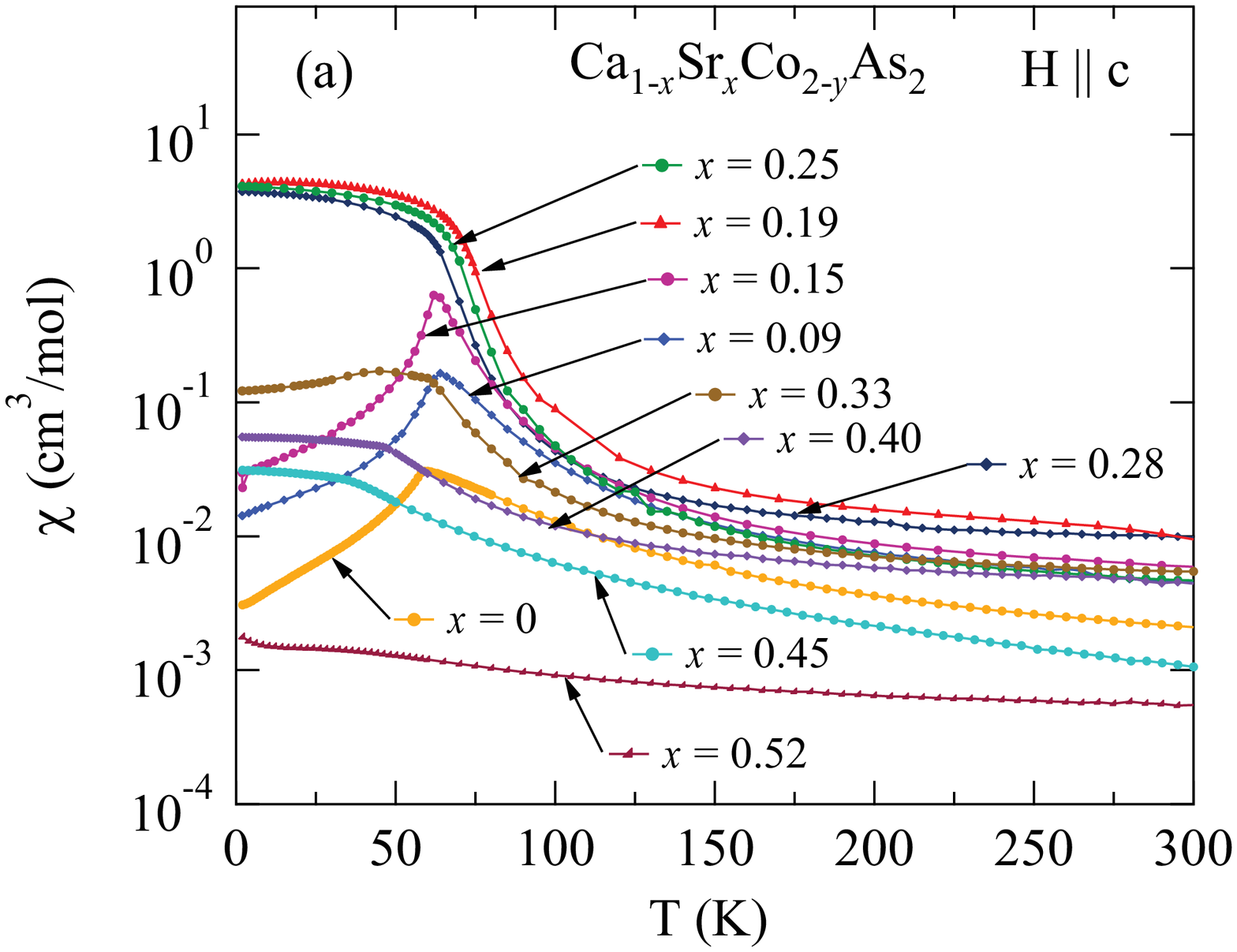}
\includegraphics[width=3.5in]{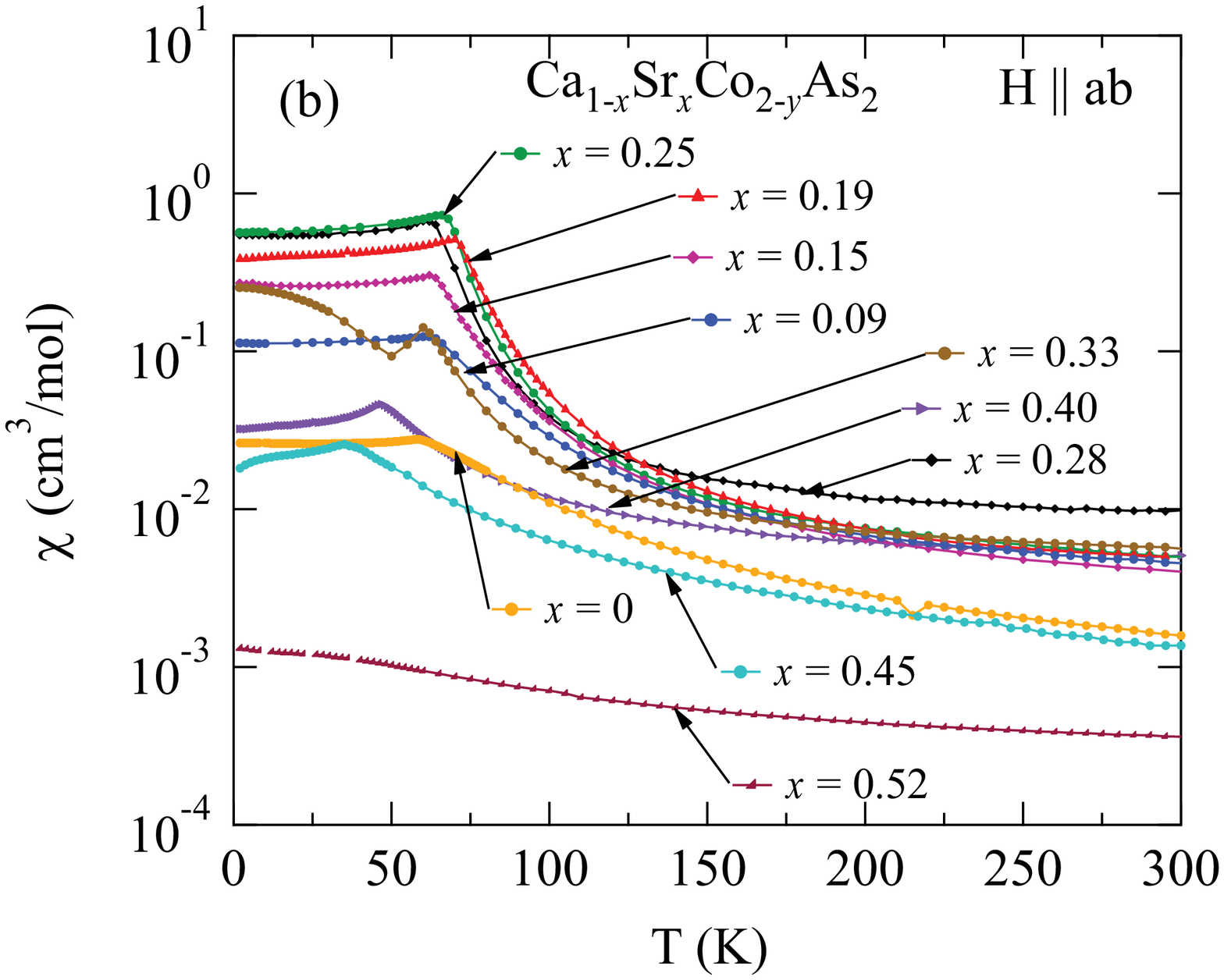}
\caption{(Color online) Plots of (a) $\log_{10}\chi_c$ and (b) $\log_{10}\chi_{ab}$ versus~$T$ of the ten \csca\ crystals with Sr compositions~$x$ studied in the main text.}
\label{Fig:ChiLog_vs_T}
\end{figure}

\begin{figure}[h]
\includegraphics[width=3.5in]{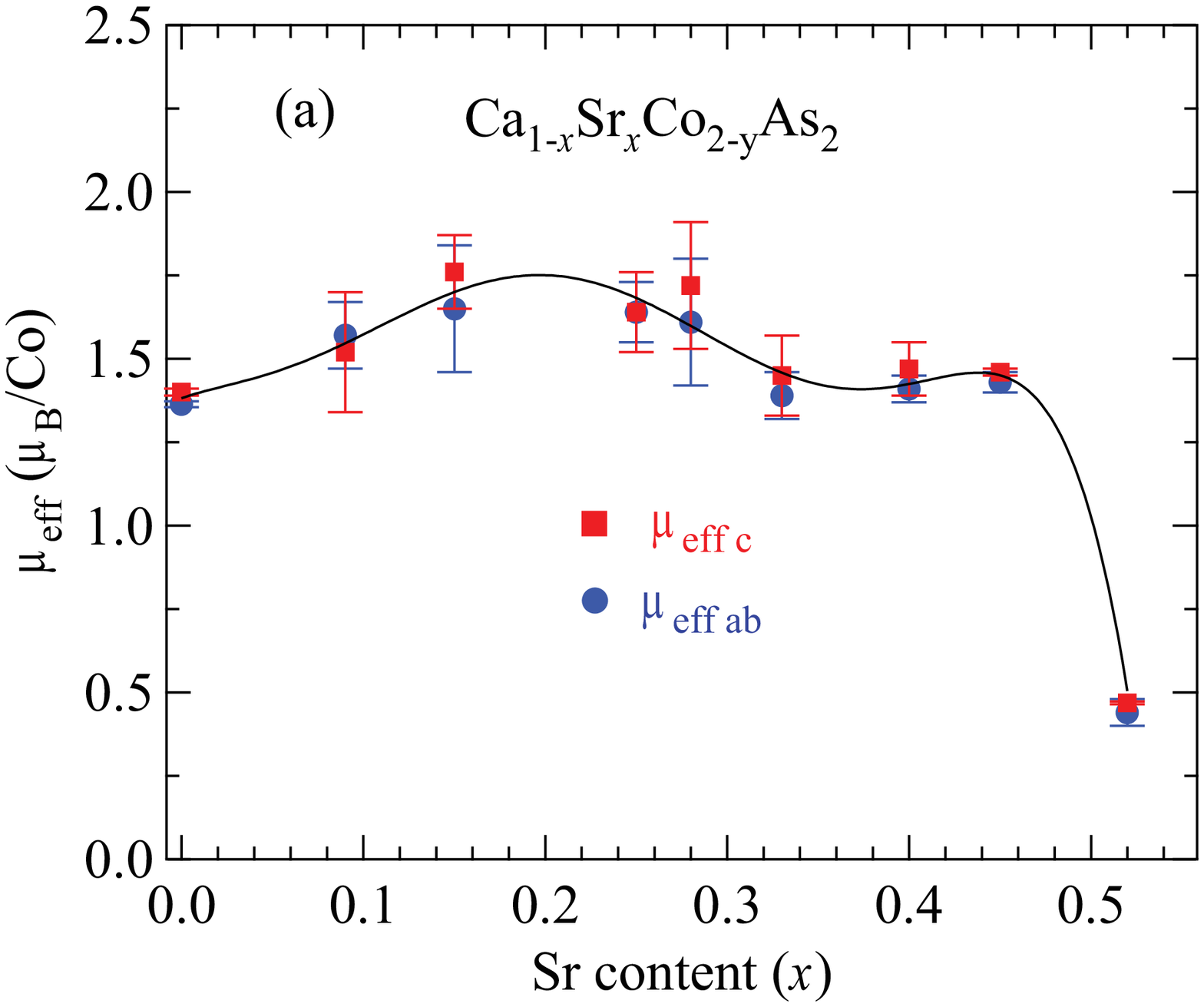}
\includegraphics[width=3.5in]{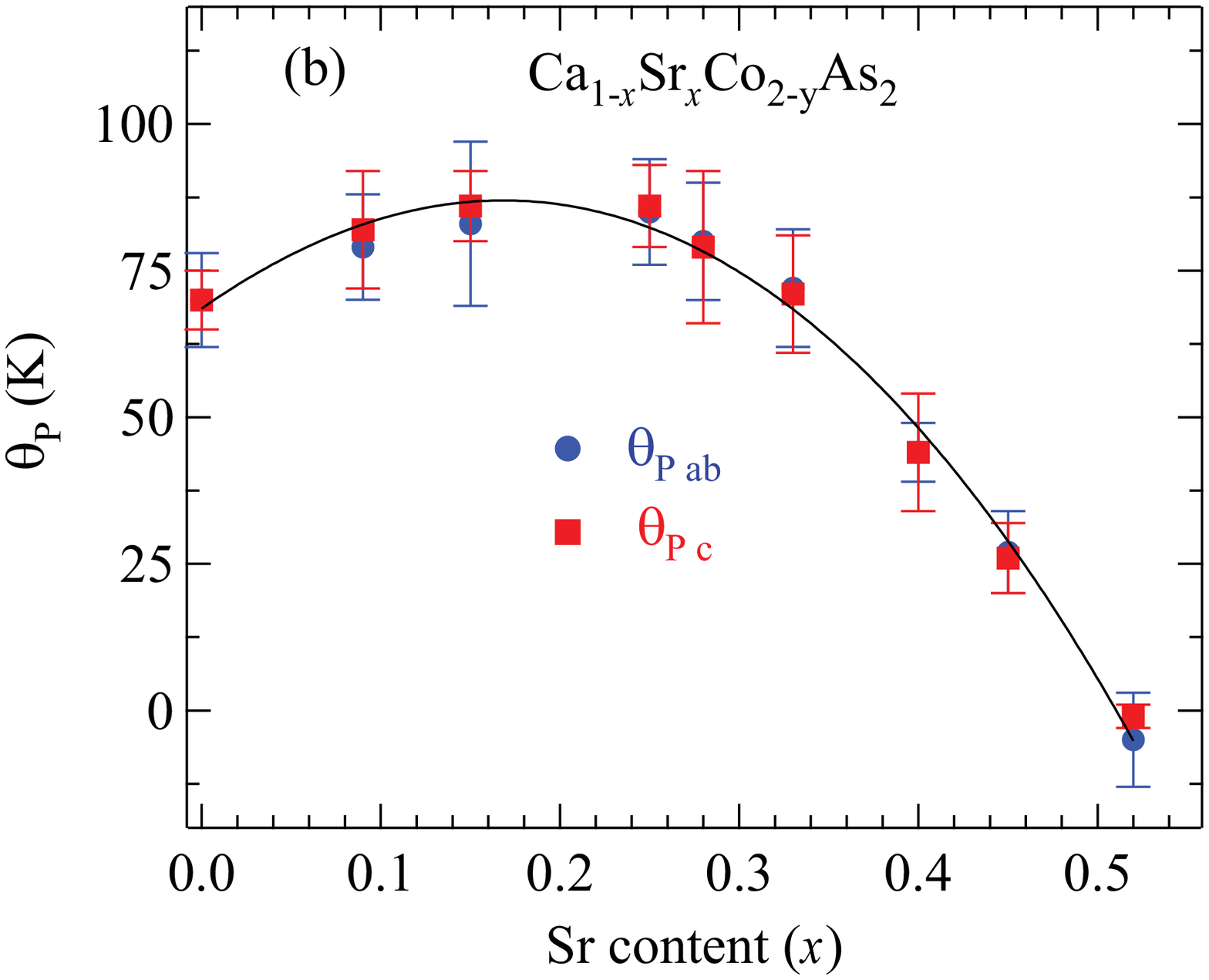}
\caption{(Color online) (a) Effective moment per Co $\mu_{{\rm eff}\alpha}$ ($\alpha = ab,\ c$) and (b)~Wiess temperatures $\theta_\alpha$ of \csca\ crystals versus composition~$x$ from fits to magnetic susceptibility data in the paramagnetic state from 100 to 300~K\@.}
\label{Fig:mu_eff_vs_x}
\end{figure}

\clearpage

\section*{${\rm\bf CaCo_{1.84}As_2}$}

\begin{figure}[h]
\includegraphics[width=4in]{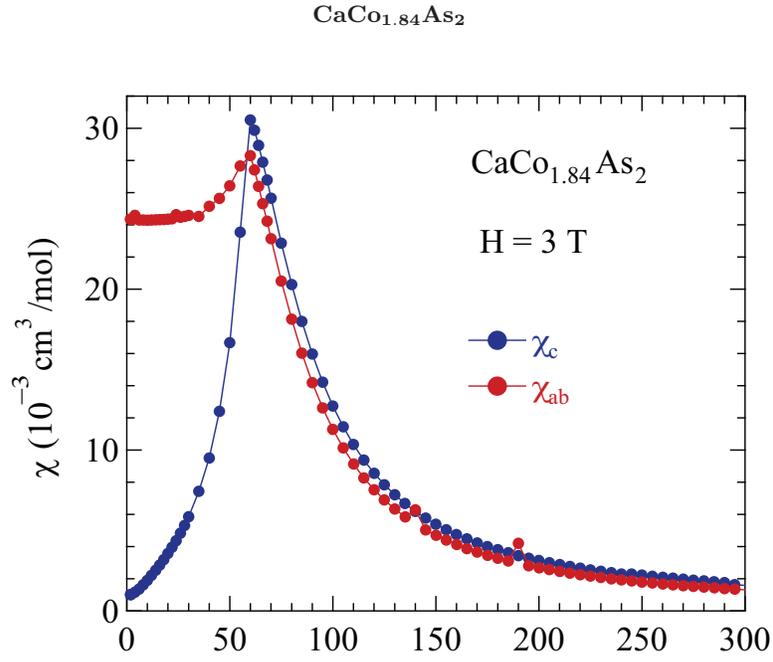}
\caption{(Color online) Magnetic susceptibility $\chi$ of a ${\rm CaCo_{1.84}As_2}$ single crystal as a function of temperature~$T$ measured in a magnetic field $H=3$~T applied in the $ab$-plane ($\chi_{ab}$) and along the $c$-axis ($\chi_c$).}
\label{Fig:MT_CaCo2As2}
\end{figure}

\begin{figure}[h]
\includegraphics[width=4in]{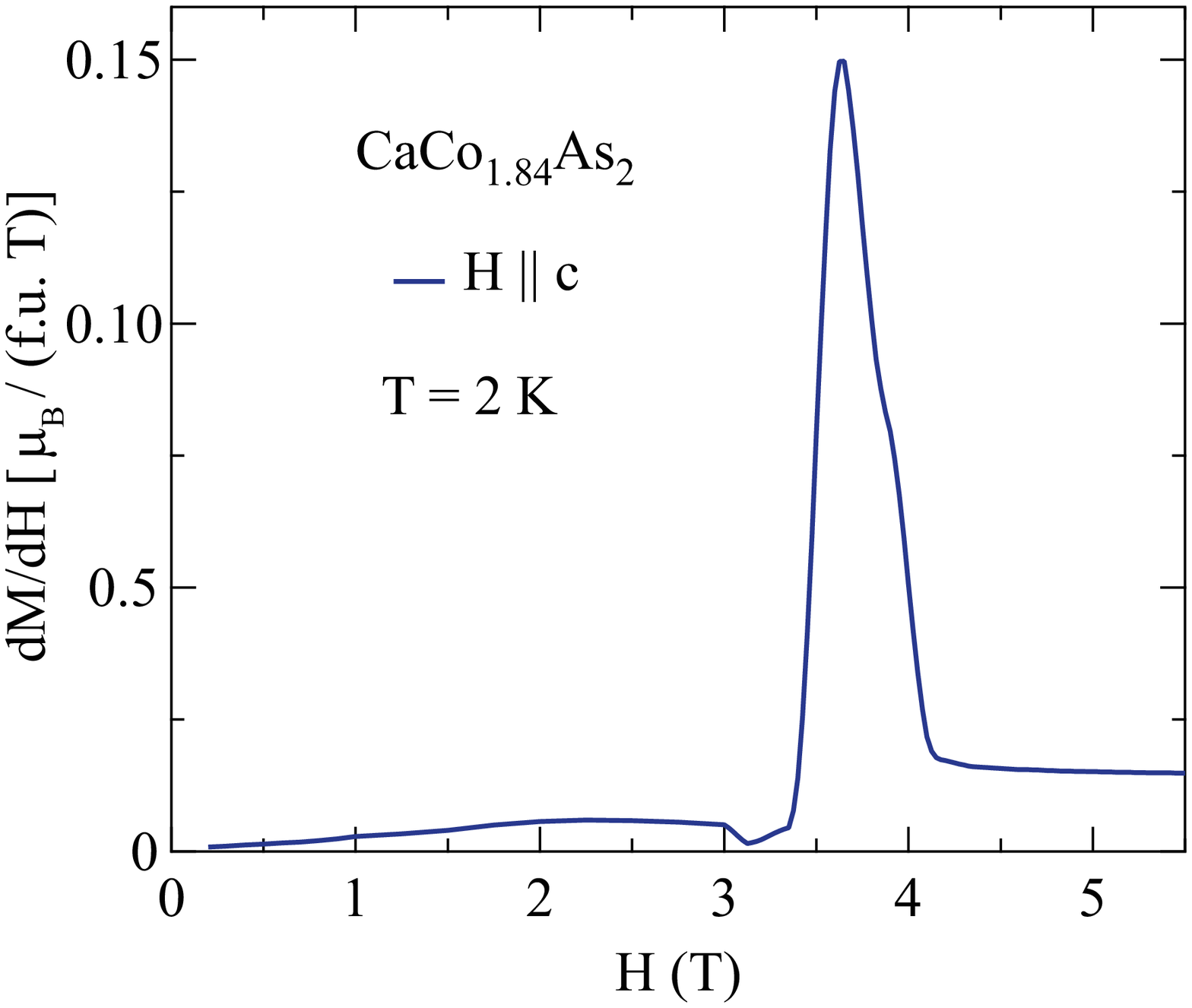}
\caption{(Color online) Derivative $dM/dH$ versus applied magnetic field~$H$ of a ${\rm CaCo_{1.84}As_2}$ single crystal at temperature $T=2$~K with the field applied along the $c$-axis.}
\label{Fig:dM_dH_Sr0}
\end{figure}

\clearpage

\section*{${\rm\bf Ca_{0.91}Sr_{0.09}Co_{1.86}As_2}$}

\begin{figure}[h]
\includegraphics[width=3.5in]{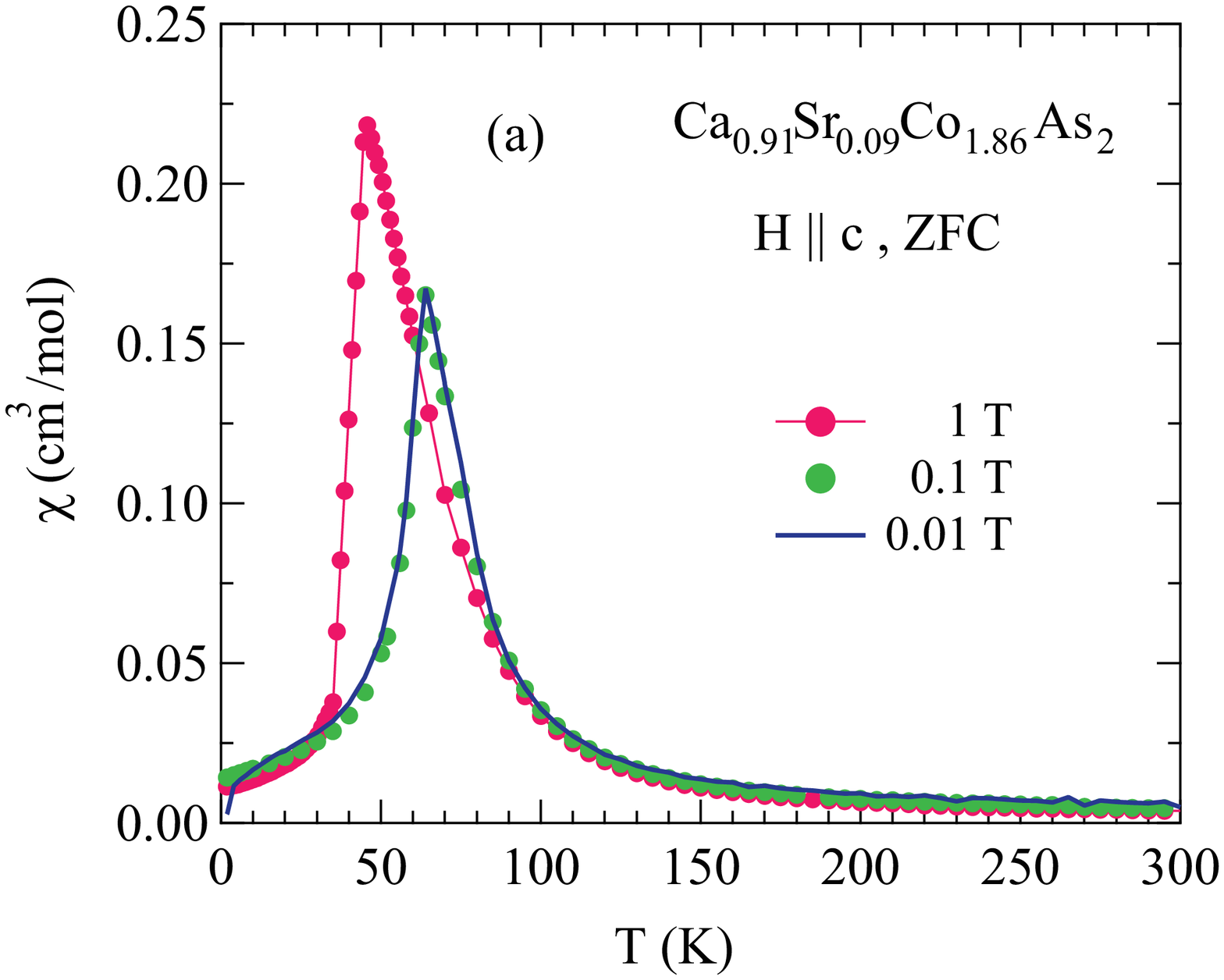}
\includegraphics[width=3.5in]{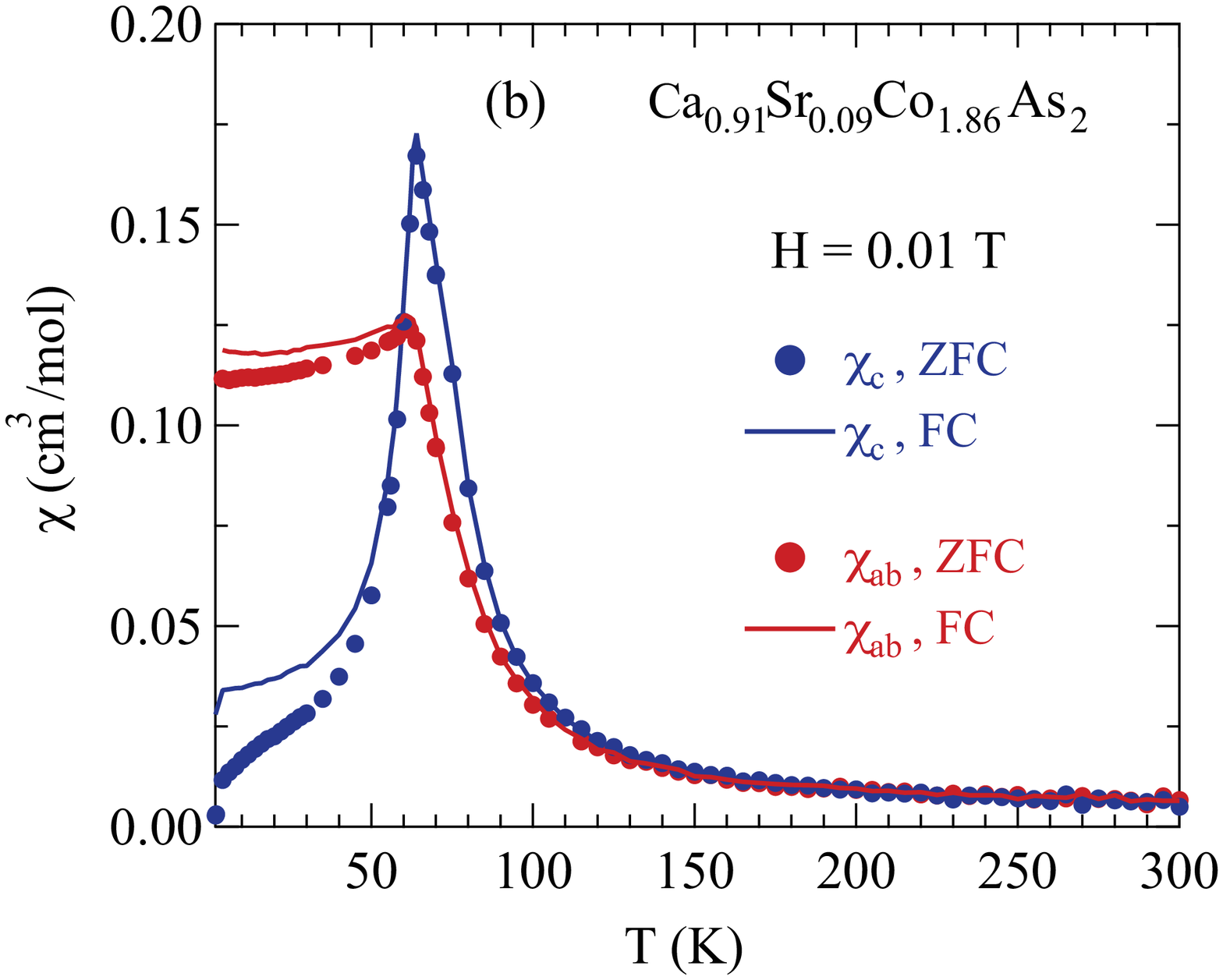}
\caption{(Color online) (a) Field dependence of the magnetic susceptibility $\chi_c\equiv M_c/H$ of a ${\rm Ca_{0.91}Sr_{0.09}Co_{1.86}As_2}$ single crystal versus temperature~$T$ for fields aligned along the $c$~axis. (b) Comparison of field-cooled (FC) and zero-field-cooled (ZFC) susceptibilities in $H=100$~Oe.}
\label{Fig:Fig_Sr10_Chi_c_axis}
\end{figure}

\begin{figure}[h]
\includegraphics[width=3.5in]{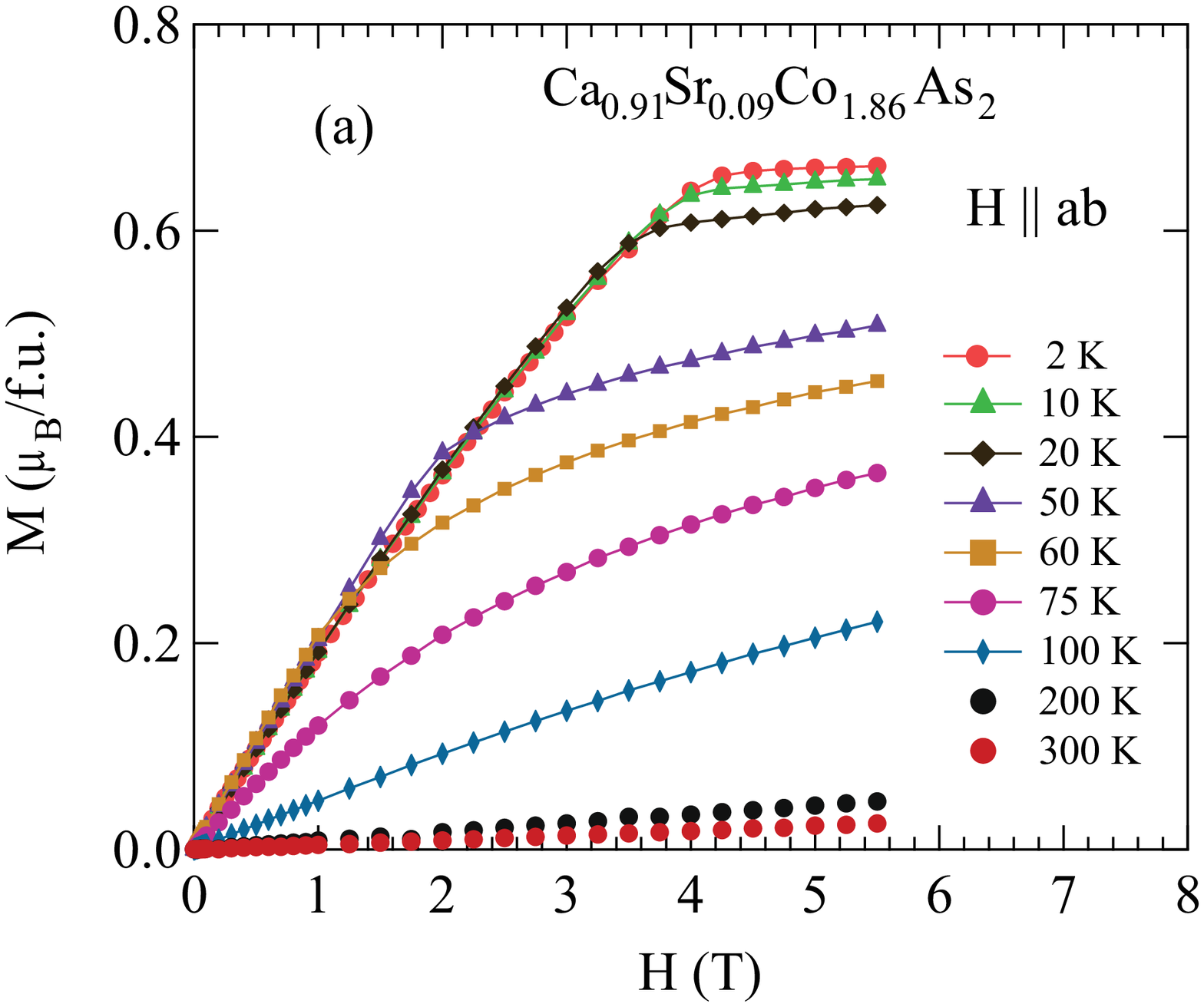} 
\includegraphics[width=3.5in]{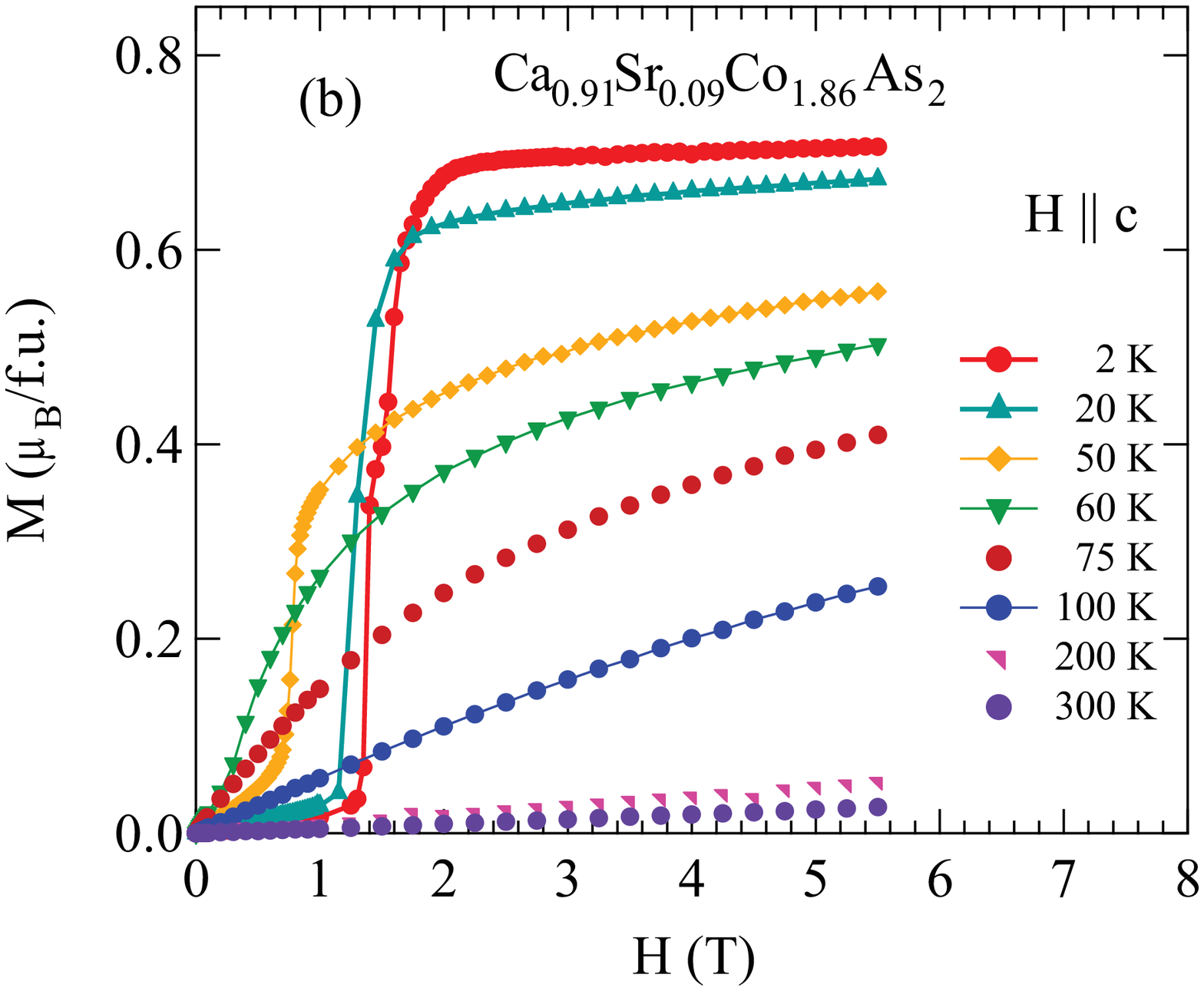} 
\caption{(Color online) Isothermal magnetization $M$ of a ${\rm Ca_{0.91}Sr_{0.09}Co_{1.86}As_2}$  single crystal as a function of applied magnetic field $H$ measured at the indicated temperatures for $H$ applied (a) in the $ab$~plane ($H \parallel  ab$) and (b) along the $c$~axis ($H \parallel c$).}
\label{Fig:Fig_Sr10_MH}
\end{figure}

\begin{figure}[h]
\includegraphics[width=4in]{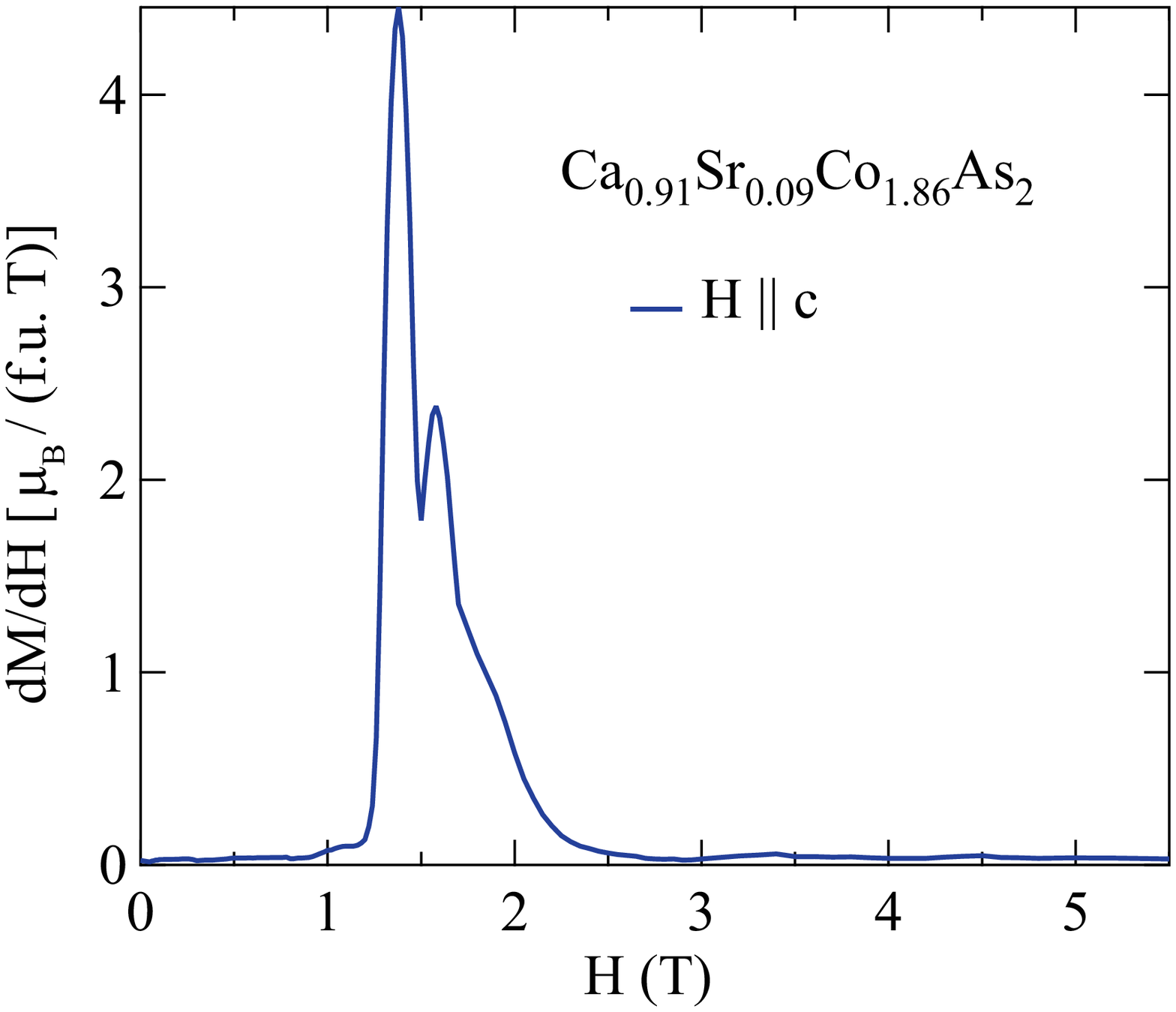}
\caption{(Color online) Derivative $dM/dH$ versus applied magnetic field~$H$ of a ${\rm Ca_{0.91}Sr_{0.09}Co_{1.86}As_2}$ single crystal at temperature $T=2$~K with the field applied along the $c$-axis.}
\label{Fig:dM_dH_Sr09}
\end{figure}

\clearpage

\section*{${\rm\bf Ca_{0.85}Sr_{0.15}Co_{1.88}As_2}$}

\begin{figure}[h]
\includegraphics[width=3.3in]{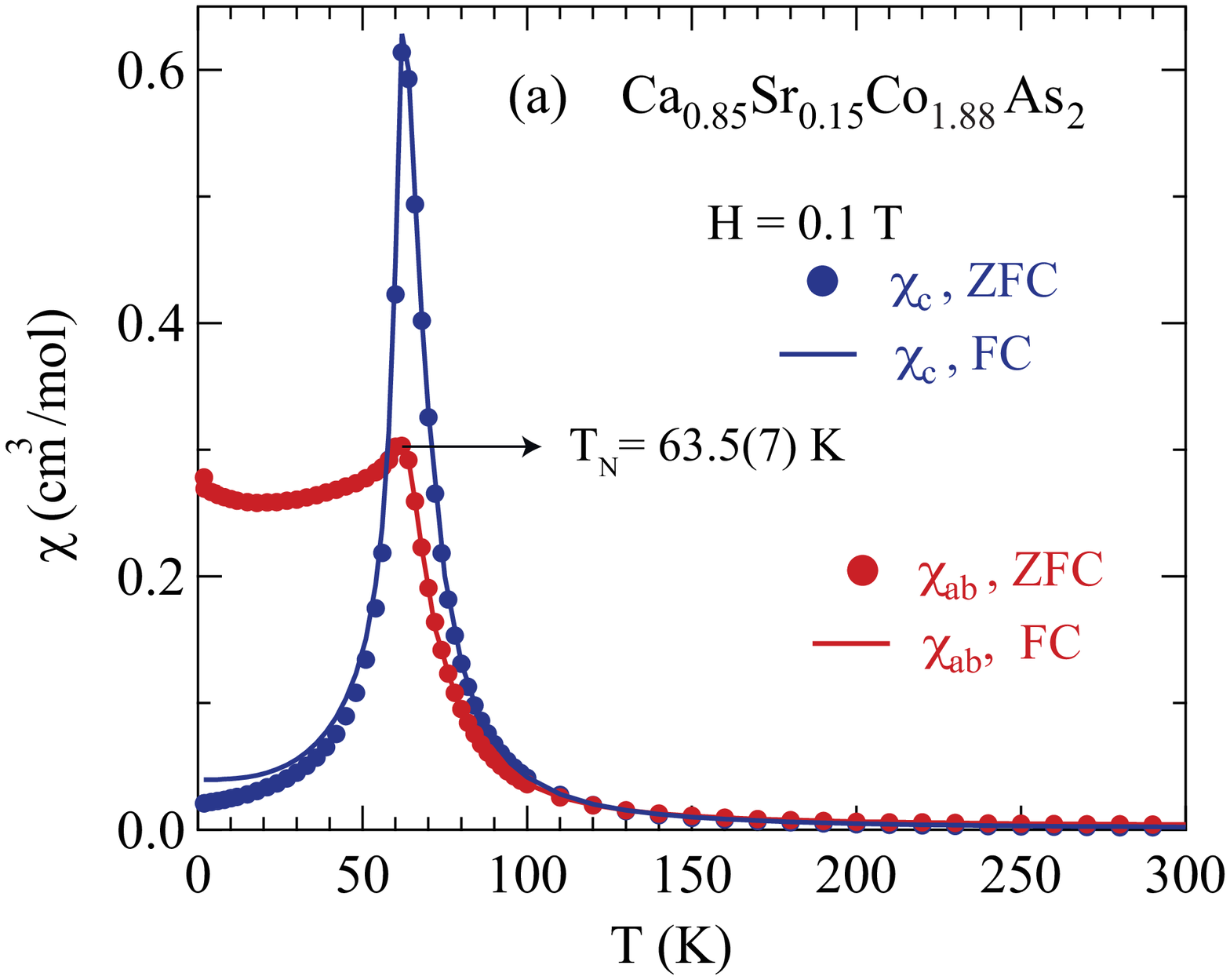}
\includegraphics[width=3.5in]{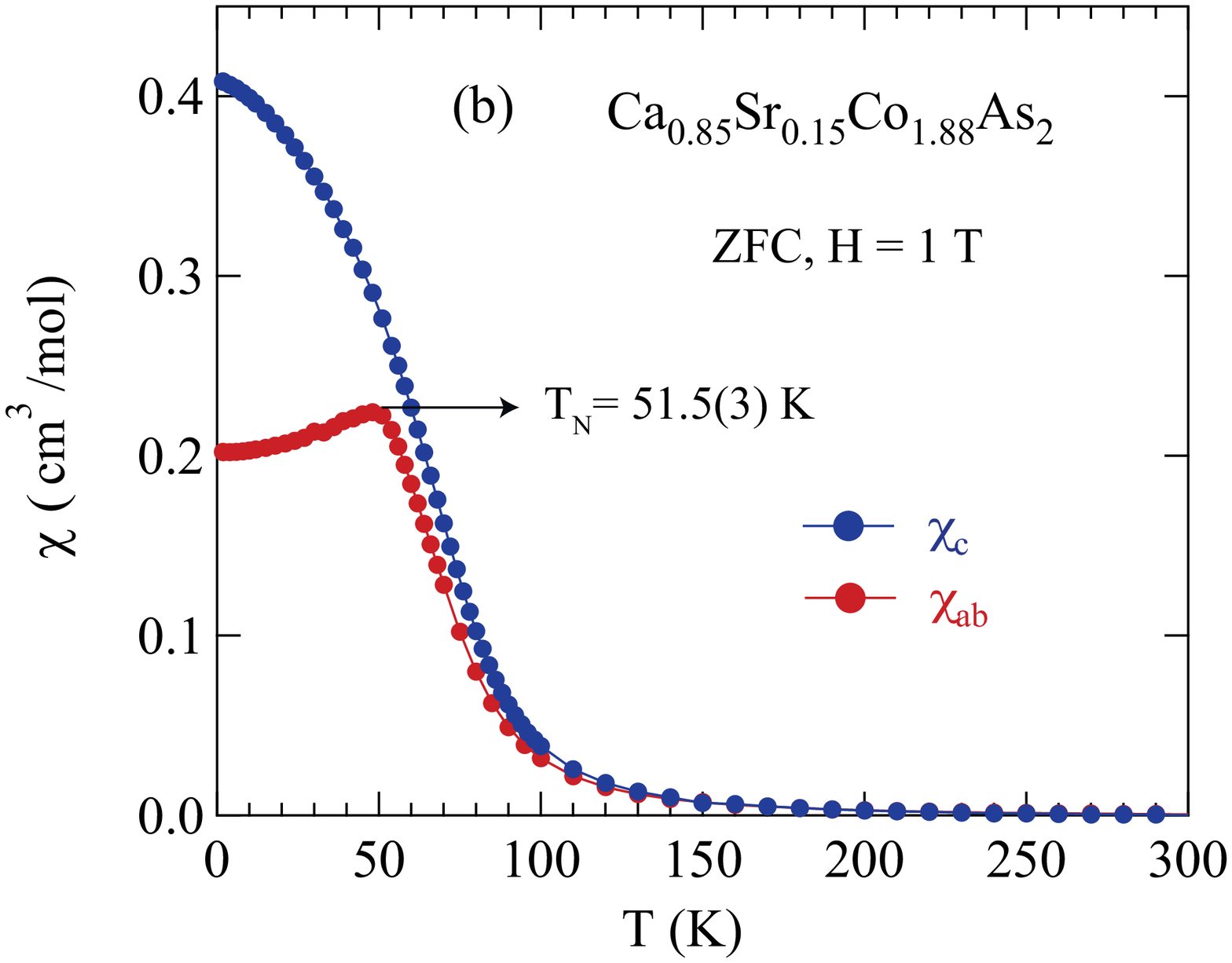}\vspace{0.2in}
\includegraphics[width=3.5in]{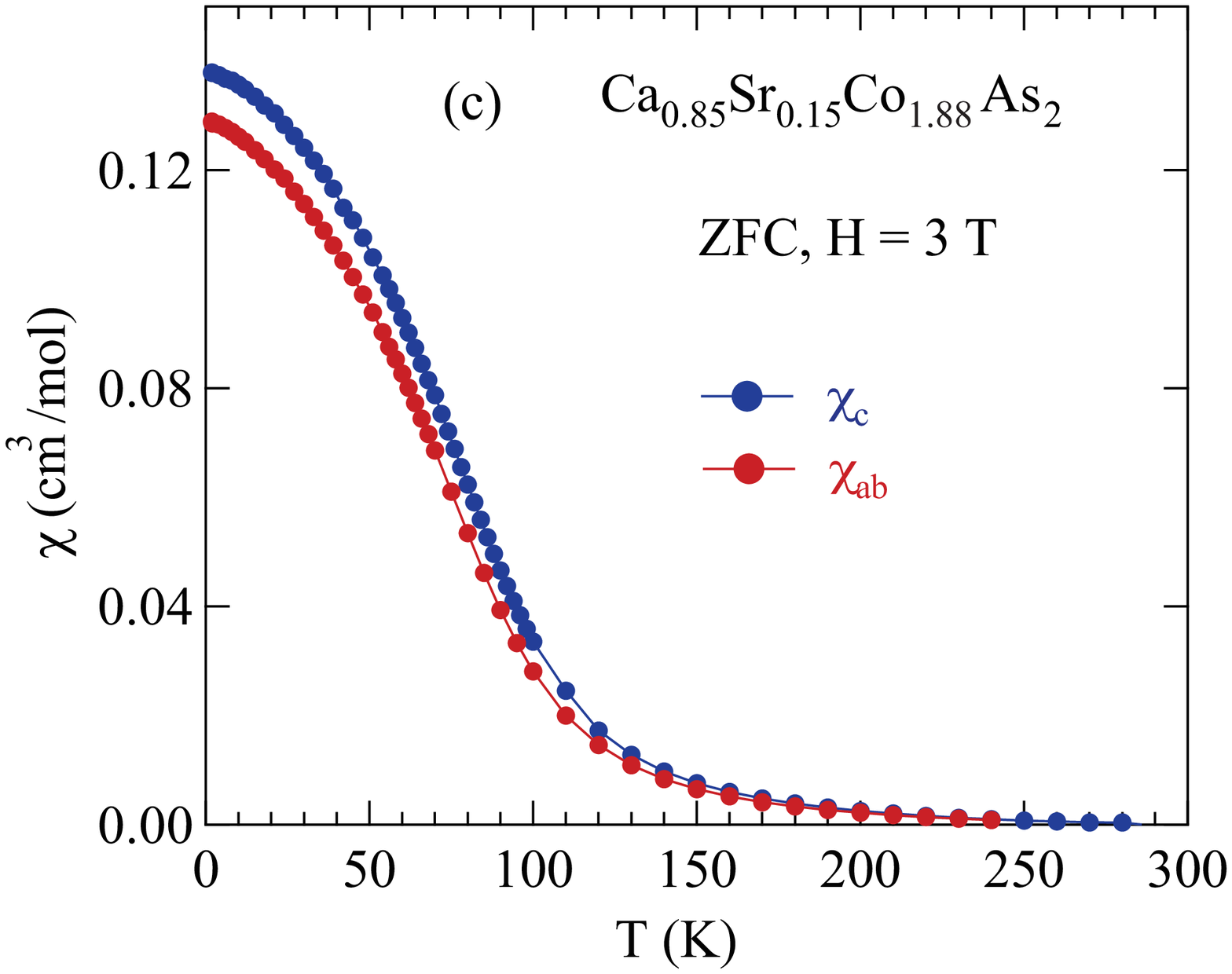}
\caption{(Color online) Zero-field-cooled (ZFC) and field-cooled (FC) magnetic susceptibilities $\chi\equiv M/H$ of a ${\rm Ca_{0.85}Sr_{0.15}Co_{1.88}As_2}$ single crystal versus temperature $T$ measured in magnetic fields (a)~$H=0.1$~T, including ZFC and FC data, (b)~$H=1$~T and (c)~$H=3$~T applied in the $ab$~plane ($\chi_{ab}$) and along the $c$~axis ($\chi_c$).}
\label{Fig:Fig_Sr15_Chi}
\end{figure}

\begin{figure}[h]
\includegraphics[width=4in]{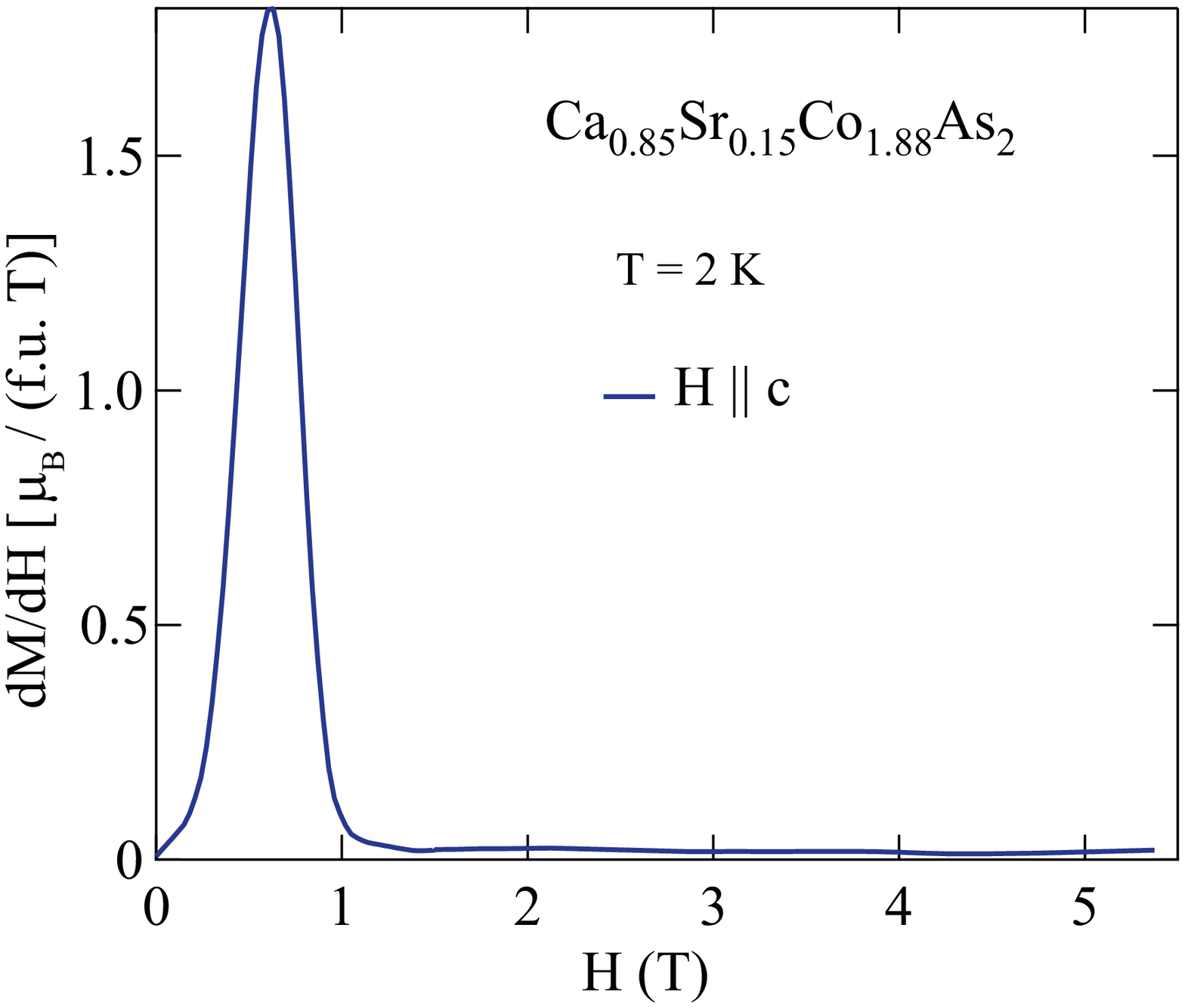}
\caption{(Color online) Derivative $dM/dH$ versus applied magnetic field~$H$ of a ${\rm Ca_{0.85}Sr_{0.15}Co_{1.88}As_2}$ single crystal at temperature $T=2$~K with the field applied along the $c$-axis.}
\label{Fig:dM_dH_Sr15}
\end{figure}

\clearpage

\section*{${\rm\bf Ca_{0.81}Sr_{0.19}Co_{1.90}As_2}$}

\begin{figure}[h]
\includegraphics[width=4in]{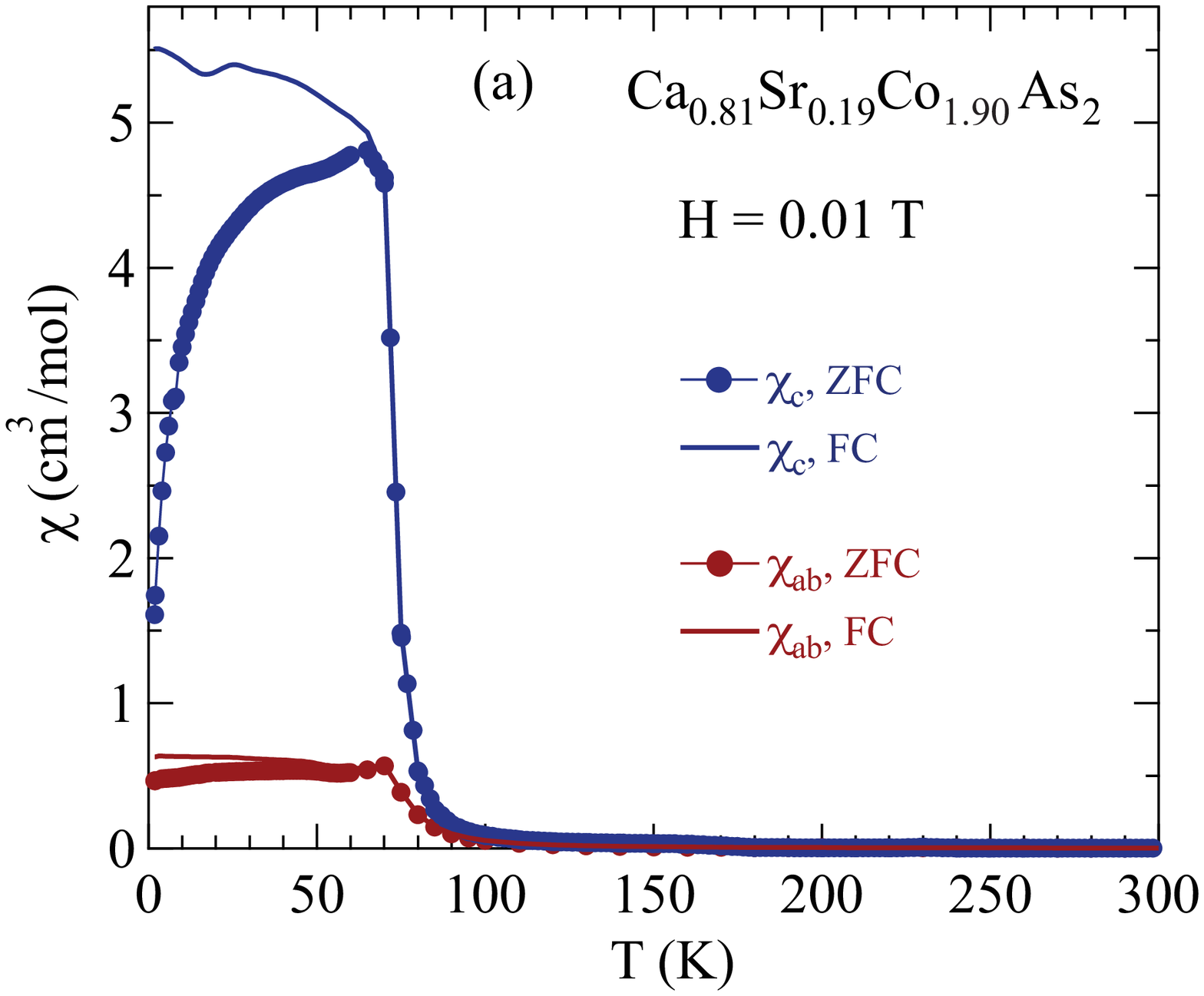}\vspace{0.2in}
\includegraphics[width=4.2in]{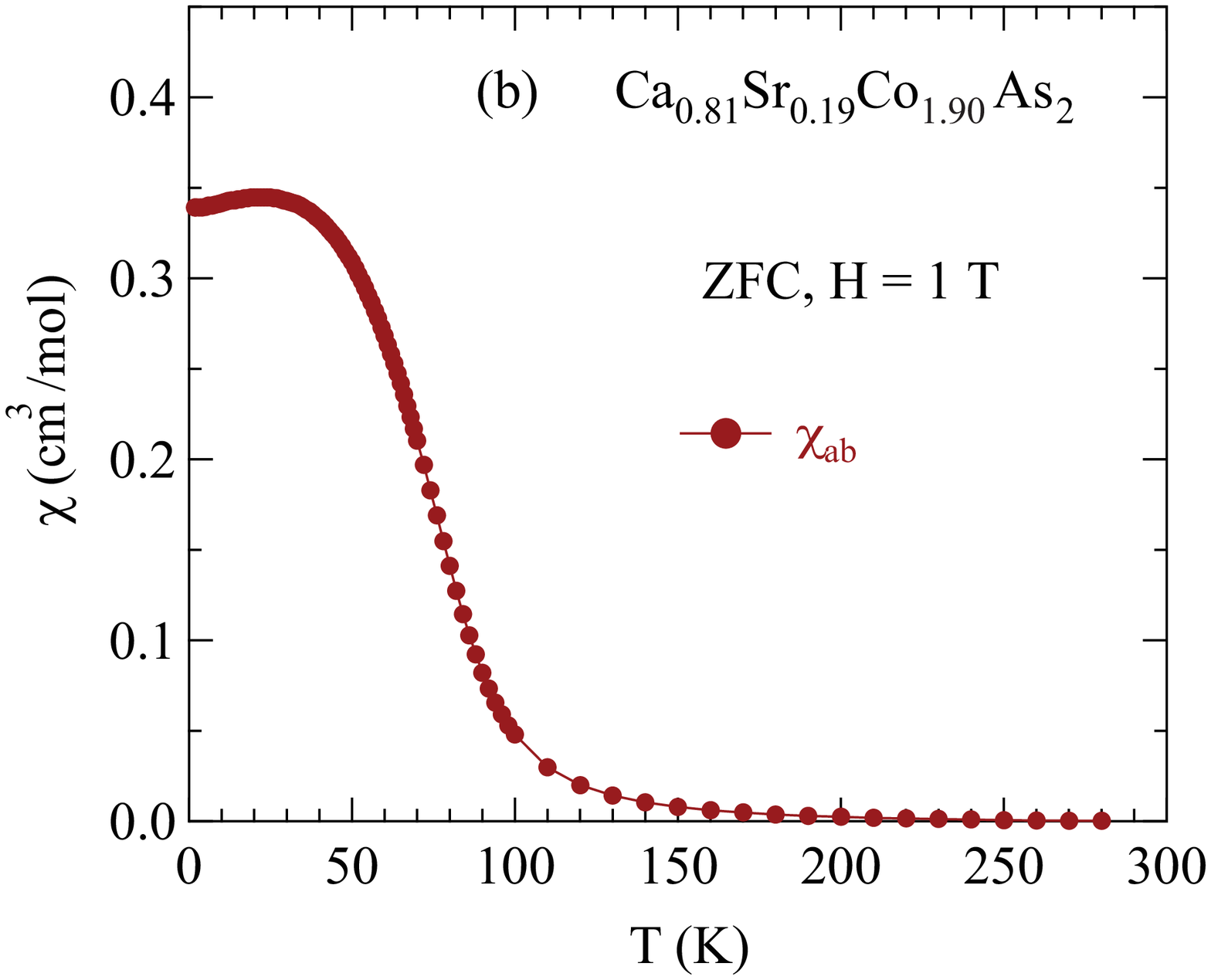}
\caption{(Color online) Zero-field-cooled magnetic susceptibility $\chi$ of a ${\rm Ca_{0.81}Sr_{0.19}Co_{1.90}As_2}$ single crystal versus temperature $T$ measured magnetic fields $H$ applied in the $ab$~plane ($\chi_{ab}$) and along the $c$~axis ($\chi_c$).  (a)~The field is $H=0.01$~T, and both field-cooled (FC) and zero-field-cooled (ZFC) data are included for both field directions.  (b)~The field is $H=1$~T\@.  The measurement is carried out with $H\parallel ab$.}
\label{Fig:MT_Sr19}
\end{figure}

\begin{figure}[h]
\includegraphics[width=4in]{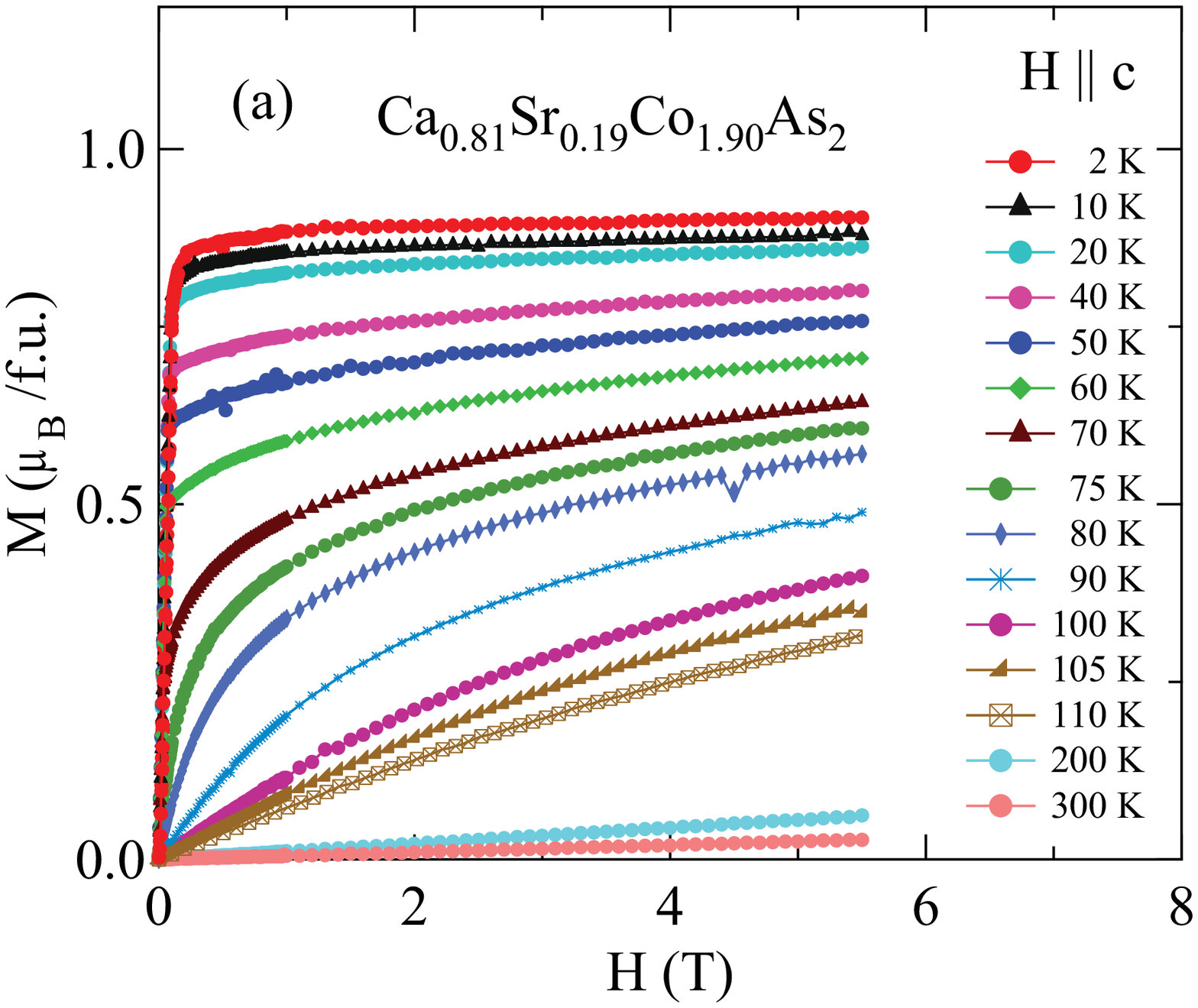}\vspace{0.1in}
\includegraphics[width=4.1in]{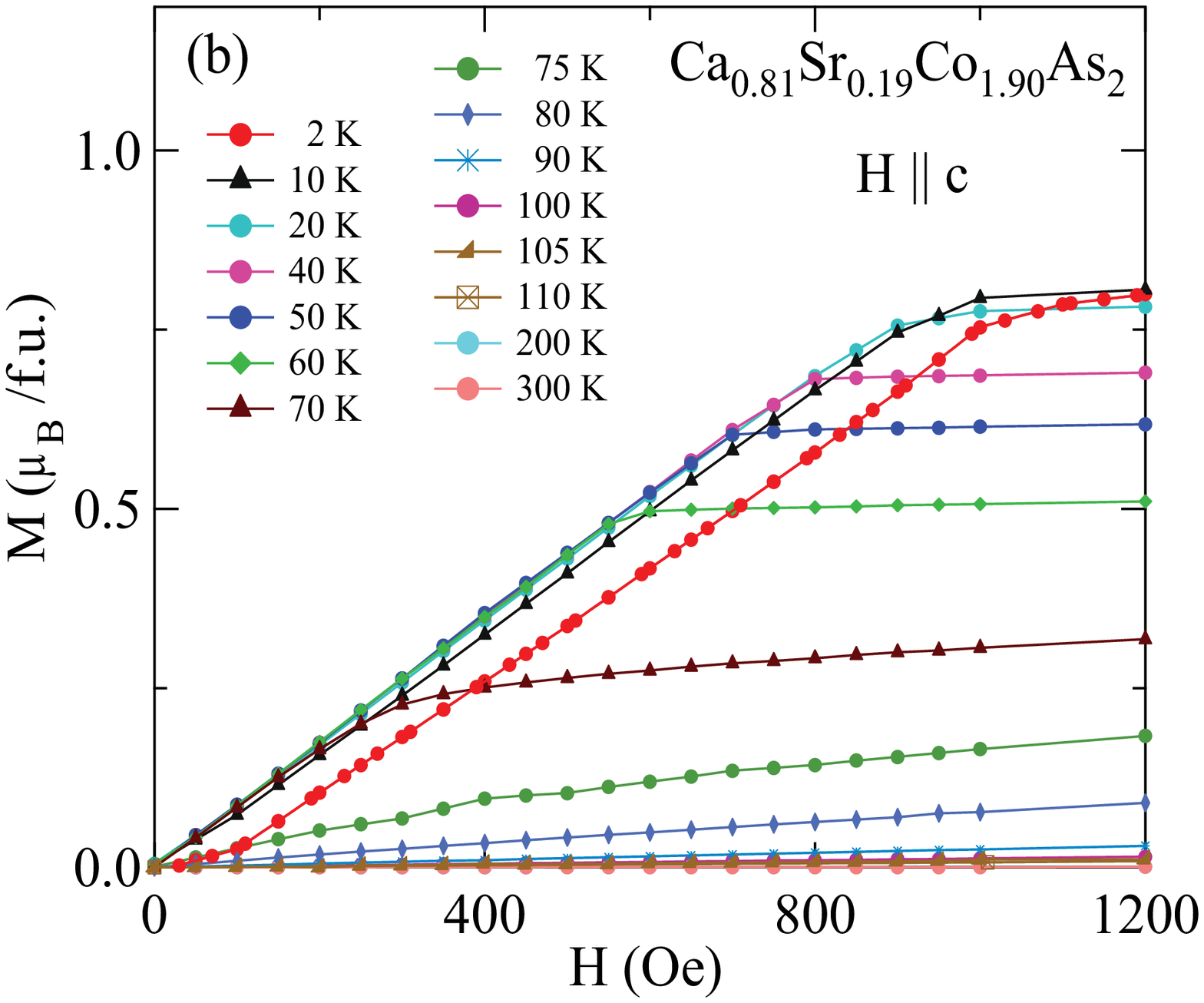}
\caption{(Color online) (a)~Isothermal magnetization $M$ of a ${\rm Ca_{0.81}Sr_{0.19}Co_{1.90}As_2}$  single crystal versus applied magnetic field $H$ for $H$ applied along the $c$~axis ($H \parallel c$) at the indicated temperatures.  (b)~Expanded plots of the data in~(a) at low fields.}
\label{Fig:MH_Sr19}
\end{figure}

\clearpage

\section*{${\rm\bf Ca_{0.75}Sr_{0.25}Co_{1.90}As_2}$}

\begin{figure}[h]
\includegraphics[width=4in]{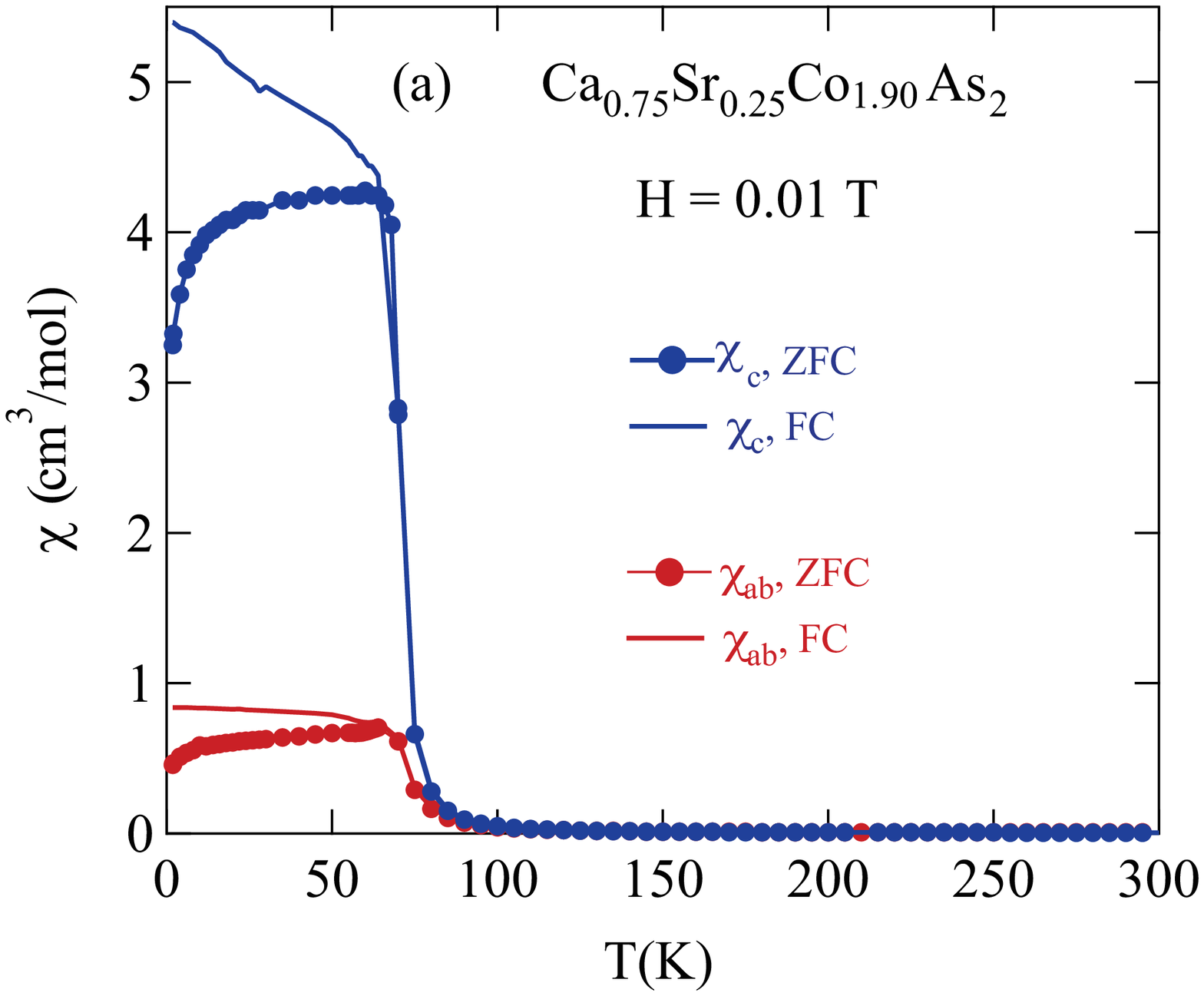}\vspace{0.2in}
\includegraphics[width=4.2in]{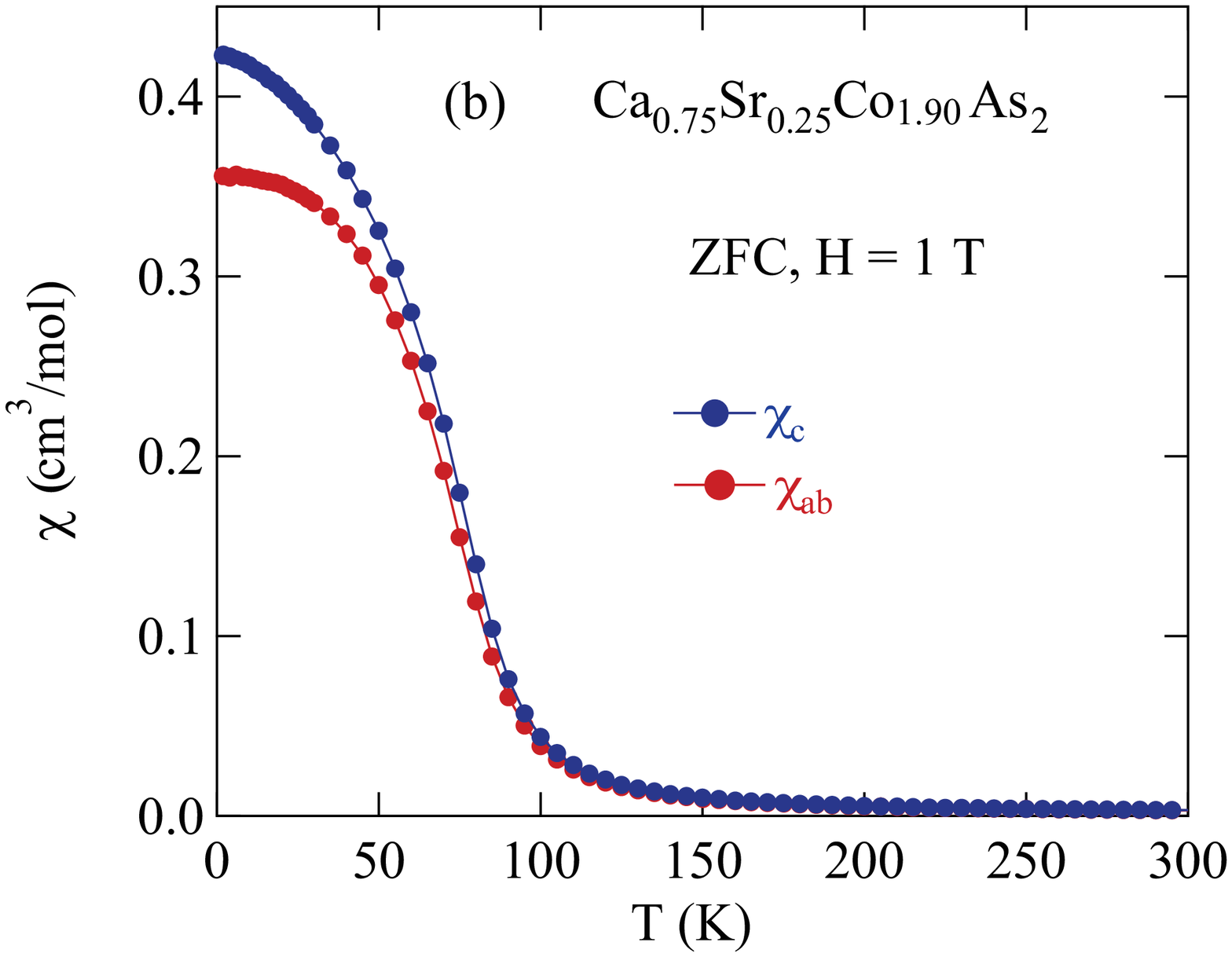}
\caption{(Color online) Zero-field-cooled (ZFC) magnetic susceptibility $\chi$ of a ${\rm Ca_{0.75}Sr_{0.25}Co_{1.90}As_2}$ single crystal versus temperature~$T$ measured in magnetic fields~$H$ applied in the $ab$~plane ($\chi_{ab}$) and along the $c$~axis ($\chi_c$).  (a)~Field-cooled (FC) and ZFC data in $H=0.01$~T for both field directions.  (b)~ZFC $\chi_{ab}$ and $\chi_c$ in $H=1$~T.}
\label{Fig:Sr19_Chi}
\end{figure}

\begin{figure}[h]
\includegraphics[width=4.in]{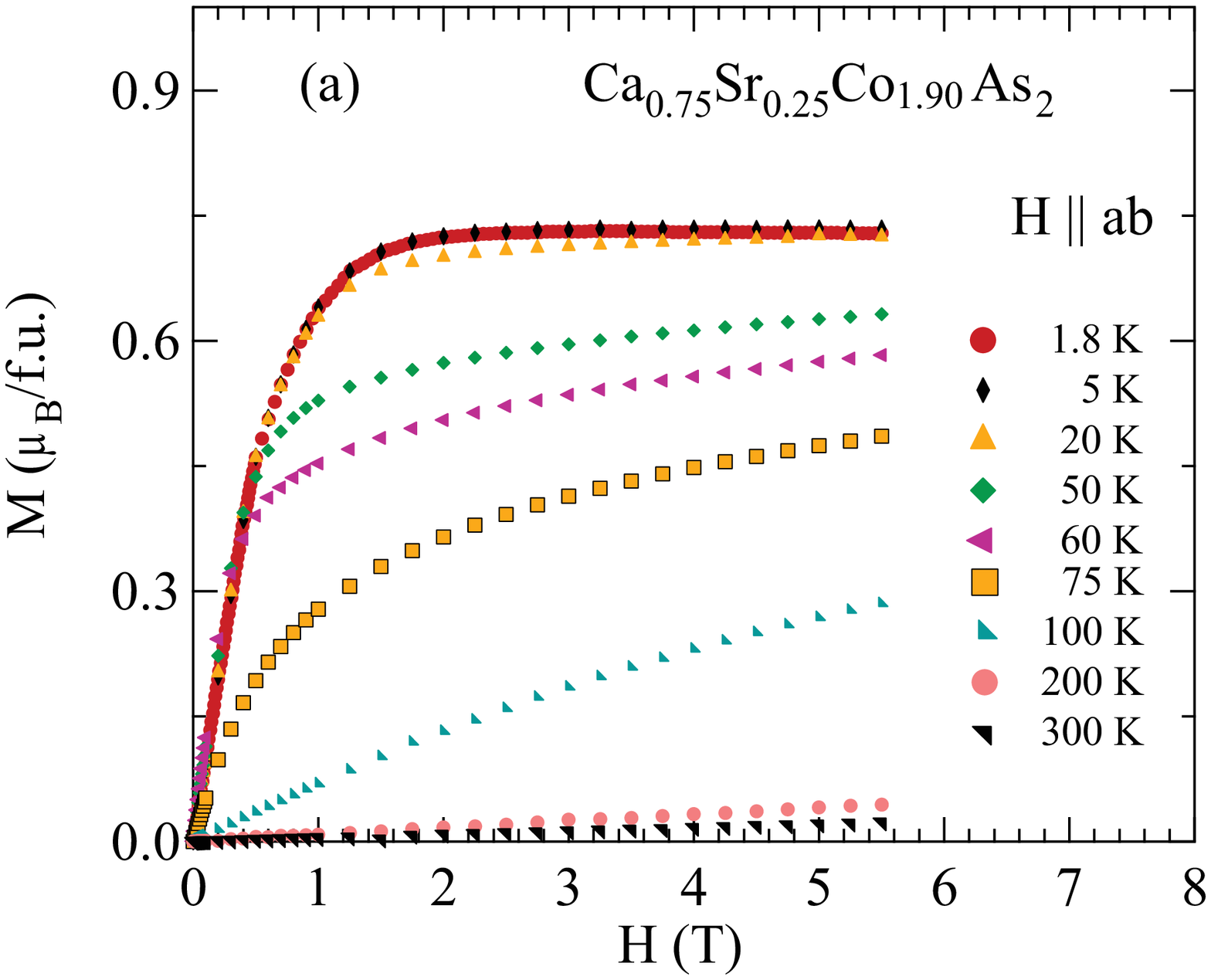}\vspace{0.2in}
\includegraphics[width=3.4in]{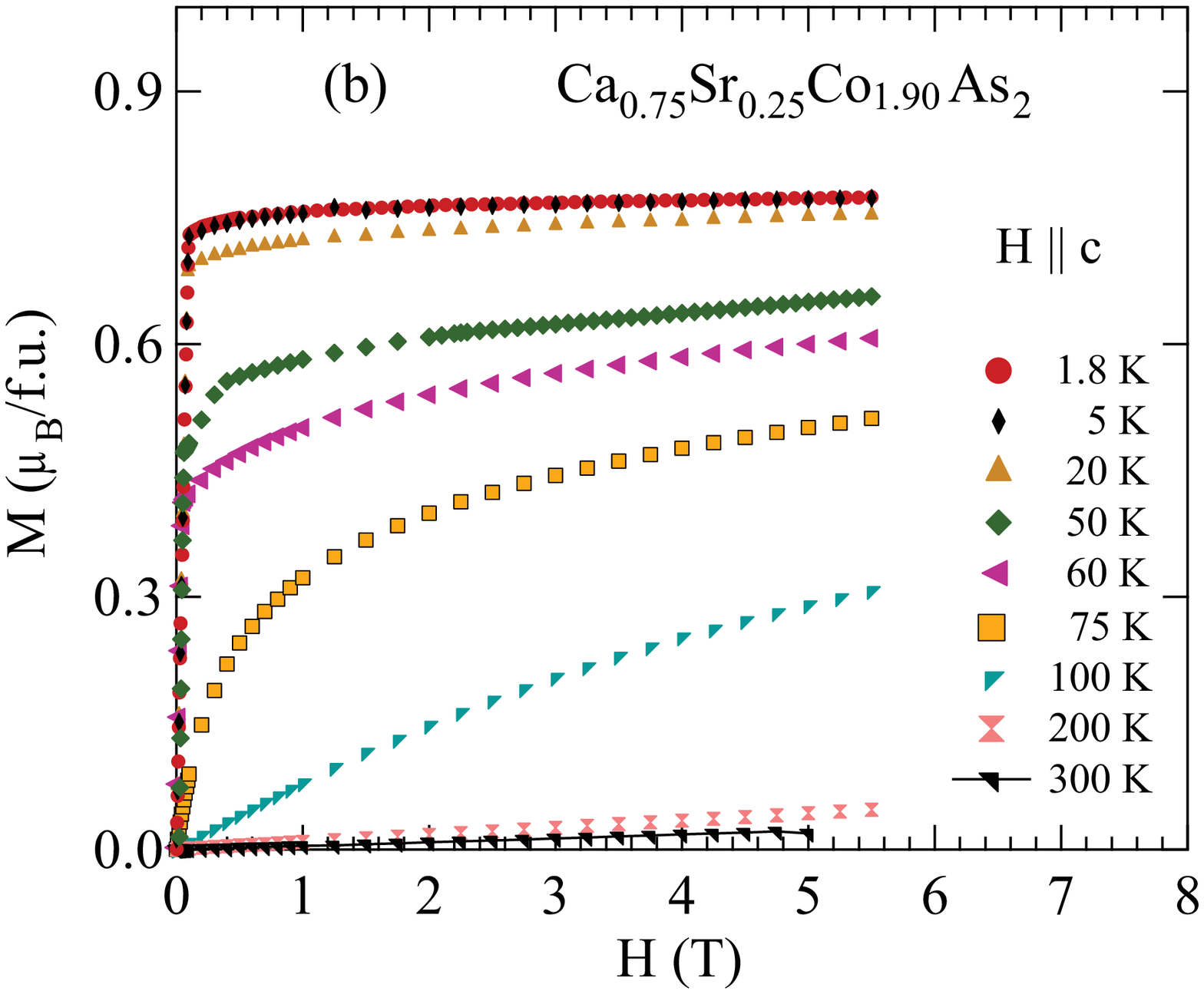}\includegraphics[width=3.4in]{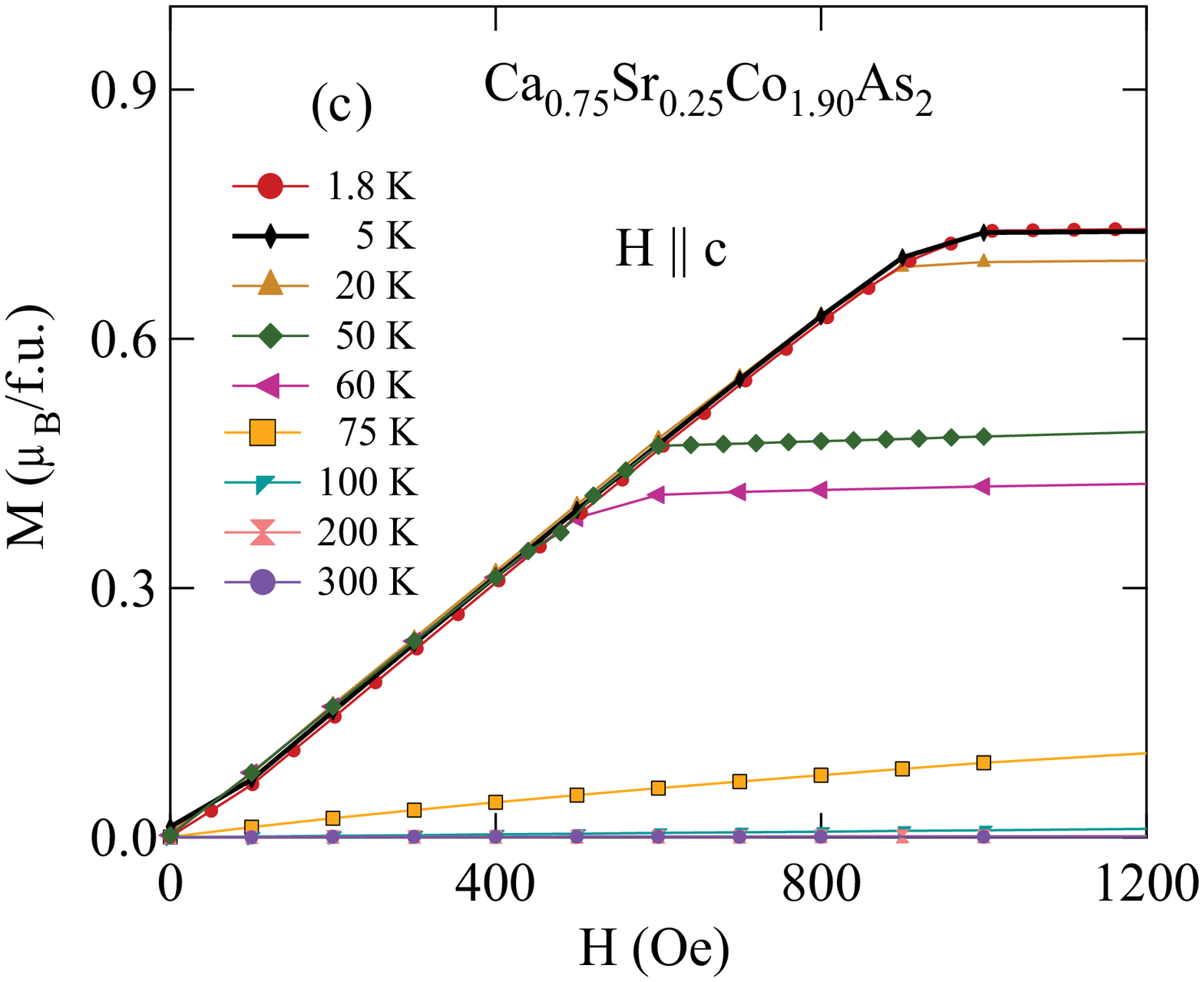}
\caption{(Color online) Isothermal magnetization $M$ of a ${\rm Ca_{0.75}Sr_{0.25}Co_{1.90}As_2}$  single crystal versus applied magnetic field $H$ measured at the indicated temperatures for $H$ applied (a) in the $ab$-plane ($M_{ab}, H \perp  c$) and (b) along the $c$-axis ($M_c, H \parallel c$), (c) Expanded plots of the data in (b) at low fields.}
\label{fig:MH_Sr25}
\end{figure}

\clearpage

\section*{${\rm\bf Ca_{0.72}Sr_{0.28}Co_{1.91}As_2}$}

\begin{figure}[h]
\includegraphics[width=4.5in]{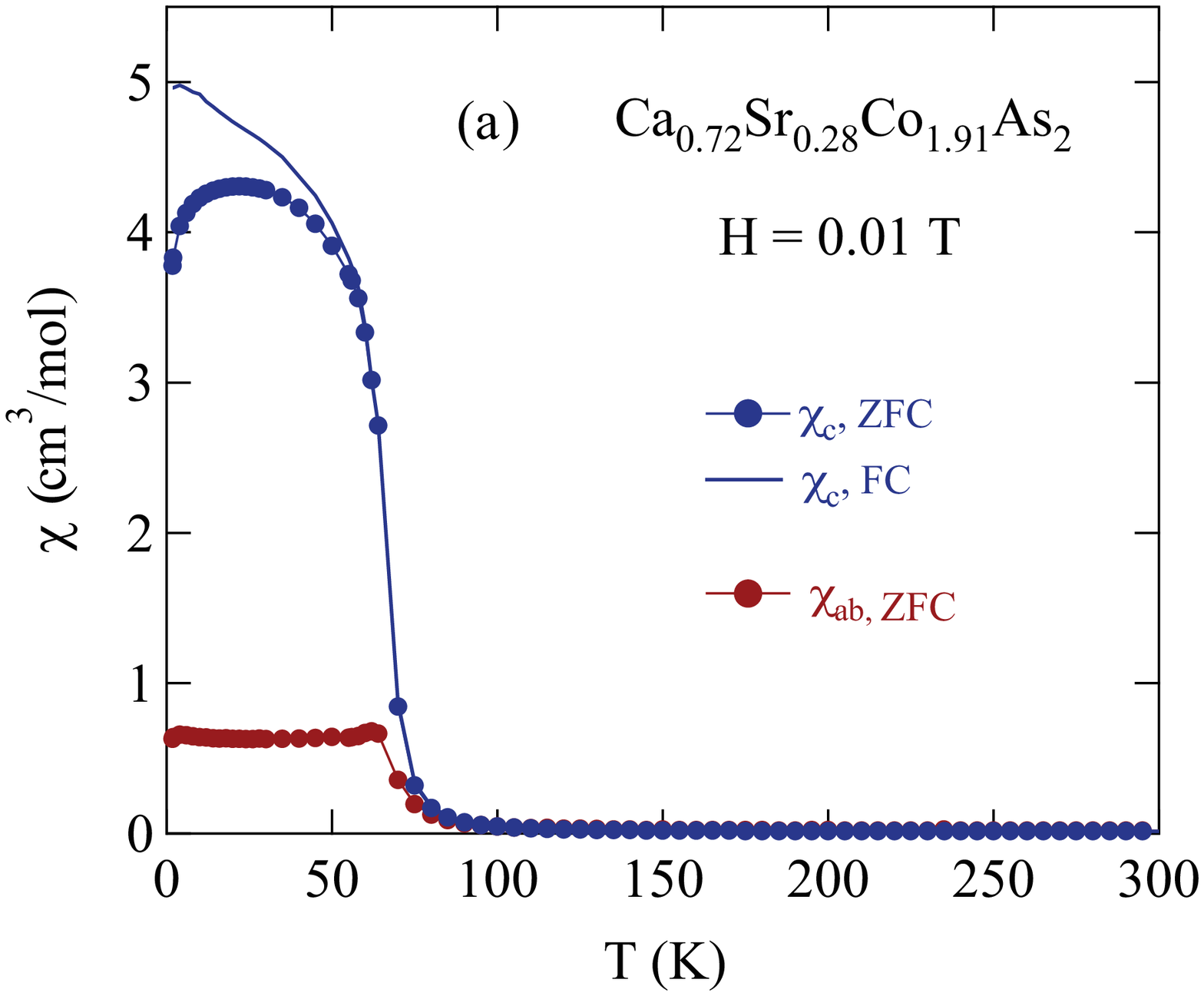}
\includegraphics[width=4.5in]{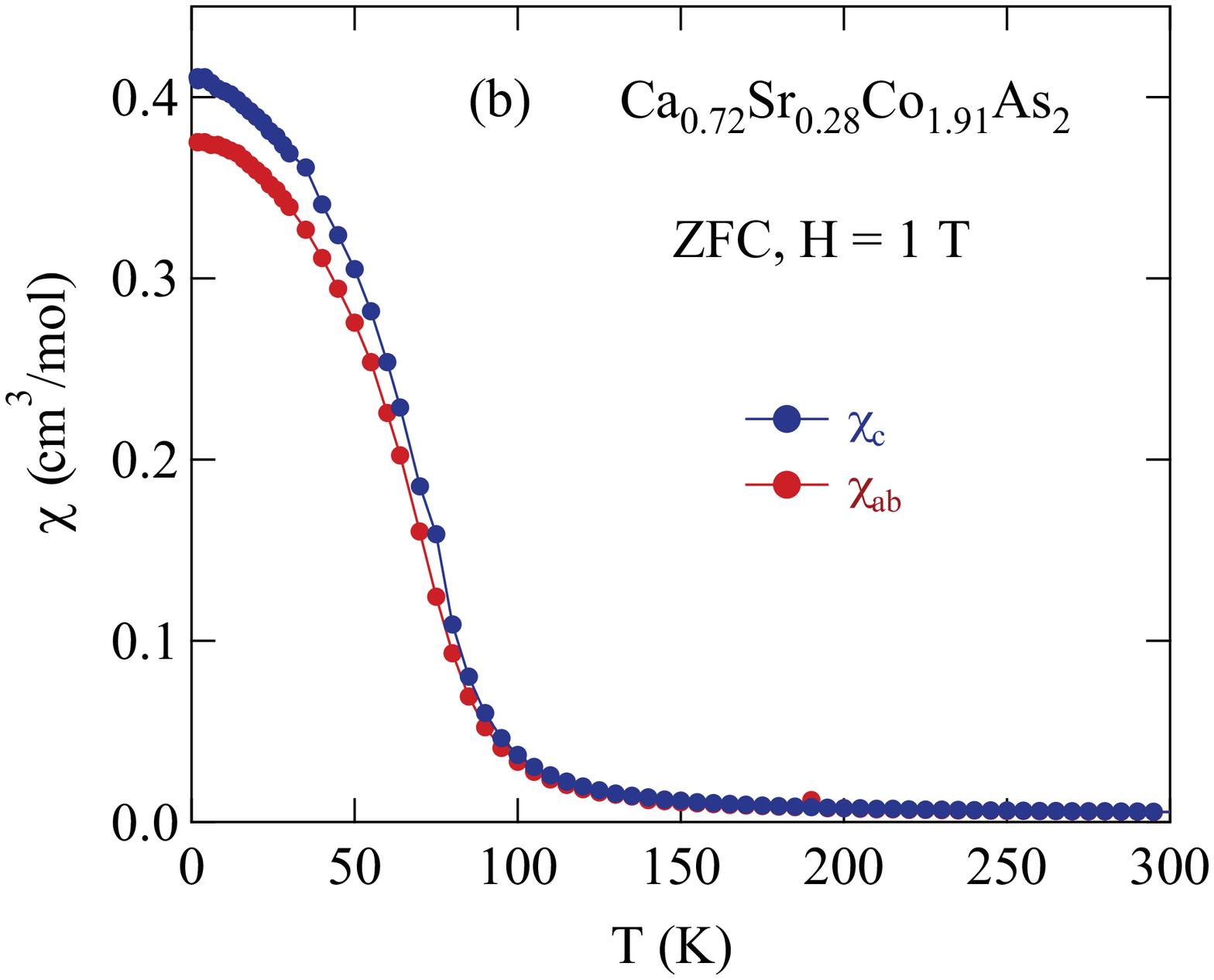}
\caption{(Color online) Zero-field-cooled (ZFC) and/or field-cooled (FC) magnetic susceptibility $\chi$ of a ${\rm Ca_{0.72}Sr_{0.28}Co_{1.91}As_2}$ single crystal versus temperature~$T$ measured in different magnetic fields~$H$ applied in the $ab$~plane ($\chi_{ab}$) and along the $c$~axis ($\chi_c$) in applied magnetic fields (a)~$H=0.01$~T and (b)~$H=1$~T.}
\label{Fig:MT_Sr27}
\end{figure}

\begin{figure}[h]
\includegraphics[width=4in]{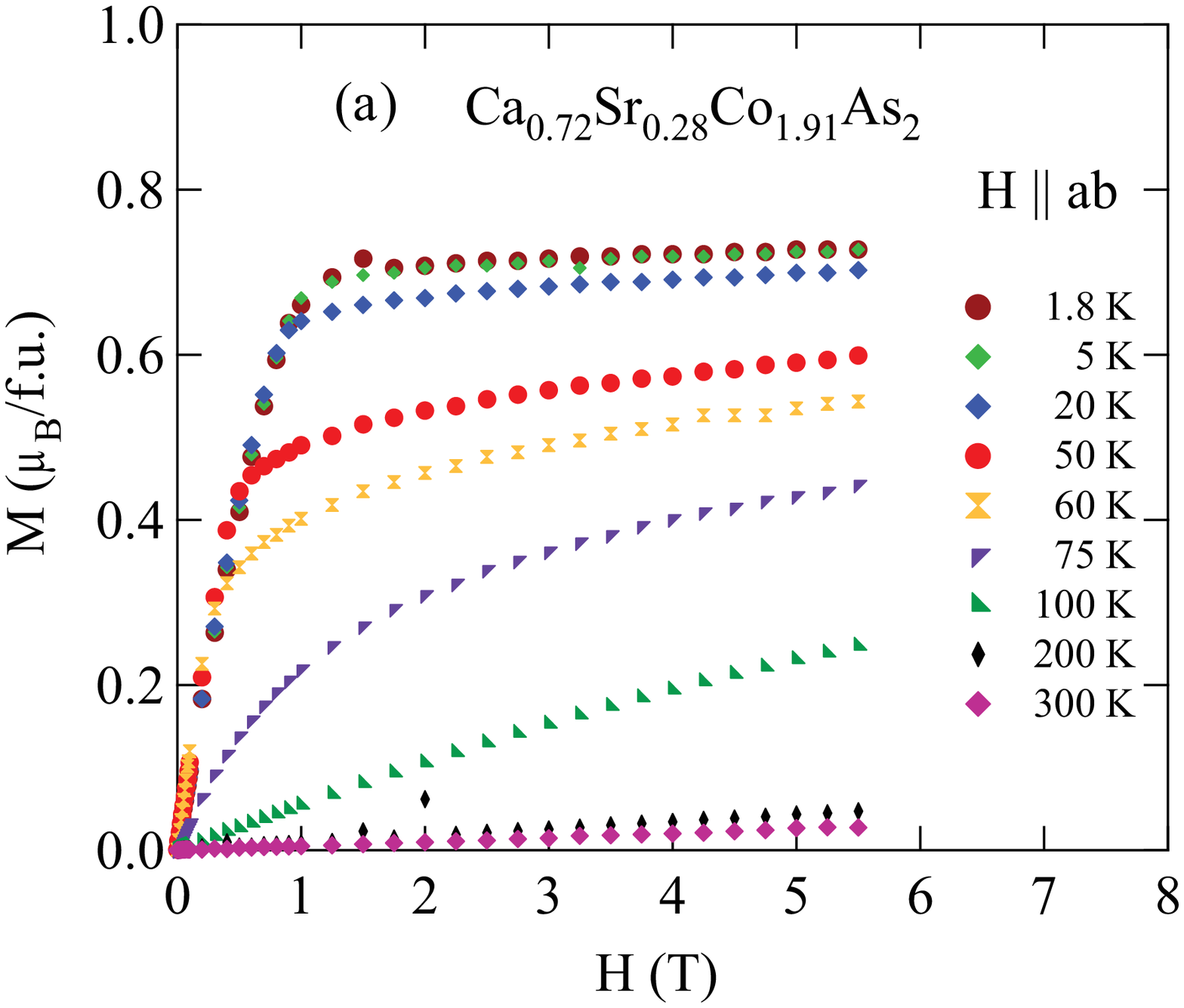}\vspace{0.2in}
\includegraphics[width=3.5in]{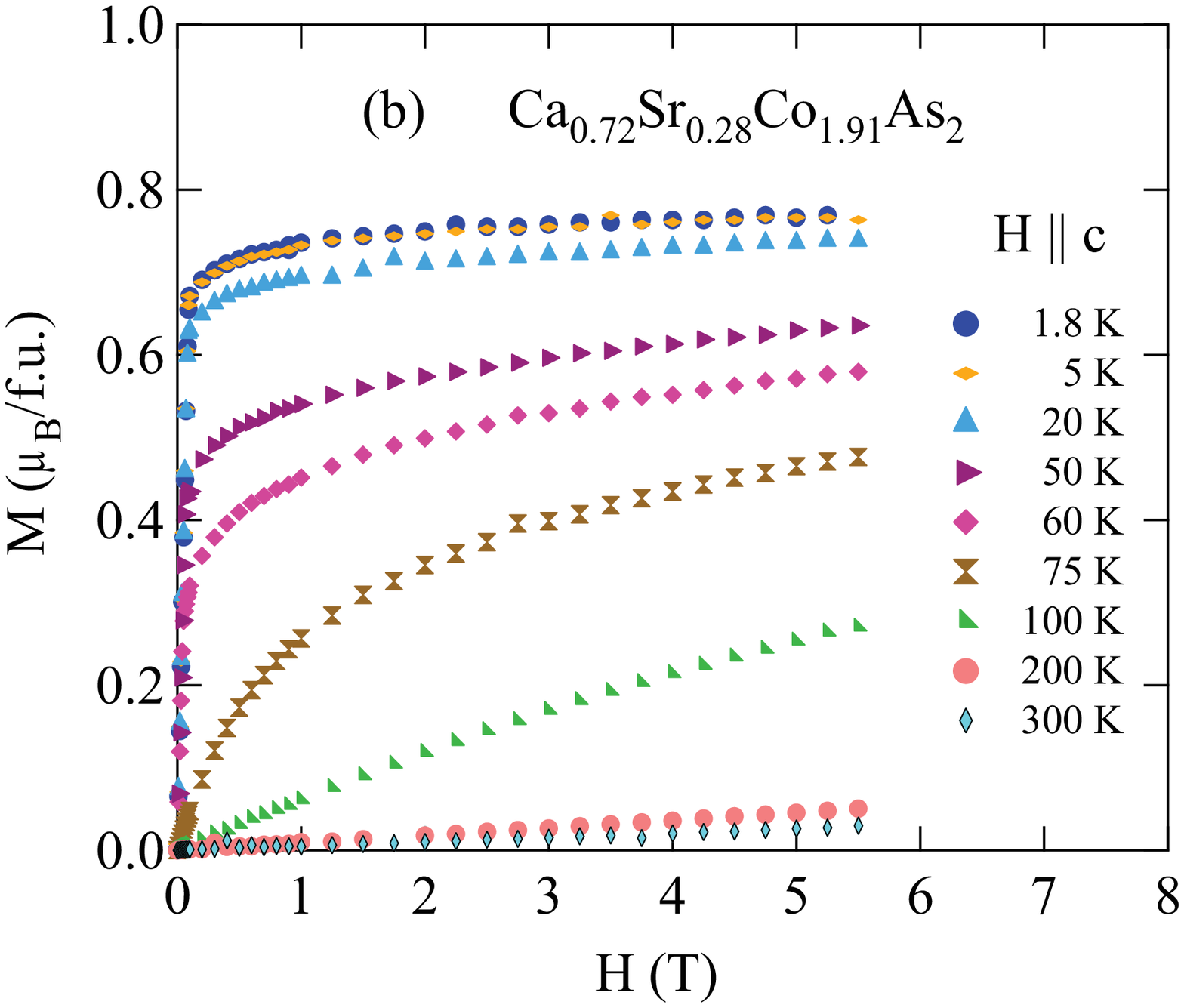}\includegraphics[width=3.5in]{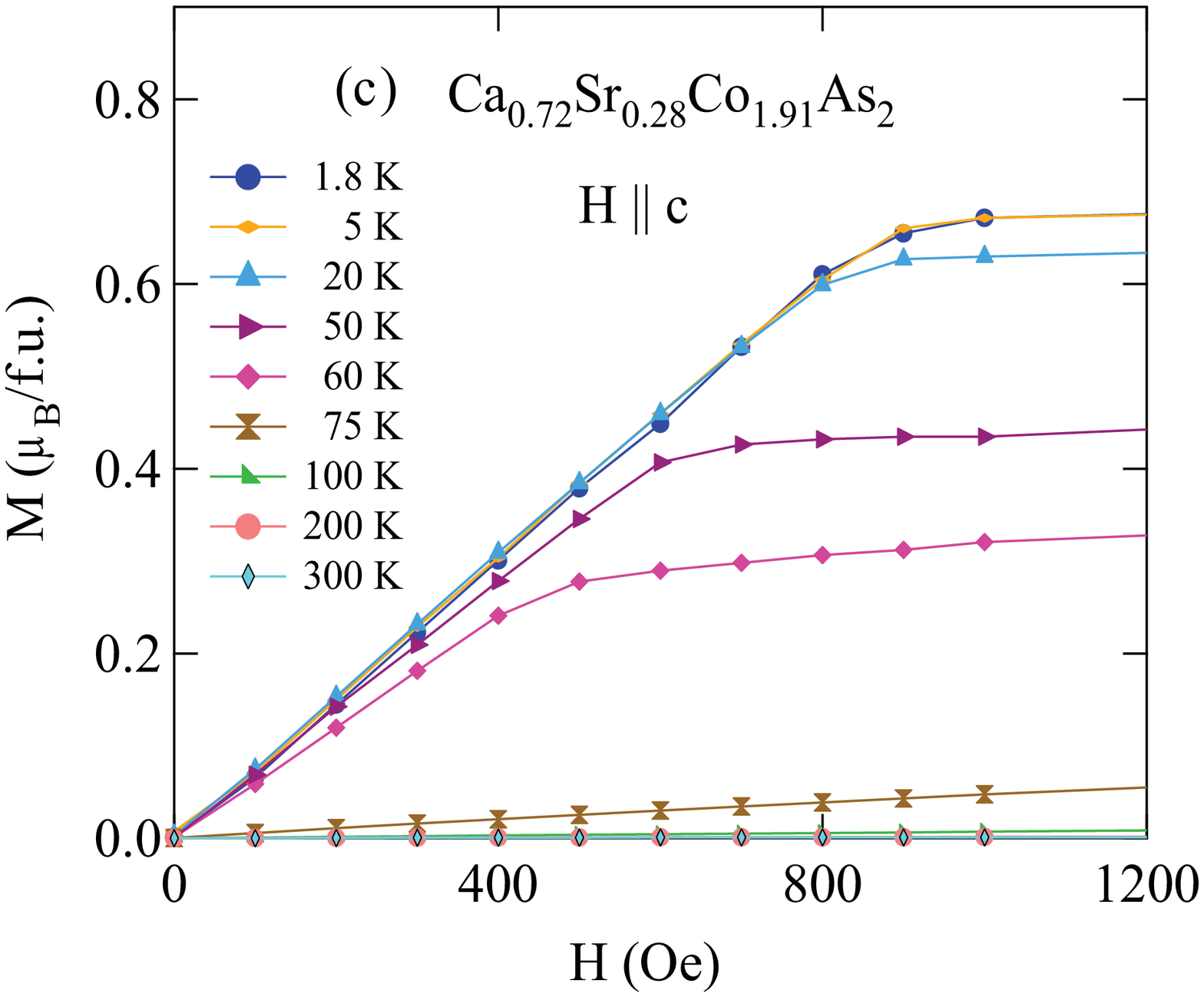}
\caption{(Color online) Isothermal magnetization~$M$ of a ${\rm Ca_{0.72}Sr_{0.28}Co_{1.90}As_2}$ single crystal versus applied magnetic field~$H$ measured at the indicated temperatures for $H$ applied (a)~parallel to the $ab$~plane ($H \parallel ab$) and  (b) parallel to the $c$~axis ($H\parallel c$), (c)~expanded plots of data in (b) at low fields.}
\label{Fig:MH_Sr27}
\end{figure}

\clearpage

\section*{${\rm\bf Ca_{0.67}Sr_{0.33}Co_{1.90}As_2}$}

\begin{figure}[h]
\includegraphics[width=4.in]{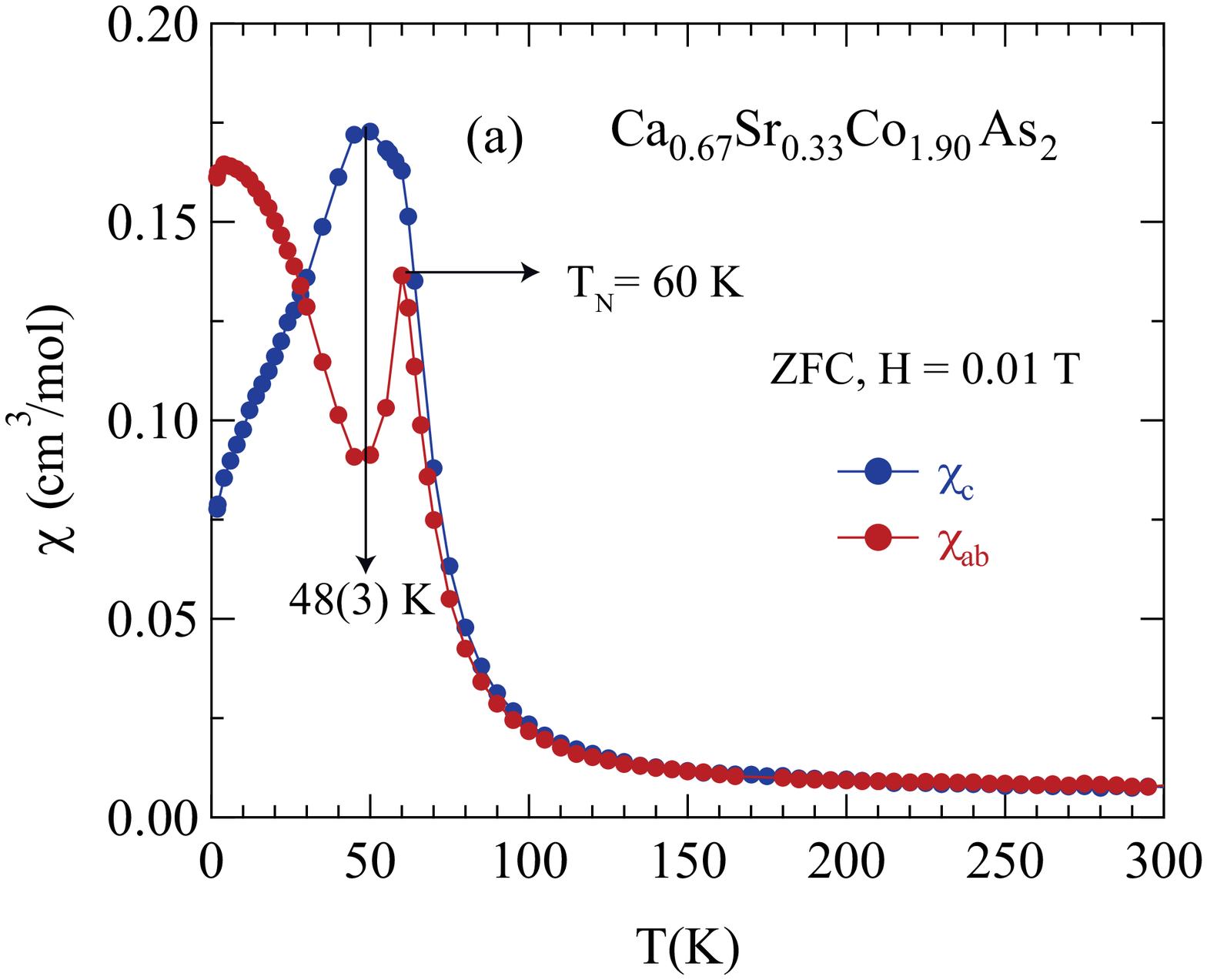}
\includegraphics[width=4.in]{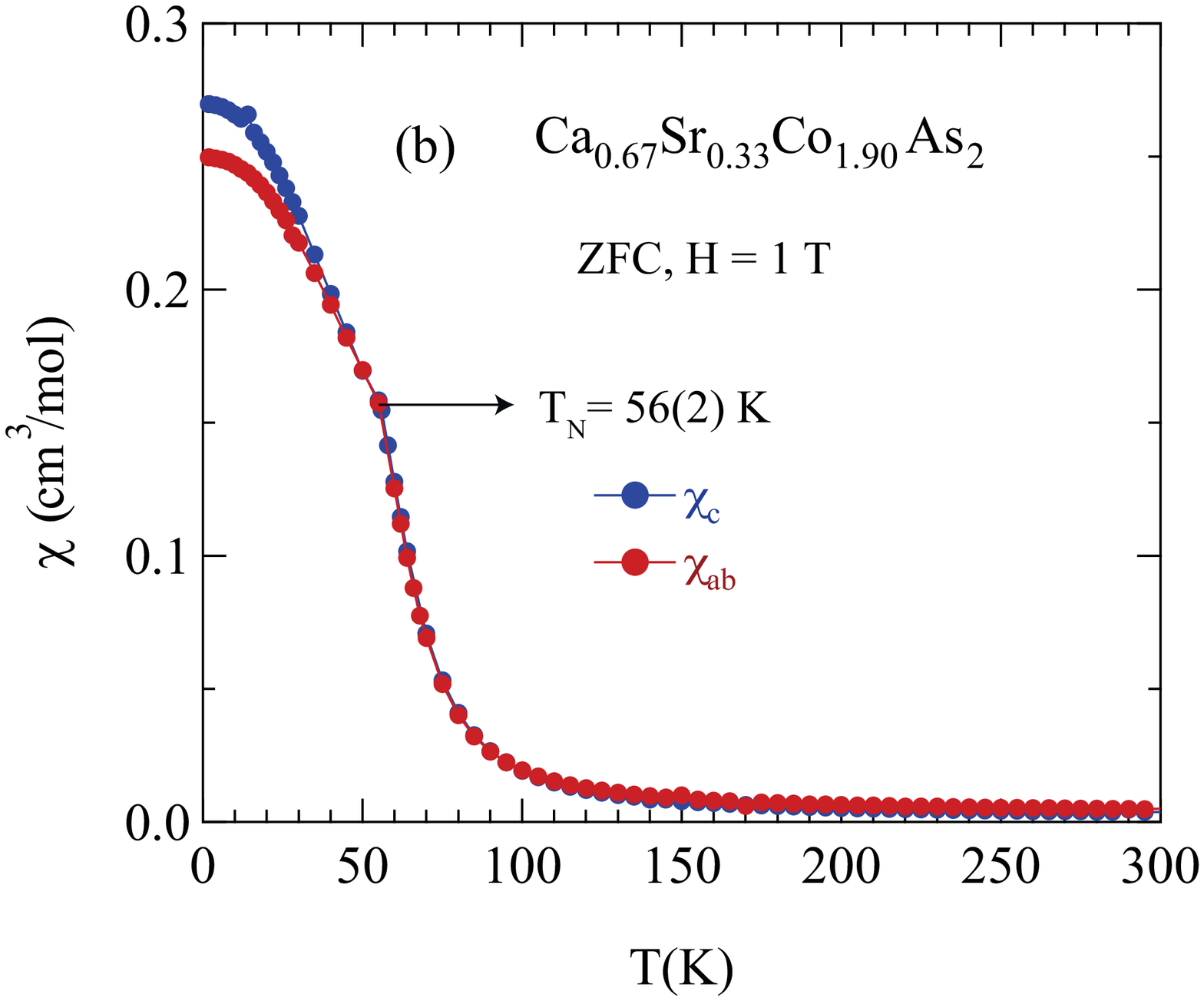}
\caption{(Color online) Zero-field-cooled magnetic susceptibility $\chi$ of a ${\rm Ca_{0.67}Sr_{0.33}Co_{1.90}As_2}$ single crystal versus temperature $T$ measured in different magnetic fields $H$ applied in the $ab$~plane ($\chi_{ab}$) and along the $c$~axis ($\chi_c$) for (a)~$H=0.01$~T and (b)~$H=1$~T\@.  The AFM ordering temperatures~$T_{\rm N}$ are indicated, which show that $T_{\rm N}$ decreases with increasing field.  In~(a), the lower temperature is the spin-reorientation temperature from $ab$-plane moment alignment to $c$-axis alignment with decreasing temperature.  This inference is confirmed by the $M(H)$ isotherm data in Fig.~\ref{Fig:Sr34_MH}.}
\label{Fig:Sr34_Chi}
\end{figure}

\begin{figure}[h]
\includegraphics[width=3.5in]{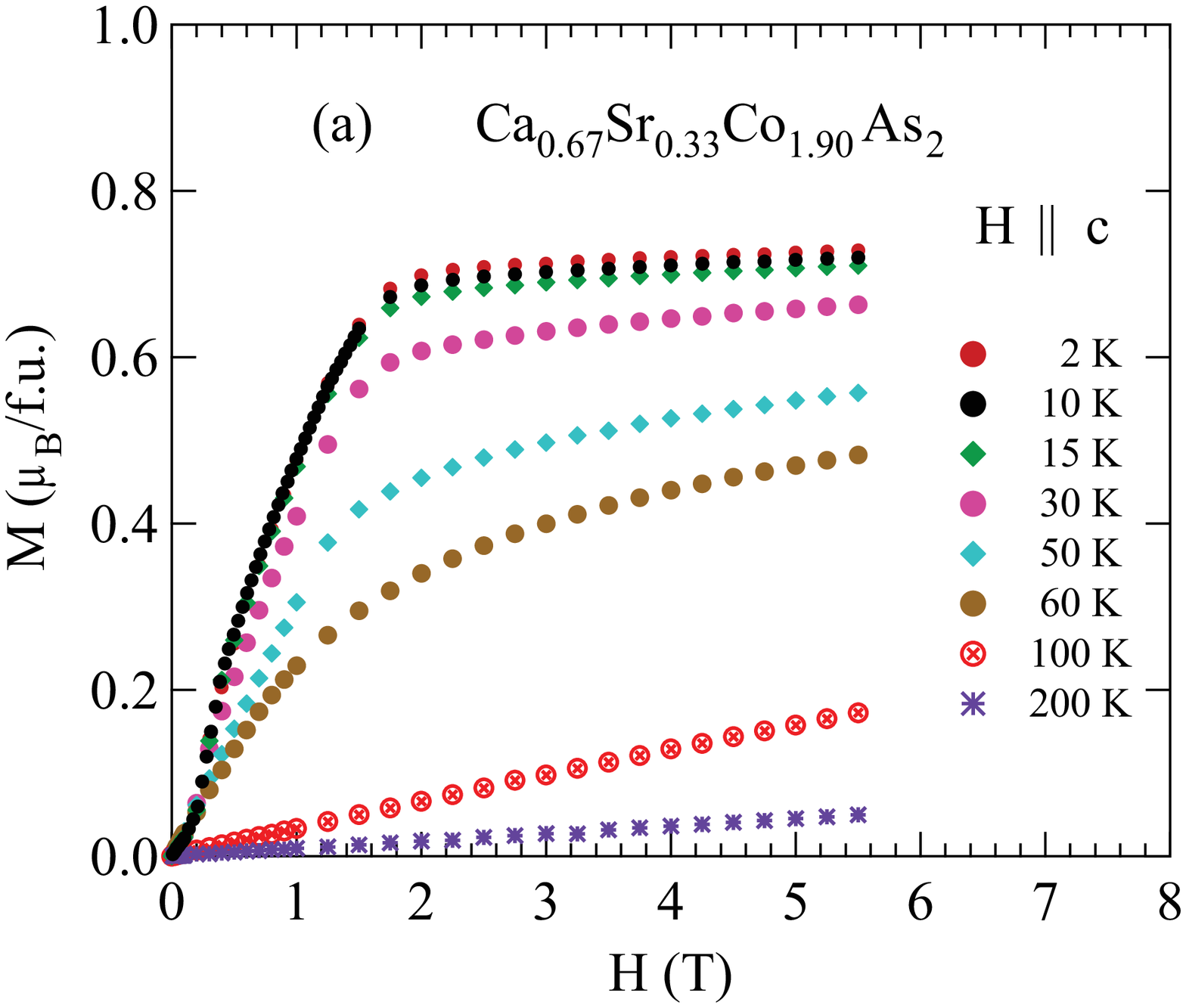}\vspace{0.1in}
\includegraphics[width=3.5in]{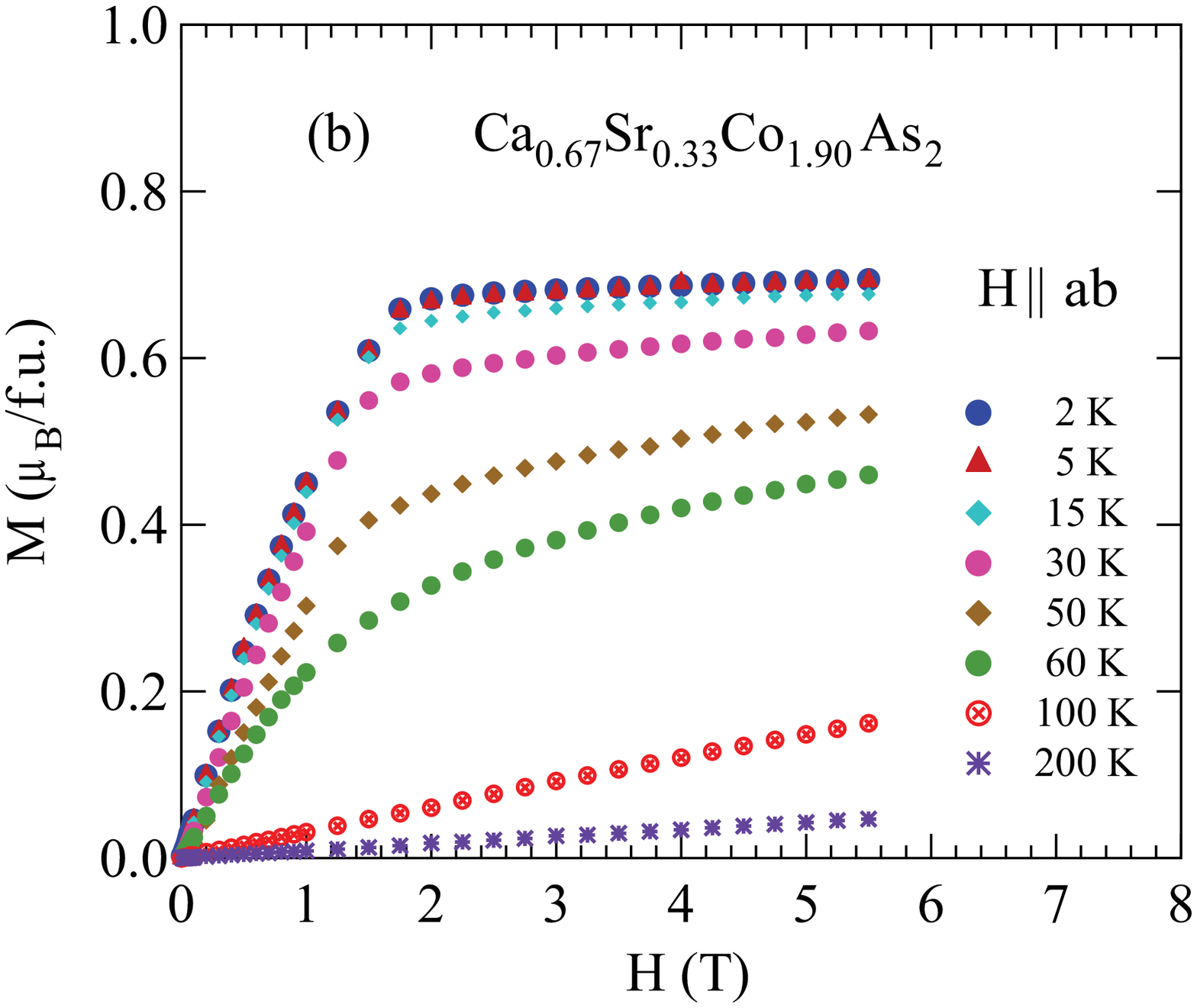}\vspace{0.1in}
\includegraphics[width=3.5in]{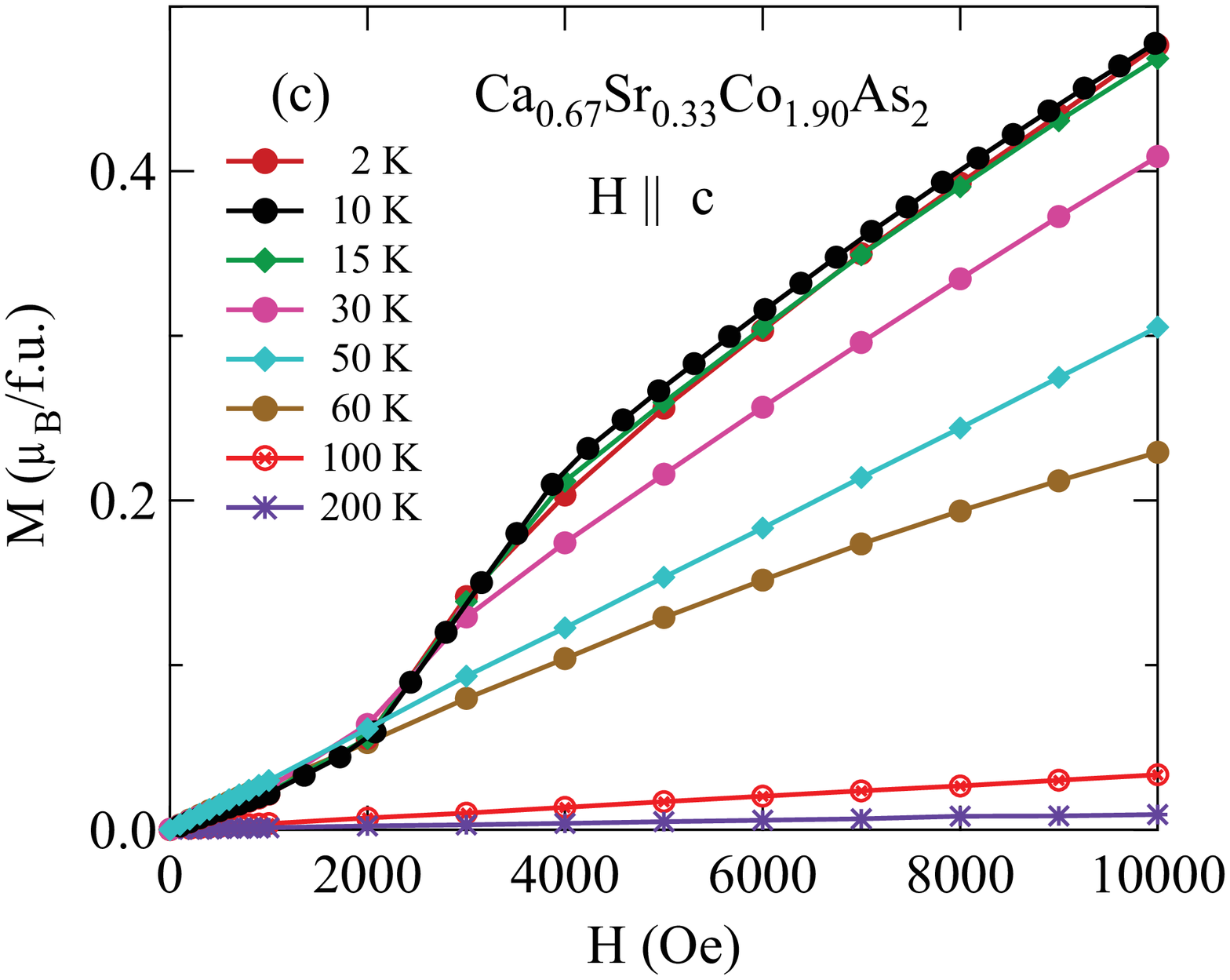}
\includegraphics[width=3.5in]{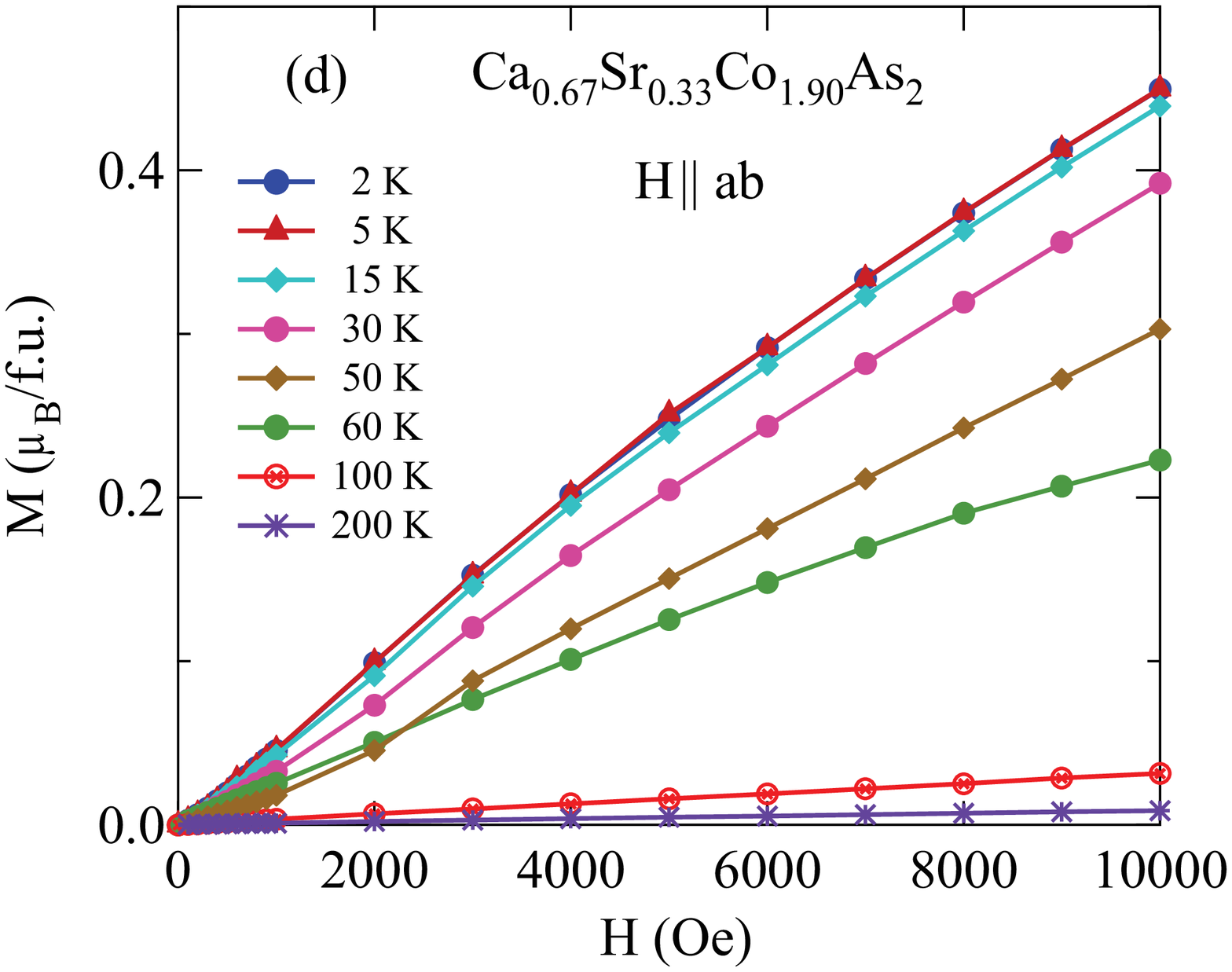}
\caption{(Color online) \label{Fig:Sr33MH} Isothermal magnetization $M$ of a ${\rm Ca_{0.67}Sr_{0.33}Co_{1.90}As_2}$  single crystal versus applied magnetic field $H$ measured at the indicated temperatures for $H$ applied (a)~along the $c$~axis ($H \parallel c$) and (b) in the $ab$~plane ($H \parallel  ab$). (c) and (d) are expanded plots at low fields of the data in (a) and~(b), respectively.  A weak metamagnetic transition occurs for $H\parallel ab$ at about 0.24~T at 50~K in~(d), which is converted to a spin-flop transition at about the same field with $H\parallel c$ at lower temperatures in~(c), consistent with the $\chi(T)$ data for $x=0.33$ in Fig.~1 of the main text. }
\label{Fig:Sr34_MH}
\end{figure}

\clearpage

\section*{${\rm\bf Ca_{0.60}Sr_{0.40}Co_{1.93}As_2}$}

\begin{figure}[h]
\includegraphics[width=4.5in]{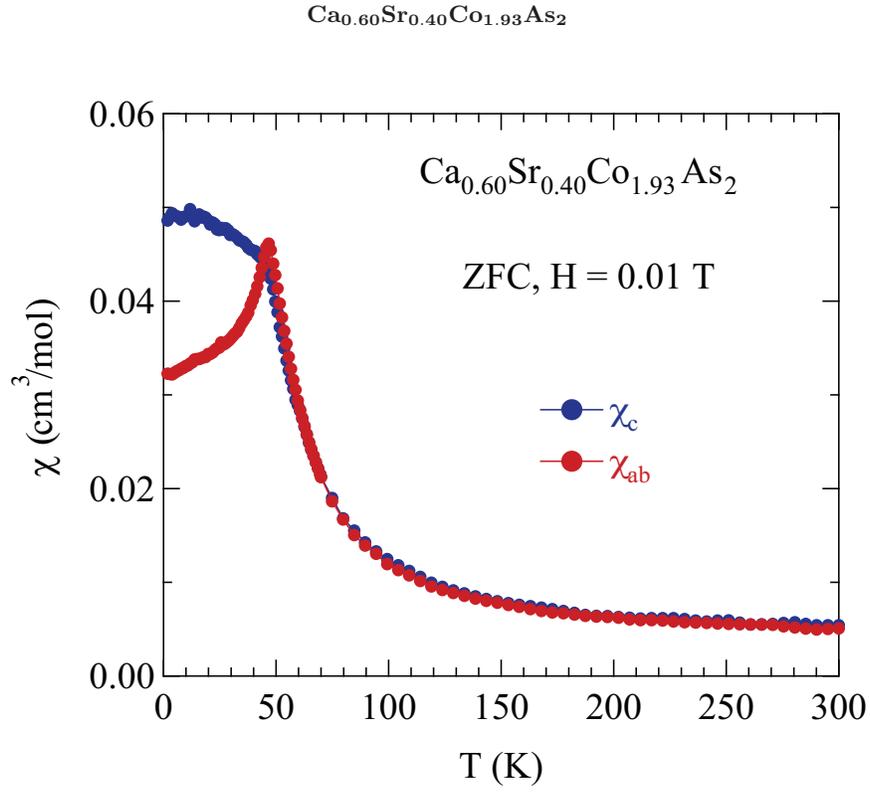}
\caption{(Color online) Zero-field-cooled magnetic susceptibility $\chi$ of a ${\rm Ca_{0.60}Sr_{0.40}Co_{1.93}As_2}$ single crystal versus temperature $T$ measured in a field $H = 0.01$~T = 100~Oe applied in the $ab$-plane ($\chi_{ab}$) and along the $c$-axis ($\chi_c$).}
\label{Fig:Sr40_Chi}
\end{figure}

\begin{figure}[h]
\includegraphics[width=3.5in]{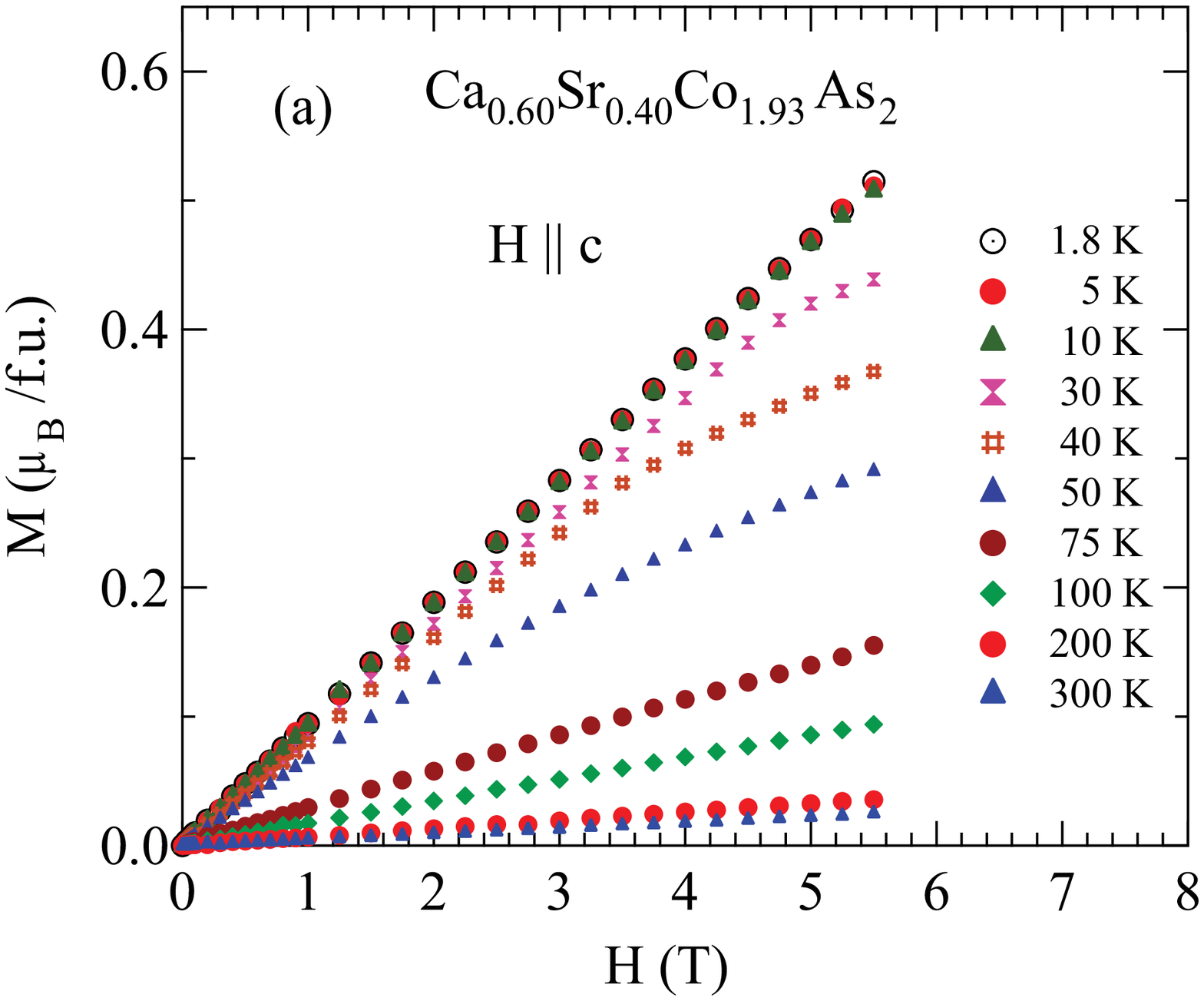}
\includegraphics[width=3.5in]{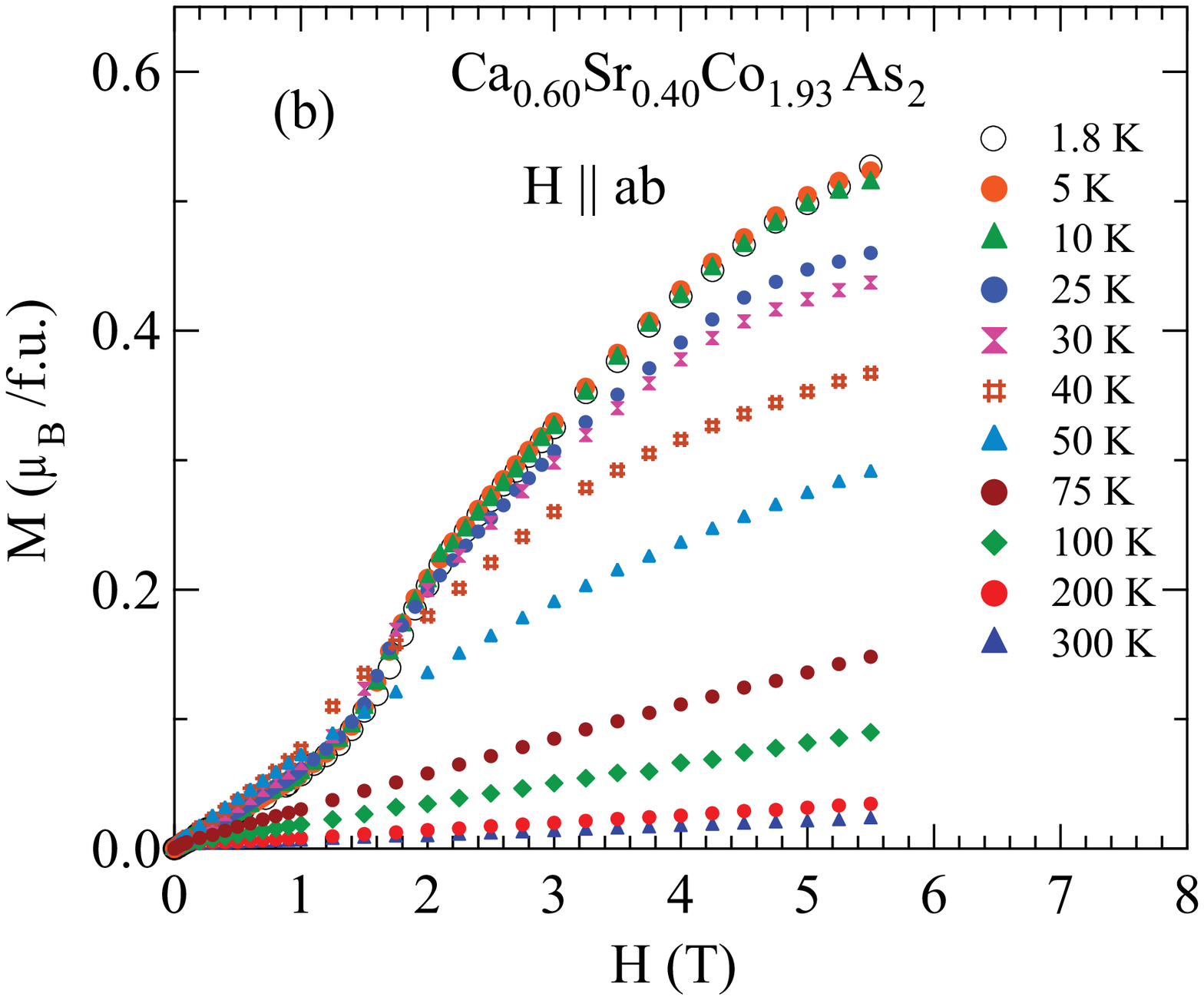}\includegraphics[width=3.5in]{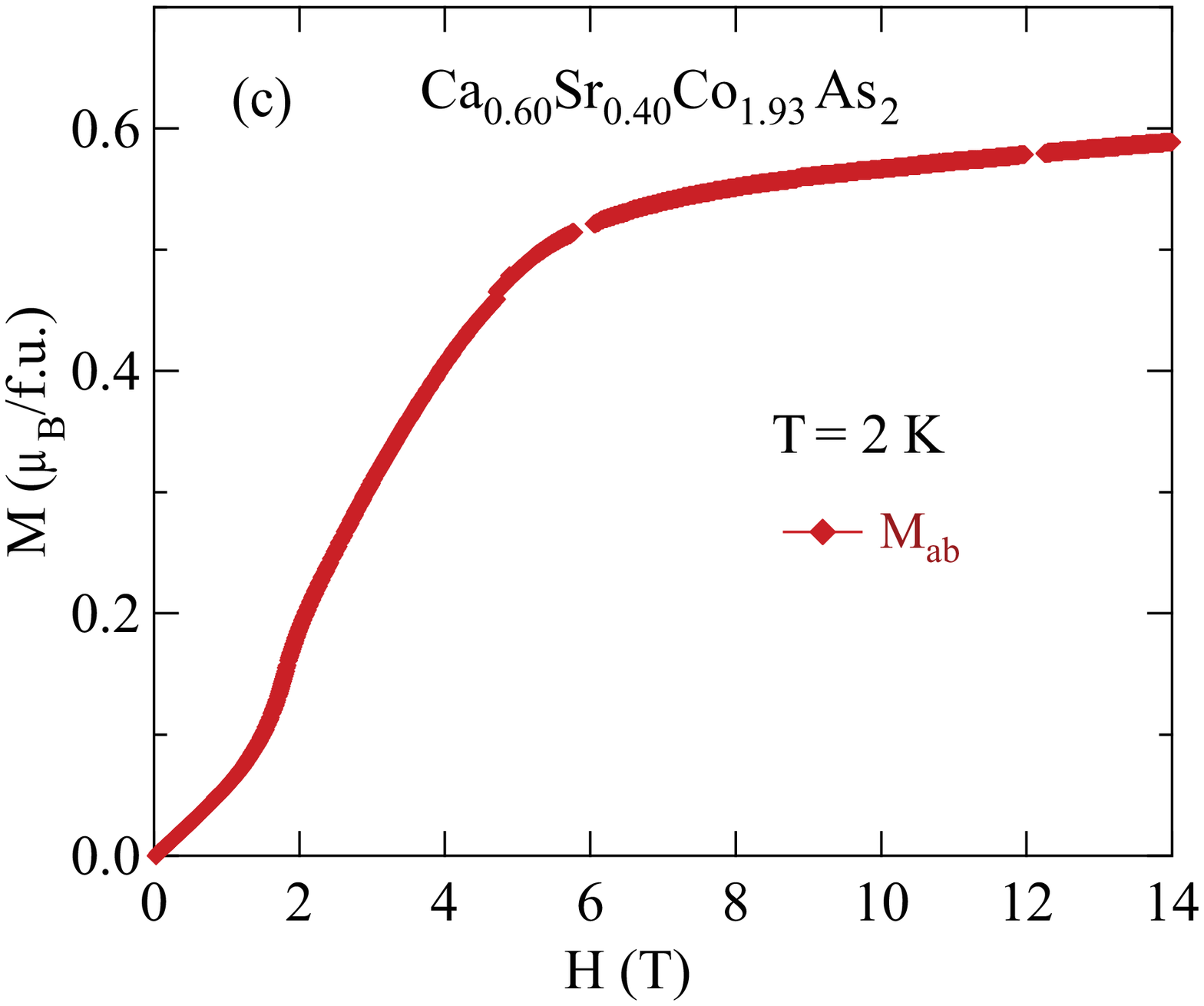}
\caption{(Color online) Isothermal magnetization $M$ of a ${\rm Ca_{0.60}Sr_{0.40}Co_{1.93}As_2}$  single crystal versus applied magnetic field~$H$ measured at the indicated temperatures for $H$ applied (a)~along the $c$~axis ($H \parallel c$) and (b) along the $ab$~plane ($H \parallel  ab$). (c)~$M$ versus $H$ measured upto 14~T at $T=2$~K for $H$ applied along the $ab$ plane.}
\label{Fig:Sr41_MH}
\end{figure}

\clearpage

\section*{${\rm\bf Ca_{0.55}Sr_{0.45}Co_{1.92}As_2}$}

\begin{figure}[h]
\includegraphics[width=4.5in]{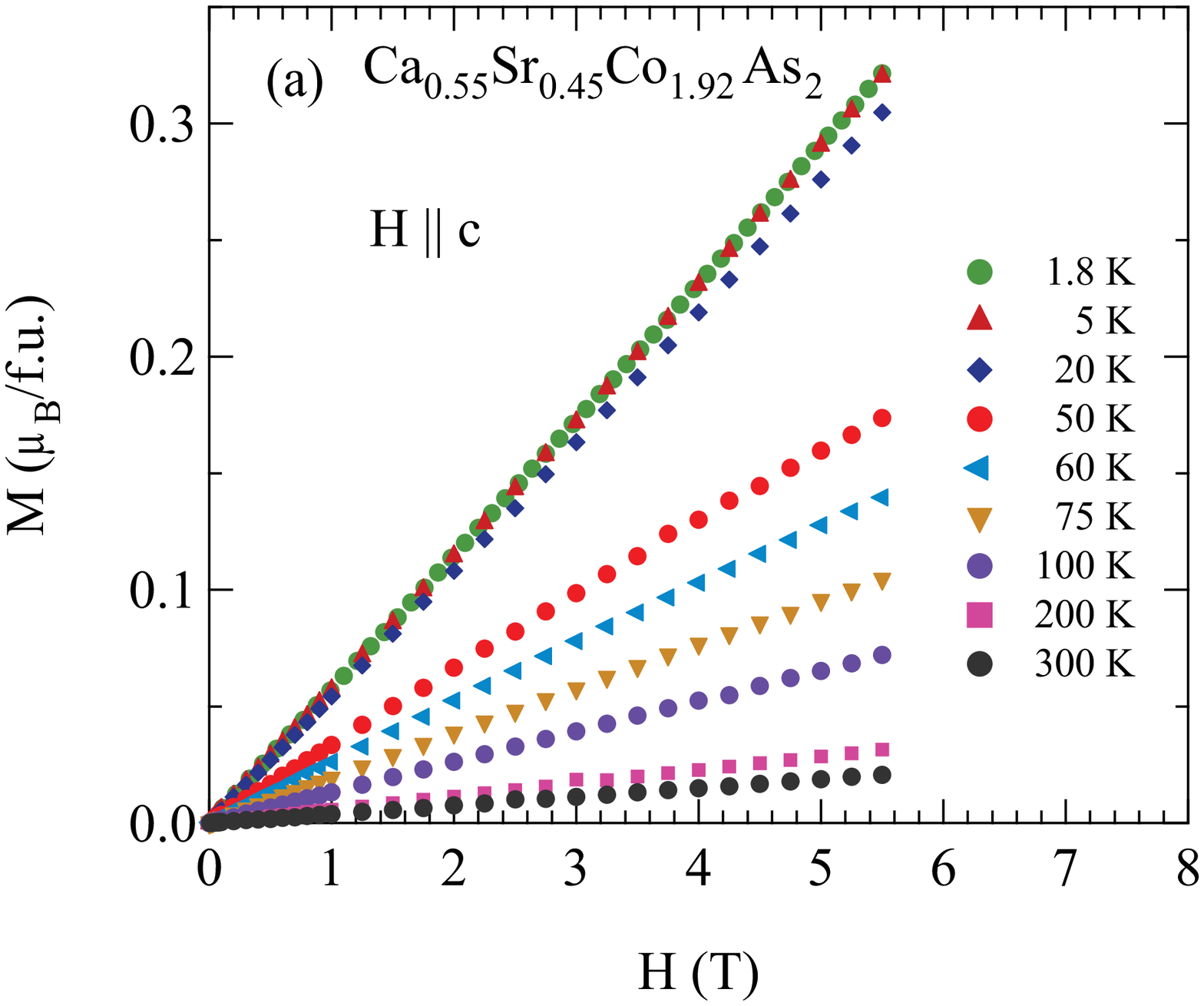}
\includegraphics[width=4.5in]{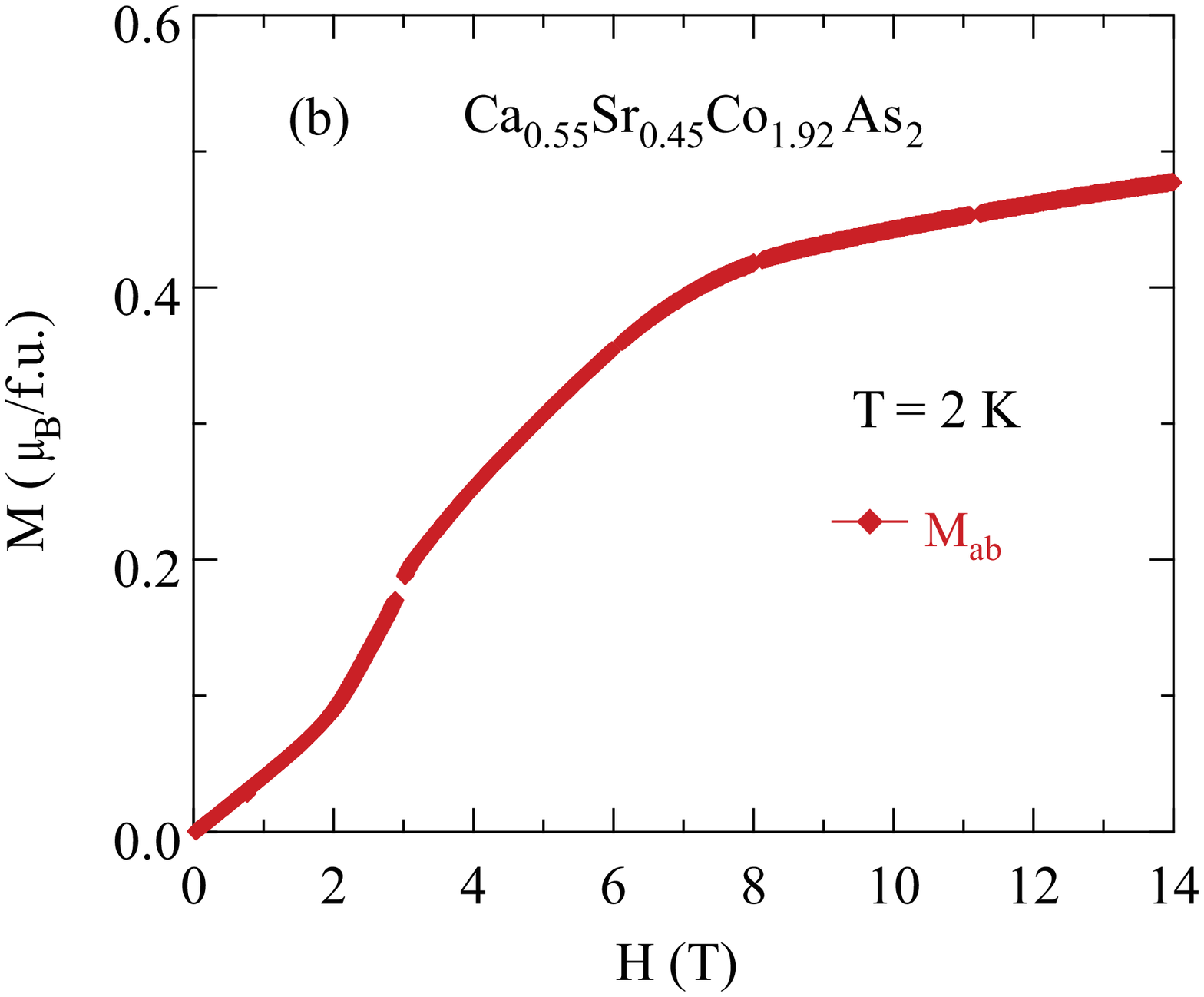}
\caption{(Color online) Isothermal magnetization~$M$ of a ${\rm Ca_{0.55}Sr_{0.45}Co_{1.92}As_2}$  single crystal versus applied magnetic field~$H$ measured at the indicated temperatures (a) for $H$ applied along the $c$~axis ($H \parallel c$). (b)~$M$ versus $H$ measured upto 14~T at $T=2$~K for $H$ applied along the $ab$ plane ($H \parallel ab$).}
\label{Fig:Sr44_MH}
\end{figure}

\clearpage

\section*{${\rm\bf Ca_{0.48}Sr_{0.52}Co_{1.92}As_2}$}

\begin{figure}[h]
\includegraphics[width=4in]{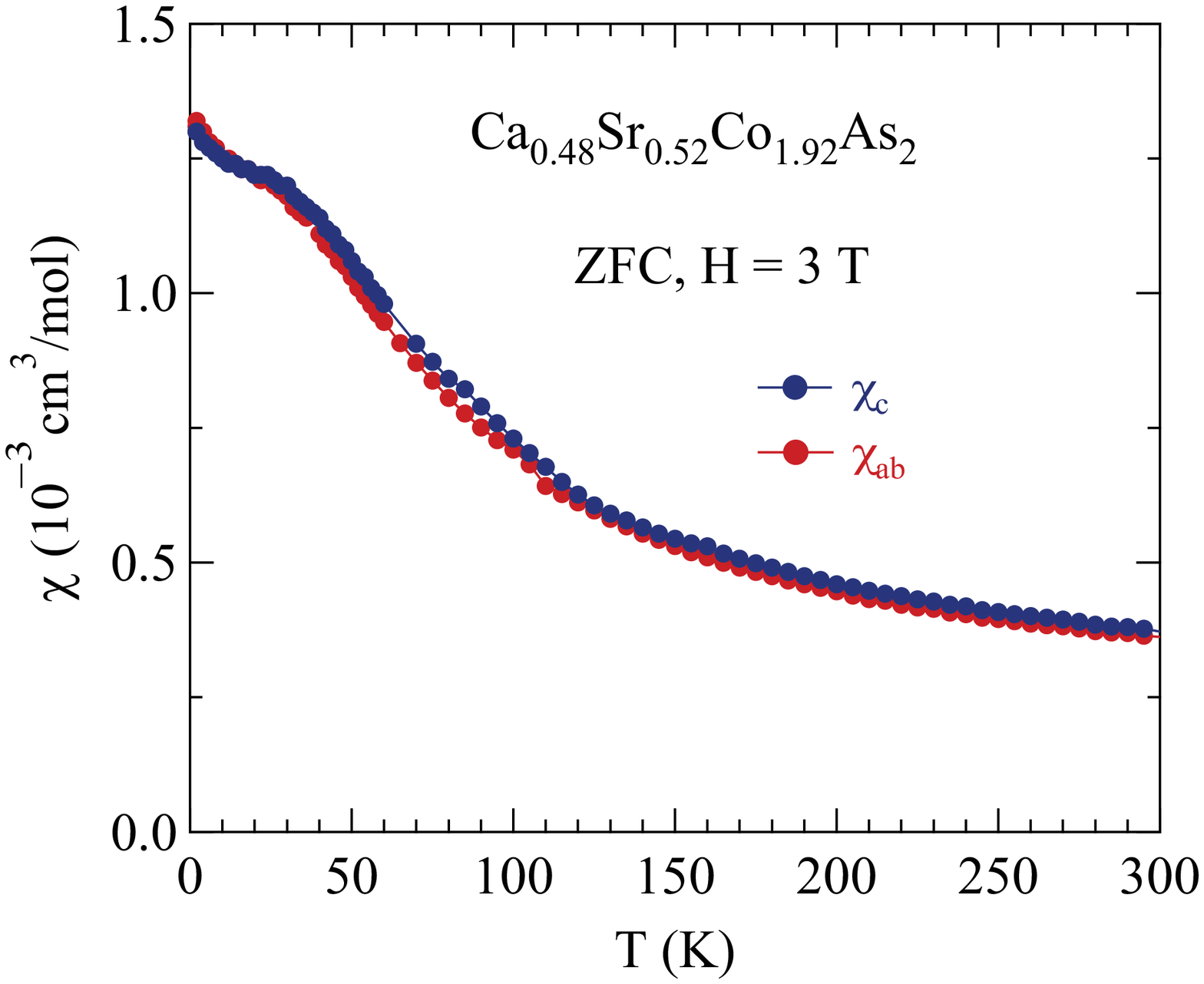}
\caption{(Color online) Zero-field-cooled (ZFC) magnetic susceptibility $\chi\equiv M/H$ of a ${\rm Ca_{0.48}Sr_{0.52}Co_{1.92}As_2}$ single crystal versus temperature~$T$ measured in a field $H = 3$~T applied in the $ab$~plane ($\chi_{ab}$) and along the $c$~axis ($\chi_c$). }
\label{Fig:Fig_Sr51_MT_3T}
\end{figure}

\begin{figure}[h]
\includegraphics[width=3.4in]{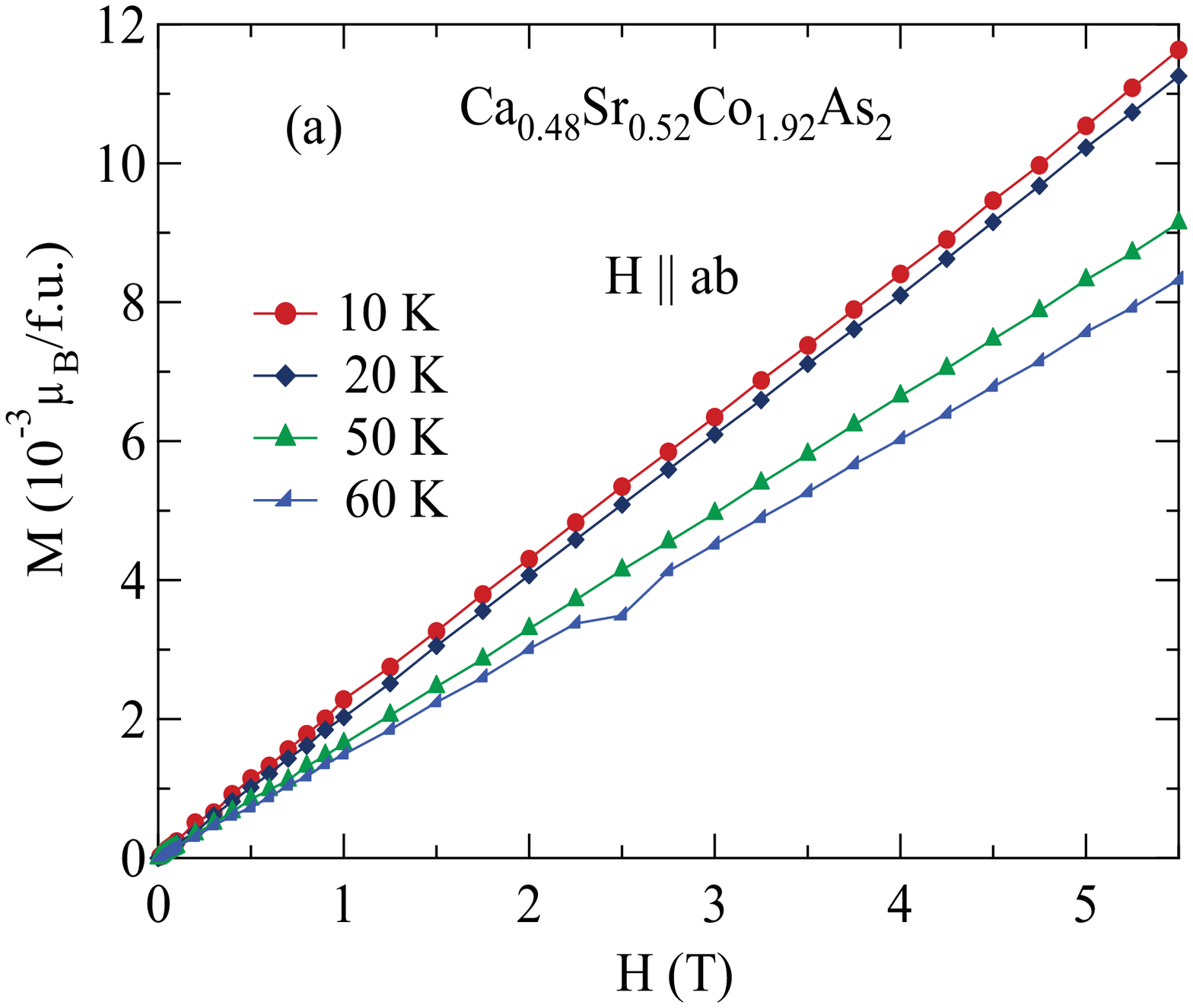}
\includegraphics[width=3.4in]{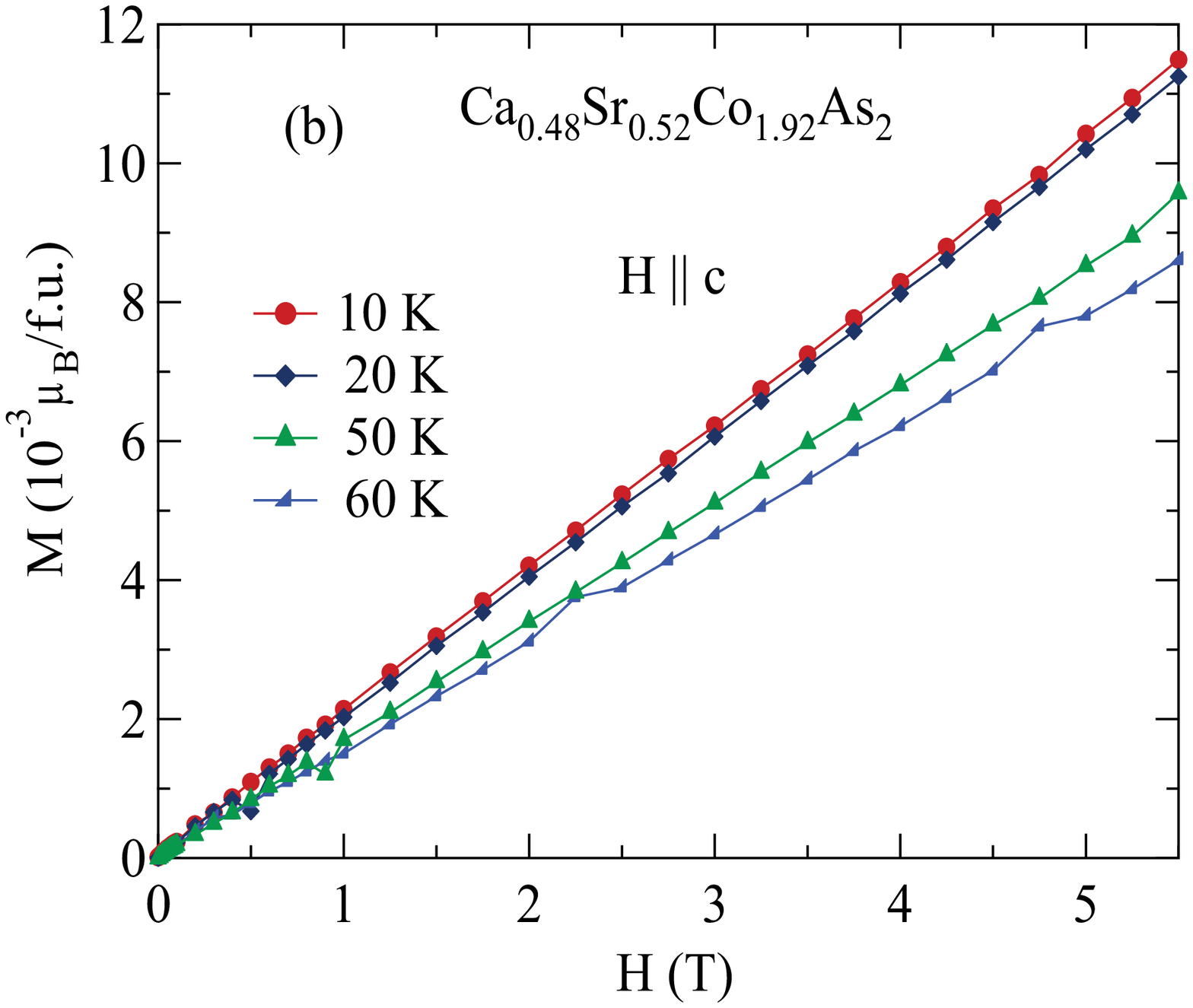}
\caption{(Color online) Isothermal magnetization~$M$ of a ${\rm Ca_{0.48}Sr_{0.52}Co_{1.92}As_2}$ single crystal versus applied magnetic field~$H$ measured at the indicated temperatures for $H$ applied (a)~along the $ab$~plane $(H\parallel ab)$ and (b)~along the $c$~axis ($H \parallel c$). }
\label{Fig:Fig_Sr51_MH}
\end{figure}

\clearpage

\acknowledgments

This research was supported by the U.S. Department of Energy, Office of Basic Energy Sciences, Division of Materials Sciences and Engineering.  Ames Laboratory is operated for the U.S. Department of Energy by Iowa State University under Contract No.~DE-AC02-07CH11358.\\

\begin{center}
\bf References
\end{center}

\vspace{0.1in}

\noindent
[40] APEX3, Bruker AXS Inc., Madison, Wisconsin, USA, 2015.

\noindent
[41] SAINT, Bruker AXS Inc., Madison, Wisconsin, USA, 2015.

\noindent
[42] L. Krause, R. Herbst-Irmer, G. M. Sheldrick, and D. Stalke, Comparison of silver and molybdenum microfocus X-ray sources for single-crystal structure determination, J. Appl. Crystallogr. {\bf 48}, 3 (2015).

\noindent
[43] G. M. Sheldrick.  SHELTX -- Integrated space-group and crystal-structure determination. Acta Crystallogr. A {\bf 71}, 3 (2015).

\noindent
[44] G. M. Sheldrick.  Crystal structure refinement with SHELXL.  Acta Crystallogr.~C {\bf 71}, 3 (2015).

\end{widetext}

\end{document}